\documentclass{aa}  
\usepackage{graphicx}
\usepackage{natbib}
\usepackage{amsmath}
\usepackage{txfonts}
\bibpunct{(}{)}{;}{a}{}{,} 
\newcommand{\zs}{Z$_{\odot}$}
\newcommand{\ms}{M$_{\odot}$}
\newcommand{\ls}{L$_{\odot}$}
\newcommand{\mpc}{M$_{\odot}$/pc$^2$}

\newcommand{\afe}{[$\alpha$/Fe]}
\newcommand{\hmol}{H$_{\rm 2}$}
\newcommand{\hatm}{H${\rm I}$}

\newcommand{\rd}{$\rm R_d$}
\newcommand{\rsol}{$\rm R_0$}

\begin{document}

   \title{Evolution of the Milky Way with radial motions of stars and gas }

   \subtitle{I. The solar neighbourhood and the thin and thick disks}

   \author{M. Kubryk
          \inst{1},          
          N. Prantzos
          \inst{1}
          \and E. Athanassoula
          \inst{2}
          }

   \institute{Institut d'Astrophysique de Paris, UMR7095 CNRS, Univ. P. \& M. Curie, 98bis Bd. Arago, 75104 Paris, France\\
              \email{kubryk@iap.fr,prantzos@iap.fr}
         \and
             Aix Marseille Universit\'e, CNRS, LAM (Laboratoire d’Astrophysique de Marseille) UMR 7326, 13388, Marseille, France\\
             \email{lia@lam.fr}
                          }

   \date{Received ; accepted }

 
  \abstract
   {We study the role of radial migration of stars on the chemical evolution of the Milky Way disk. }
   {We are interested in the impact of that process on the  local properties of the disk (age-metallicity relation and its dispersion, metallicity distribution, evolution of abundance ratios) and on the morphological properties of the resulting thick and thin disks. }
   {We use a model with several new or up-dated ingredients: atomic and molecular gas phases, star formation that depends on molecular gas, yields from a  recent homogeneous grid for low-mass and massive stars
and  observationally inferred SNIa rates. We describe  radial migration with parametrised time- and radius-dependent diffusion coefficients, based on the analysis of an N-body+SPH simulation. We also consider parametrised radial gas flows, induced by the action of the Galactic bar.}
   {Our model reproduces current values of most of the main global observables of the MW disk and bulge, and also the observed "stacked" evolution of MW-type galaxies. 
The azimuthally averaged radial velocity of gas inflow is constrained to less than a few tenths of km/s. Radial migration is  constrained by the observed dispersion in the age-metallicity relation. Assuming  that the thick disk is the oldest ($>$9 Gyr) part of the disk, we find that the adopted radial migration scheme can quantitatively reproduce the main local properties of the thin and thick disk: metallicity distributions, {"two-branch" behaviour} in the  O/Fe vs Fe/H relation and the  local surface densities of stars. The thick disk extends up to $\sim$11 kpc and has a scale length of 1.8 kpc, which is considerably shorter than the thin disk, because of the inside-out formation scheme. We also show how, in this framework,  current and forthcoming spectroscopic observations can  constrain the nucleosynthesis yields of massive stars for the metallicity range of 0.1 \zs \ to 2-3 \zs.}
   {}


   \maketitle

\section{Introduction}
\label{sec:Intro}
The Milky Way  (MW) offers the possibility of detailed observations of a large number of galactic properties, which are inaccessible in the case of other galaxies. Information on chemical composition and kinematics is now available for a few thousand stars of various ages, in the solar vicinity, across the MW disk and away from the Galactic plane. A large amount of information also exists for the gaseous content of the Galaxy (its molecular and atomic components and its chemical composition) as a function of galactocentric radius. 


The chemical properties of the MW  (local age-metallicity relation,  local metallicity distribution, abundance ratios vs metallicity, abundance profiles across the disk) have been extensively studied long before the era of large scale numerical simulations.  Such studies were performed with simple numerical models, either  for the solar neighbourhood or for the whole disk (with "independent ring" models) and revealed some key aspects of the chemical evolution of the Galaxy: the need for a supplementary source of Fe (beyond massive stars), namely SNIa, to reproduce the observed decline of O/Fe with metallicity; the need of a long-term early infall, to reproduce the early part of the G-dwarf metallicity distribution and the high present-day abundance of deuterium; the need for a radial variation in the efficiency of star formation (and/or the corresponding infall timescale) in order to obtain the observed gradients in the radial abundance profiles (e.g. \citealt{Pagel2009,Matteucci2012}). The aforementioned results are robust qualitatively but not quantitatively, because of large uncertainties in the observational data (e.g. dispersion in the age-metallicity relation, shape of the metallicity distribution) and also because of the poorly understood role of radial gaseous flows. Although  they are  justified on physical grounds 
(e.g. \cite{Lacey1985} ) radial flows 
were never shown conclusively to play an important role in the chemical evolution of the Milky Way, because of the impossibility to observe or to infer from theory the corresponding radial velocity profiles in the Galaxy.

The action of the bar can mix radially not only gas but also stars, and the effects of stellar radial motions on the abundance profiles have been studied to some extent with N-body+SPH codes by \cite{Friedli1993} and \cite{Friedli1994}. Observations in the 90s revealed that the  MW does have a bar \citep{Blitz1991}, but its origin, size and age are not well known yet; as a result, its impact on the evolution of the MW is difficult to evaluate quantitatively.

Independently of the role of the bar, \cite{SellwoodBinney2002}  show  that,  in the presence of recurring transient spirals, stars in a galactic disk could 
undergo important radial displacements. Stars 
found at corotation with a spiral arm may  be scattered to different galactocentric radii (inwards or outwards),  a process that preserves
overall angular momentum distribution and does not contribute to the radial heating of the stellar disk.
Using a simple model, they showed how this process can increase the dispersion in the local metallicity vs age relation, well above the amount due to the epicyclic motion.
This development paved the way for a large number of theoretical studies on radial migration, both with  N-body codes (e.g. \citealt{Roskar2008,Sanchez2009,Martinez2009,Sales2009,Roskar2010,minchevfamaey2010,Minchev2011,Brunetti2011,Minchev2012a, Grand2012,Baba2013,Bird2013,DiMatteo2013,KPA2013,Grand2014}) and with semi-analytical models \citep{Lepine2003, Prantzos09, SB2009, Minchev2013, Wang2013, Minchev2014}. Because of the difficulty  producing realistic MW-like disks, the former class of models focused mostly on generic properties of radial migration (origins of it and impact on some observables), while the latter focused exclusively on the properties of the MW.

\cite{Roskar2008}  investigated the implications of radial migration  for the chemical evolution of galactic disks with N-body+SPH simulations. The main effects they found and analysed are:  the  resulting dispersion in the age-metallicity relation, the broadening of the local metallicity distribution, the flattening of observed past abundance profiles and the flattening of the observed past star formation history. 

\cite{SB2009} introduced a parametrised prescription of
radial migration (distinguishing epicyclic motions 
from migration  due to transient spirals) in a semi-analytical chemical evolution code.
They suggest that radial mixing could also explain the formation of the Galaxy's thick disk, by bringing   a kinematically "hot" stellar population from the inner disk to the solar neighbourhood. 
That possibility was subsequently investigated
with N-body models, but controversial results have been  obtained up to now. While \cite{Loebman2011} find that secular processes (i.e. radial migration) are sufficient to explain the kinematic properties of the local thick disk, \cite{Minchev2012b} find this mechanism insufficient  and suggest that an external agent (e.g. early mergers) is required for that. This is still being debated,  e.g. \cite{Sales2009,Wilson2011,Navarro2011,Bekki2011,Brook2012, Forbes2012, Steinmetz2012, Liu2012,Bird2013, Kordopatis2013, Haywood2013, Roskar2013}.
 On the other hand,  using mono-abundance populations (i.e. defined in the plane of [O/Fe] vs. [Fe/H]), \cite{Bovy2012} conclude that the thick disk is not really a distinct component of the Milky Way,  as initially suggested in \cite{SB09b} (see also \cite{Rix2013} for a review).

\cite{minchevfamaey2010} suggest a different mechanism for radial migration than transient recurring spirals, namely resonance overlap of the bar and spiral structure \citep{Sygnet1988}. Tthis strongly nonlinear coupling leads to a more efficient redistribution of angular momentum in the disk  and produces a stellar velocity dispersion that increases with time,  in broad agreement  with local observations. 
This bar-spiral coupling was studied in detail by \cite{Shevchenko2011} and  \cite{Brunetti2011}. The latter study 
 found that the extent of radial migration also depends 
on the kinematic state of the disk and is reduced in the case of 
kinematically hot disks. They also show that radial migration can be assimilated to a diffusion process, albeit with time- and position-dependent diffusion coefficients. That idea was confirmed by the analysis of N-body+SPH simulations of a disk galaxy by \cite{KPA2013}: they extracted such coefficients from the simulation of an early-type barred disk and, applying them in a semi-analytical model of that same disk, they showed that all the main features of the N-body+SPH simulation can be reproduced to a good accuracy. They also showed that radial migration moves around not only "passive" tracers of chemical evolution (i.e. long-lived stars, keeping on the surfaces the chemical composition of the gas at the time and place of their birth), but also "active" agents of chemical evolution, i.e. long-lived nucleosynthesis sources (mainly SNIa producing Fe and $\sim$1.5 \ms \ stars producing s-process elements).


In this work we present a model for the evolution of the MW disk that includes  radial motions of gas and stars. 
Our treatment of radial migration of stars is a mixture of the techniques adopted in some pevious works in the field. As in \cite{SellwoodBinney2002} and \cite{SB2009} - but unlike \cite{Minchev2013,Minchev2014}  or \cite{KPA2013} - we consider separately the epicyclic motion of stars (blurring) from the true variation in their guiding radius (churning). For the former, we adopt an analytic formalism
based on the epicyclic approximation. For the latter, we are inspired by N-body+SPH simulations - as in \cite{Minchev2013} and \cite{KPA2013} -  and we  adopt a parametrised description, using time- and 
radius-dependent diffusion coefficients. In this way, we are able to quantitatively study the impact of epicyclic motion alone to the dispersion of the local age-metallicity relation and, of course, the collective impact of the two processes (blurring+churning).

The ingredients of our model are presented in Sec. \ref{sec:Model} and some of them are described in more detail in Appendices \ref{App:SFR} and \ref{App:Chem}. Some of the observational constraints are presented in Sec. \ref{subsec:ObsConstr}, while the adopted gas and SFR profiles are discussed in detail in Appendices \ref{App:Gas} and \ref{App:SFR}. The global evolution of the Galaxy (i.e. various quantities as a function of time and radius) is presented in Sec. \ref{sec:GlobEvol} and the results are compared to observations; in particular, in Sec. \ref{subsec:StarProfile} we discuss  the diffusion coefficients adopted in our model and the amount of the radial migration they produce along the disk. The results concerning the solar vicinity (age-metallicity relation and its dispersion, metallicity distribution, abundance ratios) are presented in Sects. \ref{subsec:AgeMet} and \ref{subsec:MetDist}. We then analyse the properties of the thick disk, which is  assumed here to be just the old part of the disk (age $>$9 Gyr) and show that this assumption leads to results that are in fair agreement with most of the observed chemical and morphological properties of the thick disk (Sec. \ref{subsec:ThinThick}).  A summary of the results is presented in Sec. \ref{sec:summary}.

\section{The model}
\label{sec:Model}

Our model is based on a considerably updated version of the "independent-ring" model for the MW presented in \cite{PranAub1995}, \cite{Prantzos_AubAud1996}, and \cite{BP99}. The main differences include:

- Star formation rate depending on molecular gas,  instead of total gas (App. \ref{App:SFR})

- A new routine for the SNIa rate, based on observations at various redshifts, instead of theoretical precriptions (App. \ref{App:Chem});

- Stellar yields from \cite{Nomoto2013}, extending to super-solar metallicities (App. \ref{App:Chem}), instead to those of \cite{WW95} which reach metallicities only up to solar;

-A new numerical scheme to handle chemical evolution   through single stellar populations (App. \ref{App:Chem}), instead of adopting the classical formalism (with integration over mass);

to which one should add:

-The introduction of radial gas flows (Sec. \ref{subsec:rad_gas_flow})

-The treatment of radial migration through blurring (Sec. \ref{subsub:blurring}) and churning (Sec. \ref{subsub:churning}).

\subsection{Building the Galaxy: Dark matter, baryonic infall and star formation}
 \label{subsec:MW_construction}

In our model, we construct the Galaxy "backwards", i.e. we are guided in the selection of the model parameters by  the present-day properties of the Milky Way (e.g. radial profiles of baryonic matter and rotational velocity, which also depend on the distribution of the dark matter halo).  Some of those properties play an important role in determining the extent of radial displacement of stars ; this is the role of  the rotation curve for blurring, for example (Sec. \ref{subsub:blurring}).
Unfortunately, 
determination of the structural parameters of the various components of the Galaxy still suffers from 
degeneracy problems; see e.g. the excellent summary of mass models of the Milky Way in Sec. 4 of \citet{Courteau2013}.

For the purpose of this work, we describe the Milky Way as a superposition of three components: a DM halo, a bulge and a disk. 
We note, however, that the actual distribution of the baryonic material within the inner 2 kpc is
much more complex than assumed here, because it can contain a number of components such as a boxy/peanut bulge, a disky bulge and a classical bulge as well as the inner extension of the Galactic disk. These have very different
shapes, kinematics and formation histories. 
Our model, which is essentially 1D, cannot describe all this complexity. We thus do not extend our study to  the region inwards of  two kpc, which is  designated as " the bulge" here (we use the term "disk" for the region outside 2 kpc).



The current  virial mass of the DM halo of the MW is typically estimated to $M_{DH}$=10$^{12}$ \ms, although variations by a factor greater than two around that value are found in the literature, e.g. \citet{Rashkov13} and references therein.
For the evolution of the DM halo we used the simulations of \cite{Li2007},  who calculated the growth  of DM haloes
in a $\Lambda$CDM model. We extracted  about 200 DM haloes with final mass of 10$^{12}$ \ms \ from that simulation and took an average over all masses
at each redshift. We adopted this smoothed evolution 
as a reasonable approximation for  the evolution of the  DM halo of the MW, at least for the past 8 Gyr, where no major merger is thought to have occurred. We do not account for any concentration effect due to the interaction
of baryons with the DM halo, that is, we assume that  the DM halo has, at all redshifts $z$,  
a Navarro-Frenk-White profile, with the central density  varying with time or redshift as to have the mass m$_{DM}(z)$ enclosed within the virial radius.


We assume that the MW is built gradually from gas infalling in the potential well of the DM halo.
with the radial profile for the  infall rate  a function of time.
 Once the time variation of infall in each zone $r$ is assumed (in most cases an exponential decay law, with a characteristic time scale $\tau(r)$), the gas  infall rate per unit area of the disk $F_gr,(t)$   (\ms/pc$^2$/yr) is constrained  by the requirement that its integral over time  equals the total baryonic (i.e. stars+gas) surface density $\Sigma_{Tot,Obs}(r,T)$ at present time $T$:
 \begin{equation}
 \int_0^T F_g(r,t) dt \ = \ \Sigma_{Tot,Obs}(r,T)
 \label{eq:infall}
\end{equation}  
In the case of independent-ring disk models, Eq. \ref{eq:infall} is used to fix  $F_g(r,t)$ uniquely and accurately. When radial migration is taken into account, the final baryonic profile  depends not only on the integral of $F_g(r,t)$ over time,
but also on the extent of migration. In those conditions, it becomes difficult to reproduce  $\Sigma_{Tot,Obs}(r,T)$ accurately   in the end of the simulation. 
Here we adopt - after some iterations - a profile
for $\int_0^T F_g(r,t) dt$
that depends on our migration coefficients:  the combination of the adopted infall and star formation prescriptions and the adopted radial migration scheme produces a quasi-exponential stellar profile in the end of the simulation. 

Concerning the infall timescales, we adopt $\tau$=2 Gyr for the bulge (hereafter taken as the region within $r$=2 kpc)   and a smoothly increasing function, reaching $\tau(20 \rm {kpc})$=8 Gyr for the disk ($r>$2 kpc). The adopted profile of $\tau(r)$ can be seen in the bottom right panel of Fig. \ref{Fig:TheorObsProf}.
  The composition of the infall is equally important when it comes discussing the evolution of abundances and abundance ratios in the MW disk. Observations are of little help at present: although  they generally find low metallicities for gas clouds presently falling to the MW disk ($\sim$0.1 \zs, e.g. \citealt{Wakker1999}), but they provide no information on the past metallicity of such clouds or on their abundance ratios.
Here we adopt the simplest possible (but still arbitrary) assumption, namely that the infalling gas always has primordial composition, i.e. only for $i$=H,D,$^3$He,$^4$He and $^7$Li is the term $F_{g,i}(r,t)$=$F_g(r,t) X_i(r,t)$ different from zero (where  $X_i(r,t)$ is  the mass fraction of isotope $i$). This assumption hardly affects the results for the chemical evolution of the disk,
but it allows for the existence of  disk stars with metallicities lower than [Fe/H]=-1 (see \citet{Bensby2013a} and references therein).

Most of the SFR laws adopted in semi-analytical models of galactic evolution  make use of the total gaseous profile of the disk. Based on  detailed, sub-kpc scale, observations of a large sample of disk galaxies, \cite{Bigiel08}  have found that the SFR appears to
follow the \hmol \  surface density, rather than the \hatm \ or the total gas surface density. 
Based on an updated set of observational data, \cite{Krumholz2014} concludes in his recent review that "the correlation between star formation and \hmol \ is the fundamental one". 

Following these studies, we checked whether such a correspondence 
between the adopted SFR and molecular gas profiles also holds in the MW disk. 
The comparison, as discussed in Appendix \ref{App:SFR} appears to favour that idea, as also noticed (albeit with older data sets) by \cite{BlitzRos06}. 
In view of this observational support, both for the MW disk (this work) and for external galaxies \citep{Bigiel08, Leroy08}, we adopted a star formation law that depends on the 
\hmol \ surface density. To calculate it in the model of chemical evolution, we adopted the semi-empirical method of \cite{BlitzRos06} to evalulate the  ratio $R_{mol}$=\hmol/\hatm \
in a galactic disk (see Appendix \ref{App:Gas}). We note that this method provides two more observational constraints to disk models, namely the present-day radial profiles of  atomic and molecular gas, which are not usually  considered (see, however, \cite{Ferrini1994,Molla2005}). We also notice that a SFR proportional to the surface density of molecular gas has recently been used in disk models by \cite{Kang2012}, \cite{Lagos2011} and \cite{Fu2013}.

\subsection{Stellar migration}
\label{subsec:starmigr}

The orbit of a test particle (star) in the potential of a galactic
disk is commonly described, to first order approximation, as the superposition
of a main circular motion (defining the guiding radius), and
harmonic oscillations called epicycles. Following \cite{SB2009},
we call "blurring" the radial oscillations around the
guiding radius and "churning" the modifications of the guiding
radius. Churning may occur through resonant interactions of the
star with non-axisymmetric structures of the gravitational potential
(spirals, bar), causing changes in the angular momentum of the stars.
The process conserves the overall distribution of angular momentum and does not add random motion, that is,  it does not "heat" the disk radially. In contrast, blurring conserves the angular momentum of individual stars but it heats the disk in the radial direction (the epicyclic radius increases with time).

In N-body schemes, the overall effects of both blurring and churning are naturally obtained. Semi-analytical models based on N-body simulations use the knowledge
obtained from the N-body run to describe the extent of radial migration, without distinguishing beween the
two effects, such as i  \cite{Minchev2013,KPA2013,Minchev2014}.
Here, we follow SB02 and \cite{SB2009} and we parametrise  blurring and churning separately.

\subsubsection{Epicyclic motion (blurring)}
\label{subsub:blurring}


Folowing \cite{SellwoodBinney2002}, we adopt a global
mixing scheme, where a star born at radius $r'$
at time $t'$ may be found at time $t$ (i.e. after time $\tau=t-t'$)
in radius $r$ with a probability $P(r,r',\tau)$ given
by a Gaussian function 
\begin{equation}
P(r,r',\tau)\ =\ (2\pi\sigma_{\tau}^{2})^{-1/2}\exp\left[-\frac{(r-r')^{2}}{2\sigma_{\tau}^{2}}\right]   
\label{eq:prob}
\end{equation}
 where $\sigma_{\tau}$ is the 1-$\sigma$ dispersion in the radial
displacement of the particles of age
$\tau$ at radius $r$. SB02 adopted, for illustration purposes,
 a time-independent dispersion amplitude for blurring of $\sigma_b$=0.16 $R_0$ in the the solar neighbourhood ($R_0$=8 kpc).
 We improve by adopting a time-dependent,  better  motivated and easy to implement description for the epicyclic motion 
based on the epicyclic approximation 
\citep{binney_tremaine2008}, where the oscillations of stars
around their guiding radius are described by the 
resulting dispersion in their radial position as
\begin{equation}
\left\langle \sigma_{b}^{2}\right\rangle \ = \ \frac{\left\langle \sigma_{\upsilon}(r)^{2}\right\rangle }{\kappa_r^{2}}
\label{eq:epic1}
\end{equation}
The frequency $\kappa_r$ of harmonic oscillations at radius $r$ is given as a function of the rotational velocity $V_C(r)$  by
\begin{equation}
\kappa_r \ = \ \sqrt{2} \ \frac{V_C(r)}{r}
\label{eq:epic2}
\end{equation}
 in the case of   $V_C(r)$=const. For the radial velocity dispersion  $\sigma_{\upsilon}(r)$
we adopt the presently (time $t$=T) observed one in the Milky Way \citep{Lewis1989}, parametrised as
\begin{equation}
\sigma_{\upsilon}(r,T) \ = \ 40 \ \rm e^{-(r-\rm R_{\odot})/8 \ \rm kpc} \  {\rm km/s}
\label{eq:epic3}
\end{equation}
For the time dependence of the radial velocity dispersion
we adopt
\begin{equation}
\sigma_{\upsilon}(r,t) = 
\max\left\{12, \sigma_{\upsilon}(r,\rm T) \left( \frac{t}{T} \right)^{0.33} \right\} {\rm km/s}
\label{eq:age_vdisp}
\end{equation}
which agrees with, for example, the evaluation of \cite{Holmberg2007} for local stars. The time dependence of the rotational velocity $V_C(r,t)$ is obtained self-consistently from our model.  The probability distribution obtained by Eq. \ref{eq:prob} is not
symmetric with respect to the birth radius $r'$ of the stars, because the dispersion $\sigma_{\tau}(r)$
depends on the  final radius $r$ through Eqs. \ref{eq:epic1},
\ref{eq:epic2} and \ref{eq:epic3}.

We notice that \cite{SB2009} also treated blurring and churning separately. For the former,  they adopted a more physical treatment than ours, starting from first principles. They used the distribution function of angular momentum of stars and  made plausible assumptions on the radial and temporal dependence of the stellar radial velocity dispersion  to describe its evolution.   Our heuristic formulation of blurring approximates the results of \cite{SB2009} rather well, as discussed in Sec. \ref{subsec:StarProfile} and Fig. \ref{Fig:BlurChurn} (middle panel).

\subsubsection{Radial migration (churning)}
\label{subsub:churning}

  In this work, we adopt a probabilistic description of churning, {\it \`a la} SB02, 
where the amplitude of the radial mixing of stars of age $\tau$ found at radius $r$ 
(Eq. \ref{eq:prob}) is the sum of the blurring ($\sigma_b$) and churning ($\sigma_c$) terms:

\begin{equation}
 \sigma_{\tau}=(\sigma_b^2+\sigma_c^2)^{1/2}
\label{eq:churning}
\end{equation}
 As  discussed in Sec. \ref{sec:Intro}, that parametrization  of churning as a diffusion processes appears to be supported by the analysis of numerical simulations by \cite{Brunetti2011} and \cite{KPA2013}. The method is similar to the one adopted by \cite{Minchev2013}, who sampled the stellar positions of an N-body simulation and applied the results in an independent semi-analytical model of the chemical evolution of the MW. These methods differ qualitatively from  \cite{SB2009},  who adopted a local scheme
(in which only stars from up to the second-nearest neighbouring zones can exchange
places during a time step) with a probability adjusted to reproduce
some observables in the solar neighbourhood, such as the metallicity distribution.

The coefficients describing the amplitude of radial migration are extracted 
from the N-body+SPH simulation analysed in \cite{KPA2013}.  We note that  \cite{KPA2013} extracted total (i.e. churning+blurring) coefficients, while here we extract  coefficients for churning only:  we follow the variations of the angular momentum of stars, which correspond to variations in their guiding radius. We fit the corresponding radial distributions with Gaussian functions{\footnote{The adopted gaussian functions are just convenient fitting functions to the results of N-body simulation. } of 1-$\sigma$  width as a function of birth radius $r_i$ and time $t$ since stellar birth :
\begin{equation}
\sigma_c(t,r_{i}) = a(r_{i}) t_{KPA2013}^{N(t)} + b(r_{i}) \ \ \ \ \ \  \ t_{KPA2013}> {\rm 1 Gyr}
\label{eq:new_sigma}
\end{equation}
and we find  $N\sim$0.5 for $t>$1 Gyr.  However, the disk galaxy analysed  in \cite{KPA2013} has a strong and long bar, reaching a length of $\sim$8 kpc in the end of the simulation at 10 Gyr. In contrast, the Milky Way has a small bar, not necessarily as old as 10 Gyr (see next section). Its effect on churning will then be smaller than in the aforementioned simulation. Indeed, if we adopt the coefficients of \cite{KPA2013} we obtain a large dispersion in the local age-metallicity relation in the framework of our model. We chose then to reduce these coefficients of \cite{KPA2013} while keeping their temporal and radial dependence, by applying  the transformation 
\begin{eqnarray}
\sigma_c(t,r_{i})_{this \ work} = & \sigma_c(t/5+{1 Gyr},r_{i})_{KPA2013} & t>{\rm 1 Gyr} \\
                                &  0 & t<{\rm 1 Gyr} 
\label{eq:new2_sigma}                                
\end{eqnarray} 
 i.e. we assume that the evolution at, say, $t\sim$10 Gyr in our model is similar to the one at $t\sim$3 Gyr in the simulation of \cite{KPA2013}.
Physically, this corresponds to the epoch where the bar in N-body the simulation has similar size to the one of the current MW bar  and other key properties of the disk  (surface density profile, scalelength, 
presence of spiral arms) are similar to the ones of the present day MW (see Figs. 1 and 2 in \cite{KPA2013}).  As we shall see (Sec. \ref{sec:LocalEvol}),  this transformation leads to acceptable results for the observed dispersion in the local age-metallicity relation.  We checked that slightly different values of the churning coefficients, corresponding to the transformation
t to t/6 and up to t to t/4,  produce acceptable results for the
dispersion in the age-metallicity relation.
The churning coefficients  of our model are presented, discussed and compared to those of \cite{SB2009} in Sec. \ref{subsec:StarProfile}.

The two migration mechanisms are treated independently of each other in our model, meaning that we can consider the effects of blurring or churning  alone.  This allows us to  evaluate the impact of each one of them on the radial mixing of the disk and on the resulting dispersion in  the age-metallicity relation in the solar neighbourhood (see Sec. \ref{sec:LocalEvol}). 
Of course, both of them are considered in our baseline model, .

 The mechanism of radial migration discussed in this section, is not applied to the gas in our model. In contrast to the dissipationless stellar fluid, gas is dissipative and  affected little by that mechanism. This can be seen e.g. by comparing Figs. 4 (stars)  and 5 (gas) in \cite{KPA2013}. The bar, however, drives gas inwards and we describe this radial inflow in the next section.

 \subsection{Radial gas flows}
 \label{subsec:rad_gas_flow}
 
 The Ppioneering work of \cite{Tinsley1978} and \cite{Mayor1981} emphasized the potential importance of radial gaseous flows for the chemical evolution of galactic disks. \cite{Lacey1985} presented a systematic investigation of the causes of such flows: i) viscosity of the gaseous layer of the disk, ii) mismatch of angular momentum between the gas of the disk and the gas infalling on it and iii) gravitational interactions between the gas and a bar or a spiral density wave in the disk. They found that in all cases it is difficult to predict  the magnitude and the profile of the corresponding inflow velocity and explored  the impact of such effects on the chemical evolution of the Galaxy with parametrized calculations. 
 

 Among the alleged causes of radial inflows, the impact of a galactic bar is well established both from simulations and from observations. Numerical simulations \citep{Athanassoula1992,Friedli1993} showed  that the presence of a non-axisymmetric potential from a bar can drive important amounts of gas inwards of corotation (CR)  fuelling star formation in the galactic nucleus, while at the same time gas is pushed outwards outside corotation. In a disk galaxy, this radial flow mixes gas of metal-poor regions into metal-rich ones (and vice-versa). 
Bars were  previously believed to flatten
the disk chemical abundance profile  because of the
 streaming motions they induce, as shown by \cite{Friedli1994,Zaritsky1994,Martin1994,Dutil1999}. But
recent studies with two-dimensional (2D) higher spectral and
spatial resolution integral field units show that there are negligible
differences in abundance gradients between barred and
unbarred galaxies, e.g. \cite{Sanchez2012}.  
In our recent study with a high-resolution N-body+SPH simulation of an isolated barred disk \citep{KPA2013}, we find that the bar indeed drives  large amounts of gas from corotation inwards, but no significant gas displacement occurs outside corotation  (see Fig. 5 in \cite{KPA2013}). It seems now that bars may
be changing the chemical abundance profile inside the corotation
radius but they  they only have a small impact outside the bar itself.

One of the (many) difficulties of introducing the effect of a bar on the radial gas flows of a galactic disk in semi-analytical models, is the  uncertainty on its strength and length evolution. 
The  length of the MW bar is estimated to 2.5-3 kpc \citep{Babusiaux2005,Bobylev2014}, although higher values have also been reported \citep{Cabrera2007}.

For the purpose of this work, we adopt a radial inflow velocity profile induced by a bar of current size $R_B$=3. kpc, having a corotation radius at $R_C$=1.2 $R_B$,  which is a typical relation between bar length and  corotation radius (e.g. \citealt{Athanassoula1992}).
The flow outside corotation is outwards and we assume it extends up to
the outer Lindblad resonance (OLR), located
at radius $R_{OLR}\sim$ 1.7 $R_C$  (e.g. \citealt{Athanassoula1982}), which corresponds to $\sim$6.2 kpc today. Those values are somewhat less than those adopted in the recent work of \cite{Monari2013,Monari2014}. The flow velocity profile $\upsilon_f(r)$ has azimuthally averaged absolute velocities of a few tenths of  km/s,  similar to the one of \cite{Portinari2000}.  The adopted gas velocity profile is displayed in Fig. \ref{Fig:RadFlow}.

Our treatment of infall and radial flow is not self-consistent: 
as \cite{Mayor1981} and \cite{Lacey1985} have pointed out, the two are coupled through conservation of angular momentum. However, a quantitative treatment of the effect implies that the angular momentum of the infalling gas is known, but this is not the case, since the rotational profile of the infalling gas is unknown. A parametrised exploration of that effect is performed in the recent work of \cite{Bilitewski2012}. Here, we simply ignore that effect, implicitly assuming that the infalling gas has the same angular momentum as the disk gas at the accretion radius.

\subsection{Chemical evolution}
 \label{subsec:Chem_evol_model}

 Radial migration introduces important modifications to the treatment of chemical evolution in semi-analytical models. As already emphasized in \cite{KPA2013}, long-lived stars have  enough time to migrate away from their birth place before dying and releasing their gas there. This concerns not only the "passive" tracers of chemical evolution (low-mass stars carrying  the chemical composition of their birth place on their surface), but also "active" agents of chemical evolution: interesting metal producers like SNIa (for Fe-peak nuclei) and stars of $\sim$1.3-2 \ms \ ( producers of s-nuclei);  "sinks" of isotopes (i.e. long-lived stars ejecting material poor in a given isotope), which deplete  deuterium and dilute the other metal abundances. This imposes a coupling of the various radial zones of the model, through the transfer probabilities described in Sec. \ref{subsec:starmigr}.
Furthermore, while in the independent-ring models for disk evolution, one may work with surface densities of all extensive properties as functions of galactocentric distance $r$, any coupling of the rings (through gaseous radial inflows or migration of stars) makes it necessary to work with  properties integrated over the whole ring of radius $r$. In the following we shall consider  the  equations of chemical evolution by integrating all extensive quantities over the surface area of the ring centered at $r$
\begin{equation}
A(r) \ = \ \pi \ [ (r+dr/2)^2-(r-dr/2)^2]
\end{equation}
The mass of gas in the ring is, obviously, $m_g(r)=A(r) \Sigma_g(r)$, where $\Sigma_g(r)$ is the gas surface density,  and similar expressions hold for all other extensive quantities. The evolution of the mass (in \ms) of a given isotope $i$ with mass fraction $X_i$ in the  zone $r$  is given by
\begin{equation}
\frac{d(m_gX_i)}{dt} \ = \ -\psi X_i \ + \varepsilon_i \ + f_i \ + \ s_i
\end{equation}
where 

- $\psi(r,t)$  is the star formation rate in the whole ring ; 

- $\varepsilon_i(r,t) $ is the rate  of isotopic mass release in $r$ by stars produced in all previous epochs everywhere in the disk (including  zone $r$) and are found in $r$ at time $t$; 

- $f_i(r,t)$ is the $net$ rate of isotope $i$ entering zone $r$ from outside the disk (i.e. any infalling minus outflowing material); and 

- $s_i(r,t)$ is the     $net$ rate of isotope $i$  brought in zone $r$ (i.e. gas entering minus gas leaving that zone) because of  radial flows from adjacent disk regions.

The first term of the right-hand member  $\psi(r,t)$=$A(r) \Psi(r,t)$ has already been discussed in Sec. \ref{subsec:MW_construction}: the star formation rate density $\Psi(r,t)=\alpha \Sigma_{H2}(r,t)$ is proportional to the surface density of the molecular gas (Appendix \ref{App:SFR}). 

Since we do not consider outflows from the disk in this work, the term $f_i(r,t)$=$A(r) F_{g,i}(r,t)$ simply represents the infalling material in the disk and it is always positive. 
The characteristic timescale $\tau(r)$ is displayed graphically in Fig. \ref{Fig:TheorObsProf} (bottom right panel).

The term $\varepsilon_i(r,t)$ is calculated as a sum over all zones $N_{r'}$ (including zone $r$) and all times $t'\leq t$ of the ejection rate of isotope $i$, $E_i(r',t-t')$ from a single stellar population SSP formed at time $t-t'$ in zone $r'$ and found in zone $r$ at time $t$ with a probability $P$:
\begin{equation}
\varepsilon_i(r,t)  \ =  \ \int_0^R \int_0^t dt' \psi(r',t') E_i(r',t-t') P(r,r',t-t') dr'  
\end{equation}
where $\psi(r',t') dt'$ is the stellar mass created in the annulus $r'$ at time $t'$ during the timestep $dt'$ and  $E_i(r',t-t')$ is the ejection rate of isotope $i$ at time $t$ from a SSP of unit mass formed at time $t'$ with the metallicity $Z(r',t')$ of the gas of zone $r'$ at that time. The latter quantity depends on the adopted stellar initial mass function and the adopted stellar yields and  is extensively discussed in Appendix \ref{App:Chem}, where we provide information about the SNIa rates and yields. We consider 82 isotopes of all elements from H to Ge, summing their abundances at each timestep in order to calculate the abundances of the corresponding 32 elements. The term $E_i(r',t-t')$ also includes the contribution from SNIa.

Radial migration is implemented by multiplying the mass of stars created  in time $t'$ in zone $r'$ ($M_s(t',r')=\Psi(t',r') dt$ that are still alive after time $\tau=t-t'$ with the probability $P(r,r',\tau)$ in order to find the fraction of those stars that have migrated to zone $r$ after $\tau$. 
They carry  the chemical composition $X_i(r',t')$ but also release material corresponding to that composition and to the age $t-t'$. The probabilities $P(r,r',\tau)$ are normalized to $\int_0^R 
P(r,r',t-t') dr'$=1 to insure mass conservation. 

The use of the SSP formalism leads to the creation of  "star particles" of variable mass $dm_s(r,t)=\psi(r,t)dt$ endowed with the set of  chemical abundances $X_i(r,t)$ corresponding to the place and time of their formation. One may  then use  the same tools for the analysis of the results as in the case of the N-body simulations, and obtain a self-consistent comparison of results between semi-analytical models and N-body simulations, as in \cite{KPA2013}.

Although we use two phases of the ISM, the ejecta of the stars are uniformly and "instantaneously" mixed locally in the total amount of gas. 
The reason is that the timestep in our model ($\sim$30 Myr) is  longer than the 
timescales of the mixing of the ejecta (a few Myr, typically) or the timescales for the 
survival of molecular clouds. Thus, the ISM in each zone (\hatm \ and \hmol) is characterised by a unique composition at each timestep.

\subsection{Observational constraints for the Milky Way}
\label{subsec:ObsConstr}

\begin{table}
\caption{\label{Tab:Table_GlobalObs}{Global observational constraints for the MW}
}
\begin{tabular}{lcc}
 \hline \hline
   &   Disk ($>$2 kpc) &    Bulge ($<$2 kpc) \\
   \hline
   
  Stellar mass (10$^{10}$ \ms) & 3.-4.  (1) & 1.-2. (1)\\
  Gas mass$^a$ (10$^9$ \ms) & 8.1$\pm$4.5 (2) & 1.1$\pm$0.8 (3)  \\
\hatm  \ mass (10$^9$ \ms)& 4.9$\pm$2.5 (2) & 0.005 (3) \\
\hmol  \ mass (10$^9$ \ms)& 0.9$\pm$0.4  (2) & 0.05  (3) \\
SFR (\ms/yr) & 0.65-3  (4) &  --- \\
Infall rate (\ms/yr) & 0.6-1.6 (5,6) &  --- \\
CCSN rate  (per 100 yr) & 2$\pm$1 (7) &  0.23$\pm$0.1  (7) \\
SNIa rate  (per 100 yr) &  0.4$\pm$0.2 (7) & 0.14$\pm$0.06 (7) \\
\hline
 
\end{tabular}
  
  a: Total includes 0.28 of He by mass fraction. \\
  {\it References}:(1) \cite{Flynn2006}; (2) This work (Appendix \ref{App:Gas});   (3) \cite{Ferriere2007}; (4) 
  \cite{Robitaille2010,Chomiuk11}; (5) \cite{Marasco2012}; 
  (6) \cite{Lehner2011};
  (7) \cite{Prantzos2011}.
\end{table}

A successful model of the chemical evolution of the Milky Way, especially one involving a large number of parameters (such as all semi-analytical 
models that feature radial motions of stars and gas), should satisfy a large number of 
observational constraints, both global (concerning the whole Galaxy) and local ones (concerning the solar neighbourhood). Unfortunately, the values of most of the observational constraints 
depend heavily on underlying model assumptions. Thus, the mass of the bulge is estimated to be as low as 9$\times$ 10$^9$ \ms \  \citep{McMillan2011} or as high as 2.4$\times$ 10$^{10}$ \ms \ \citep{Picaud2004}. Similar uncertainties affect the stellar disk,  \citep{Flynn2006,McMillan2011}, concerning its scalelength (from $\sim$2 to more than 3 kpc) and total mass (from $\sim$3 to more than 5$\times$ 10$^{10}$ \ms. For the purpose of this work, we adopted the results of the analysis of \cite{Flynn2006} (their Fig. 15), showing that for a local stellar surface density of 39 \mpc \ (35.5 \mpc \ "counted", plus an assumed  10\% enhancement from azimuthal average of spiral arms) and for scalelengths in the range 2.2-3.5 kpc,  the mass of the disk lies in the range 3-4 10$^{10}$ \ms and the one of the bulge in the range 1-2 10$^{10}$ \ms. The total stellar mass (bulge+disk) is much better constrained, $\sim$5$\times$ 10$^{10}$ \ms.

In Table \ref{Tab:Table_GlobalObs} we present the basic observational facts concerning the current amounts of gas and stars in the bulge and the disk of the Milky Way. 
We also present adopted values and references for the rates of star formation, infall (fairly uncertain), core collapse supernovae (CCSN) and thermonuclear supernovae (SNIa). Radial profiles for all those quantities (except the uncertain profile of the infall rate and the unknown ones
 of supernovae rates) also
constitute important constraints for the models and  are discussed  in Appendices \ref{App:Gas} and \ref{App:SFR}. We include the rotational velocity of the gas in the constraints, since radial migration changes the distribution of the baryonic component and thus affects the rotation curve.
 Radial abundance profiles
of stars and gas constitute equally important constraints, but they also depend on adopted nucleosynthesis yields. Here we consider only the O and Fe profiles and  leave a detailed discussion of all other elements for an accompanying paper.

\begin{table}
\caption{\label{Tab:Table_LocallObs}{Local observational constraints 
(surface densities of gas and stars - in  \mpc \  - at R$_0$=8 kpc)}
}
\begin{tabular}{lcr}
 \hline \hline
  Stellar mass$^a$  & 38$\pm$2  &  {\it (1)} \\
  Gas mass$^b$  & 10.3$\pm$3.& {\it (2)} \\
\hatm  \ mass & 6.1$\pm$2.& {\it (2)} \\
\hmol  \ mass & 1.3$\pm$0.7 & {\it (2)}  \\

\hline 
\end{tabular}  

  a: Total stars, includes stellar remnants \\
  b: Total gas, includes 0.28 of He by mass fraction. \\
  {\it References}:(1) \cite{Flynn2006,Bovy2013}; (2) This work (Appendix \ref{App:Gas}).
\end{table} 

The observational constraints for the solar neighbourhood include the "classical" ones: local amounts of gas and stars (Table \ref{Tab:Table_LocallObs}), age-metallicity relation, metallicity distribution, and abundance ratios vs metallicity. 
The introduction of stellar radial migration makes it possible to use some new constraints:  \cite{SB2009} suggest that this ingredient may help explain quantitatively the observed 
"two-branch" behaviour of  the [O/Fe] vs [Fe/H] relation, namely the local thick vs thin disk dichotomy. We confirm this here, using large data sets from recent surveys. We also use the observed (albeit yet uncertain) dispersion in the local age-metallicity relation
as a supplementary constraint of our model. We  refer to all those constraints in Sec. \ref{sec:LocalEvol}.  We would like, however, to draw attention to the fact that comparing model results to observations is not a straightforward enterprise, because of various selection biases affecting the latter (magnitude-limited, kinematic etc.). In our case,
all  local observables of  this model concern the so-called "solar cylinder", which  
is all stars found
in the end of the simulation in a cylinder of radius 0.25 kpc (half the size of our radial bin), perpendicular to the Galactic plane and centred on the solar position, at Galactocentric distance $R_0$=8 kpc. In those conditions, a successful comparison to observations does not  imply that the model is necessarily correct, only that it possesses potentially interesting features. The wealth of current and forthcoming data (with e.g. RAVE, LAMOST, GAIA etc.) will make it necessary to adopt in the models  the same selection criteria as the observational surveys, 
to draw meaningful and quantitative conclusions.

Finally, we also compare our results to the "average histories" of mass building of Milky Way type galaxies 
(total stellar mass of 5 10$^{10}$ \ms) from the recent analysis of HST+SDSS data from \cite{Dokkum2013}.

\begin{figure}
\begin{center}
\includegraphics[width=0.49\textwidth]{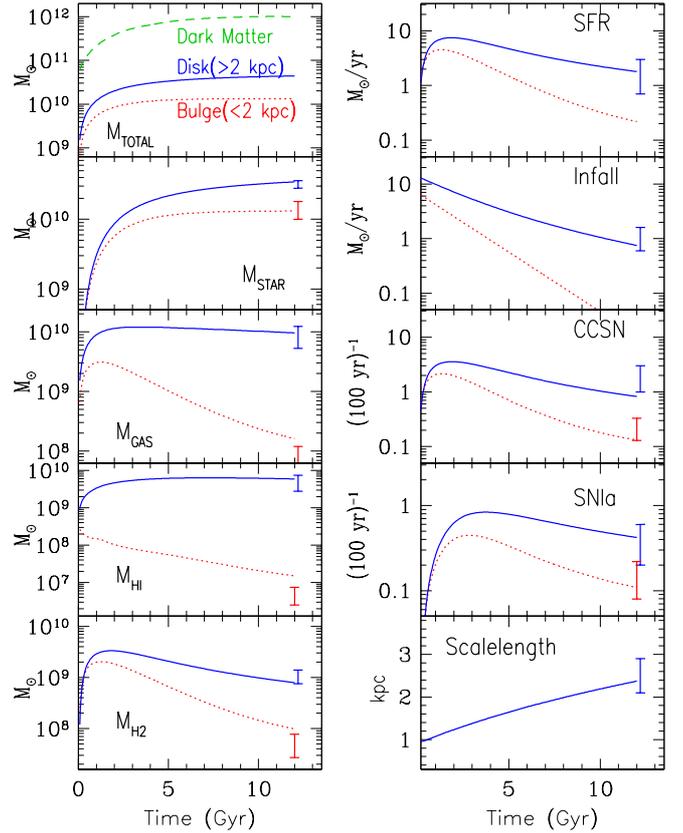}
\caption[]{Evolution of various quantities in the baseline scenario. In all panels, {\it solid blue } curves correspond to the evolution of the disk ($r>$2 kpc) and {\it dotted red } ones to that of the bulge ($r<$2 kpc), respectively. Vertical bars at 12 Gyr represent observational constraints (see text and Table \ref{Tab:Table_GlobalObs}).
}
\label{Fig:Evol0}
\end{center}
\end{figure}

\begin{figure}
\begin{center}
\includegraphics[width=0.49\textwidth]{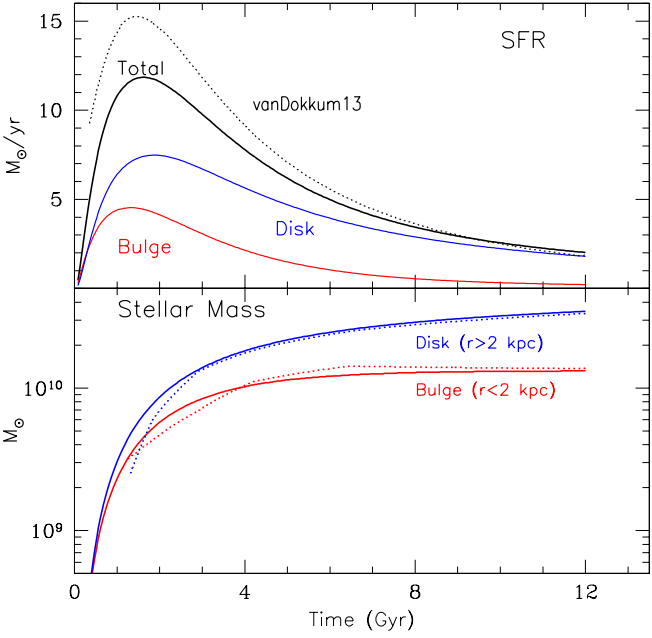}
\caption[]{Comparison of the model results to the observed  "stacked" evolution of disk galaxies of stellar mass 5 10$^{10}$ \ms \ ({\it dotted} curves) from  \cite{Dokkum2013}.
{\it Top:} Evolution of the total SFR (black solid curve), decomposed into bulge and disk contributions.
{\it Bottom:} Evolution of the stellar mass of the bulge ($r<$2 kpc) and of the disk ($r>$2 kpc) of the model ({\it solid} red and blue curves, respectively).
}
\label{Fig:Evol01}
\end{center}
\end{figure}

\begin{figure}
\begin{center}
\includegraphics[width=0.49\textwidth]{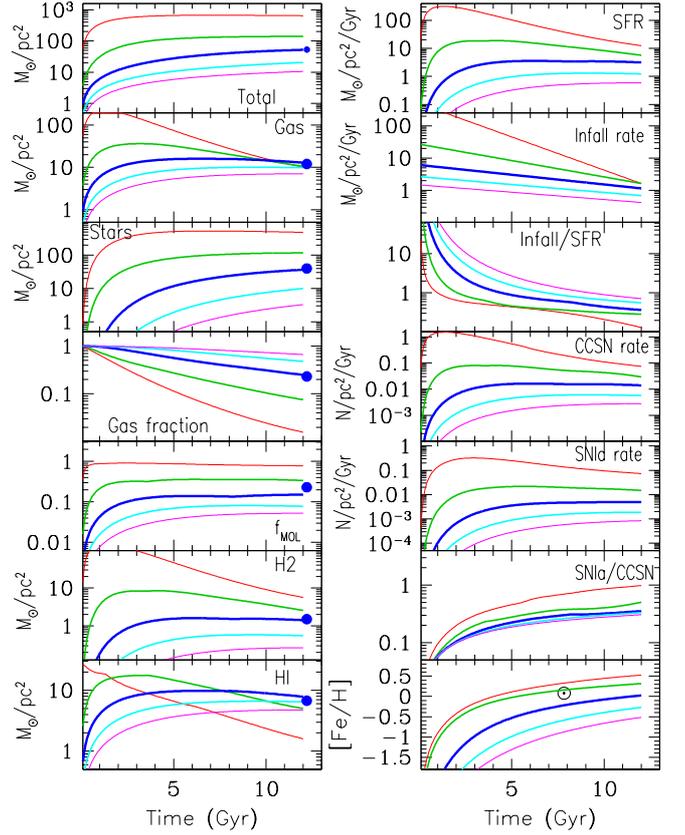}
\caption[]{Evolution of various quantities in the baseline scenario as a function of time, for five different radial zones, at 2(red), 5 (green), 8 (thick blue), 11 (cyan) and 14 (magenta) kpc. Observational data for local surface densities of gas and stars (from Table \ref{Tab:Table_GlobalObs}) are indicated with filled circles at T=12 Gyr.
}
\label{Fig:Evol_new}
\end{center}
\end{figure}

\section{Global evolution}
\label{sec:GlobEvol}

The results of our model for all the main outputs (total mass, mass of stars and gas, either atomic or molecular,
rates of star formation, infall, CCSN and SNIa, as well as the scalelength of the stellar disk) as a function of time are displayed in Fig. \ref{Fig:Evol0}. 

The results reproduce  the aforementioned present day observational constraints fairly well, with the exception of the gas amounts in the bulge, which our model overpredicts. We notice, however, that a significant fraction of the bulge gas is in the form of ionised gas ($\sim$3 10$^7$ \ms,  \citealt{Ferriere2007}, 
increasing the total gas amount of the bulge by $\sim$30\%; if this is taken into account, the discrepancy with our model (which  does not account for ionized gas) is reduced considerably.  

Regarding the evolution of the gas, we find that its mass in the disk remains quasi-constant after the first $\sim$2 Gyr, while it decreases by almost a factor of 30 in the bulge. It is interesting to note that the adopted prescriptions for the evaluation of the molecular fraction and the corresponding SFR (Sec. \ref{subsec:GasProfileSFR}  and   Appendix \ref{App:SFR}) lead to a different evolution between the atomic and molecular contents of the bulge and the disk: the bulge is dominated today by \hmol, while the disk is dominated by \hatm, as observed. Also, the total disk SFR, $\sim$2  \ms/yr, corresponds fairly well to the observed one. This  was not obvious  a priori, because we assume that the  SFR depends on the molecular gas, which is concentrated in the inner Galaxy (see discussion below).

Finally, the scalelength of the stellar disk increases steadily from $\sim$1.5 kpc at 2 Gyr to 2.3 kpc at 12 Gyr. The latter value is in reasonable agreement with  values in the literature: 2.1$\pm$0.3 kpc from the TMGS survey \citep{Porcel1998}, 2.25 kpc from COBE/DIRBE data analysis \citep{Drimmel2001}, and 2.15$\pm$0.14 kpc from the dynamical analysis of SEGUE G dwarfs \citep{Bovy2013}. In Sec. \ref{subsec:ThinThick} we shall discuss further the issues of thin and thick disk scalelengths. 

The question of whether the Milky Way is a typical spiral galaxy or not is an open one, since it appears underluminous for its rotational velocity of 220 km/s \citep{Flynn2006,Hammer2007}. One may also ask whether the Milky Way evolved as an average disk galaxy of the same present-day mass of 5 10$^{10}$ \ms. In a recent study \cite{Dokkum2013} provide relevant data by studying 
progenitors  galaxies of that mass out to redshift $z$=2.5, using data from the 3D-HST and CANDELS Treasury surveys. They find that  $\sim$90\% of the stellar mass of those galaxies was built since $z$=2.5, with most of the star formation occurring before $z$=1. Furthermore, the mass in the central 2 kpc of those galaxies increased by a factor of $\sim$3 between $z$=2.5 and $z$=1, implying that  bulges  likely formed in lockstep with disks during that period. However, after $z$=1 the growth in the central regions gradually stopped but  the disk continued to be built up.

In Fig. \ref{Fig:Evol01} we compare our results to those of \cite{Dokkum2013} (their Fig. 4, where we converted redshifts in lookback time assuming a $\Lambda$CDM cosmology with parameters from the recent PLANCK analysis).  It can be seen that our
result for the total SFR (upper panel) lie slightly below the SFR of \cite{Dokkum2013}
(by $\sim$20\% at the peak of SFR, the difference gradually decreasing with time.
At late times, there is a fairly good agreement between our model and the data.
Concerning the stellar masses (lower panel), our
 results, both for the bulge (assumed to be the region of radius $r<$2 kpc, as in \cite{Dokkum2013}) and the disk ($r>$2 kpc) are in fairly good agreement with the observed  "stacked" evolution of MW-type galaxies.
Overall, we conclude that our results are compatible with the idea that the MW evolved as a typical disk galaxy of  stellar mass $\sim$5$\times$ 10$^{10}$ \ms at the present epoch.

In Fig. \ref{Fig:Evol_new} we present the evolution of several key observables for selected radial zones (2, 5, 8, 11, 14 kpc) of our model. The main common feature is the more rapid evolution of the inner disk, reflected in the earlier rise of the stellar  surface density of the inner zones and  in the more rapid decrease in the corresponding gas fractions. In the cases of the stellar and gaseous surface densities, the local observations at $r$=8 kpc 
(Table \ref{Tab:Table_LocallObs})  are nicely reproduced. 


The net gas depletion timescale (i.e. taking both the SFR and the infall rates into account) is shorter
in the inner Galaxy than in the outer one: the ratio of  the infall rate to the SFR becomes smaller than unity 
within $\sim$1 Gyr at $r$=2 kpc but only after 10 Gyr at $r$=14 kpc. This gives a measure of the radially varying star formation efficiency of the model, due to the adopted SFR dependence on the molecular gas.

Finally, the evolution of [Fe/H] in the local gas (bottom right panel of Fig. \ref{Fig:Evol_new})   never 
saturates, even in the inner zones: there is always a steady albeit small increase in metallicity, even at late times. This result agrees with the work of \cite{Minchev2013}, but it contrasts with the results of \cite{SB2009}, who find little evolution in metallicity for all the zones of their model after the first couple of Gyr. 

\subsection{Gaseous profiles and the SFR}
\label{subsec:GasProfileSFR}

\begin{figure*}
\begin{center}
\includegraphics[angle=-90,width=0.99\textwidth]{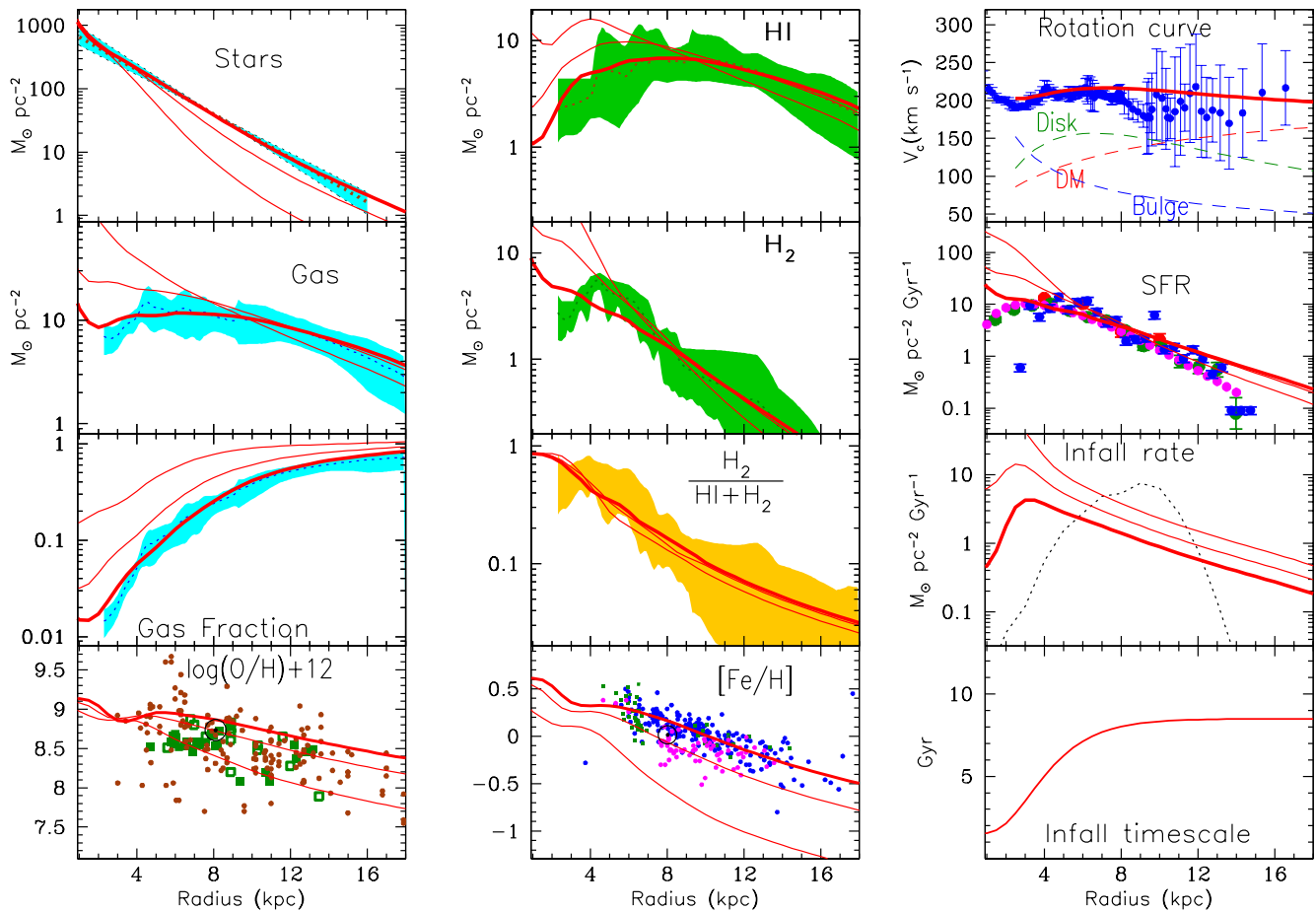}
\caption[]{Model profiles ({\it red solid} curves) at 4, 8 Gyr ({\it thin} curves) and at 12 Gyr ({\it thick} curves) of various quantities and  comparison of the latter with observational data concerning the present-day disk of the Milky Way. For all gaseous profiles (gas, \hatm,  \hmol, gas fraction and molecular fraction), observational constraints are those discussed in Appendix \ref{App:Gas} and presented in Fig. \ref{Fig:GasFrac}. For stars, the shaded area is bounded by two exponential curves with scalelengths $R_d$=2.1 and 2.7 kpc, respectively, which fix the range of plausible values for the stellar disk; they are normalised to a surface density of $\Sigma_*$(R=8 kpc)=38$\pm$2 \mpc \ in the solar neighbourhood.  Data for the MW rotation curve are from  \cite{Sofue2012}; the {\it dashed} curves represent the contributions of the bulge, disk and dark matter (blue, green and red, respectively ; see text for the bulge contribution). Data for the star formation rate (SFR) profile are discussed in Appendix \ref{App:SFR} and displayed in Fig. \ref{Fig:SfrProf}. The observed 
present-day profile of infall rate ({\it dotted} curve) is based on estimates of \cite{Marasco2012} (see text).
The {\it bottom right} panel displays the adopted timescales for exponential infall.
Data for oxygen are from \hatm \ regions \citep{Simpson1995,Afflerbach1997} and OB stars \citep{Smartt1997,Daflon2004}  and for iron from Cepheids \citep{Luck2011,Lemasle2013,Genovali2013}.
}
\label{Fig:TheorObsProf}
\end{center}
\end{figure*}

Figure \ref{Fig:TheorObsProf} presents a synthetic view of the  main results of our model, compared  to observations. Except for the top right  (velocity curve) and bottom (infall timescale) 
panels, all other panels display three profiles, at 4, 8 and 12 Gyr; the last (thick curve) is to be compared with observational data.

The infall timescale of the disk as a function of radius (bottom right panel) is tailored to smoothly match the bulge timescale of $\tau$($r$<2 kpc)$\sim$1.5 Gyr to the timescale of the outer disk while going through the value of $\tau$($r$=8 kpc)$\sim$7-8 Gyr  for the solar neighbourhood. The latter has been shown to provide a good fit to the local metallicity distribution in simple (independent-ring) models of the MW chemical evolution (e.g. \cite{Chiappini1997,BP99}) 
and we show that this is also the case here (Sec. \ref{sec:LocalEvol}), although the adopted SNIa rate and the effects of radial migration also play a role. The resulting profiles of infall rate appear in the right panel just above the bottom. Because of the shorter infall timescales in the inner disk, the final infall profile
peaks around 3 kpc. 

In a recent work,  \cite{Marasco2012}  use a model of the Galactic fountain to simulate the neutral-hydrogen emission of the Milky Way. Their model was developed to account for external
galaxies with sensitive HI data. For appropriate parameter values, their 
model reproduces the observed HI emission of the MW. They find a global current value of $\sim$2 \ms/yr for the MW, and they derive the infall profile displayed in the  right panel above the bottom  (dashed curve), which peaks at $\sim$9 kpc.
In our model we obtain a similar value for the total present-day infall rate  ($\sim$1 \ms/yr, see Fig. \ref{Fig:Evol01}), but our infall profile peaks at $\sim$3 kpc. 
We  notice that the existence of a peak in the current infall profile  appears naturally in our model, albeit not in the claimed position. 
Further 
studies and understanding  of the properties of the gas accreted onto the MW, 
including its velocity profile, will provide 
much stronger  constraints in evolution models of the Galaxy.

The radial profiles of gas (total, \hatm \ and \hmol) are in rather good agreement with observations,
for the largest part of the disk. In the inner disk,  the presence of the
bar plays an important role, inducing  radial  flows of both gas (see below) and  stars. Had the migrating long lived stars, had they remained in place, they could return  their H-rich ejecta in the local ISM at late times.
The overall result of those motions is a total gas profile going through a broad maximum at $\sim$7 kpc, still
compatible with the observations.  This is also true for the \hatm \ profile . In contrast, the "molecular ring" in the inner disk is not reproduced well; it is not clear whether this is due to the poor fit of the BR2006 prescription to MW data (see their Fig.4) or to some combination of the various relevant ingredients adopted here (SFR, infall, radial inflow, evaluation of \hatm \ profile).


The model rotation curve at $t$=12 Gyr compares fairly well to the data of \cite{Sofue2013} (right, 1st row panel in Fig. \ref{Fig:TheorObsProf}). It peaks at $\sim$212 km/s in the solar neighbourhood, where the contribution of the disk exceeds slightly the one of dark matter. This is slightly lower than the canonical IAU value of 220 km/s and clearly lower than some values recently proposed in the literature (see \cite{Bhattar2013} and references therein). The rotation curve is used in all time-steps to evaluate the epicyclic motion and the extent of blurring in the disk (see Sec. \ref{subsub:blurring}).

\begin{figure}
\begin{center}
\includegraphics[width=0.49\textwidth]{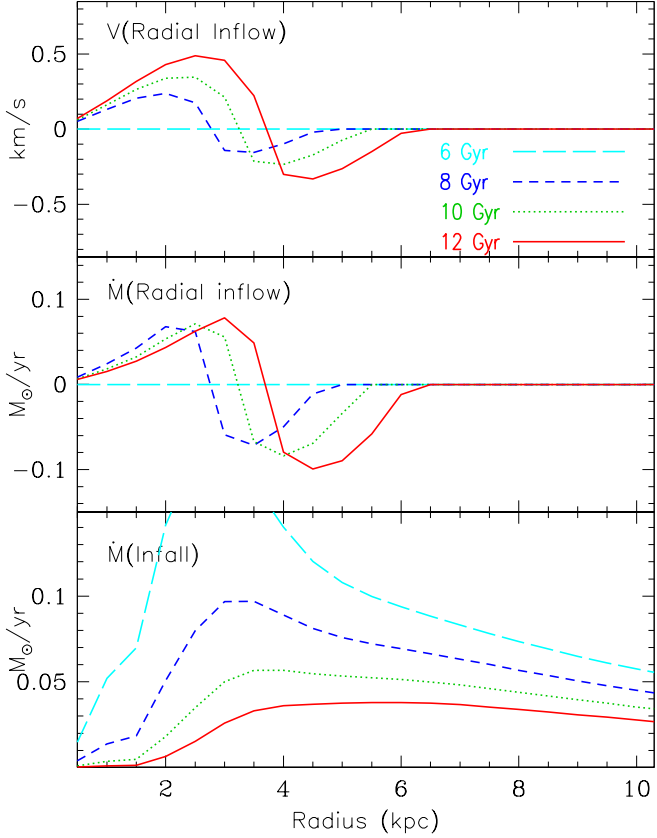}
\caption[]{{\it Top:}  Adopted velocity profile for radial inflow. Positive velocities indicate flow inwards (towards the Galactic centre) and negative ones an outward flow. The flow velocity is everywhere  zero before the assumed appearance of the bar at $t$=6.5 Gyr. Profiles are given at $t$= 6, 8, 10 and and 12 Gyr (cyan long-dashed, blue short-dashed, green dotted  and red solid, respectively). As corotation moves outwards (see text), the flow pattern moves also. {\it Middle:} Corresponding mass flow profiles through radial zones $r$ 
{\it Bottom:}
Corresponding mass infall profiles (rate of mass infalling over the ring of radius $r$ and width $\Delta r$=0.5 kpc, at $t$=6, 8, 10 and and 12 Gyr.
}
\label{Fig:RadFlow}
\end{center}
\end{figure}

\begin{figure}
\begin{center}
\includegraphics[width=0.49\textwidth]{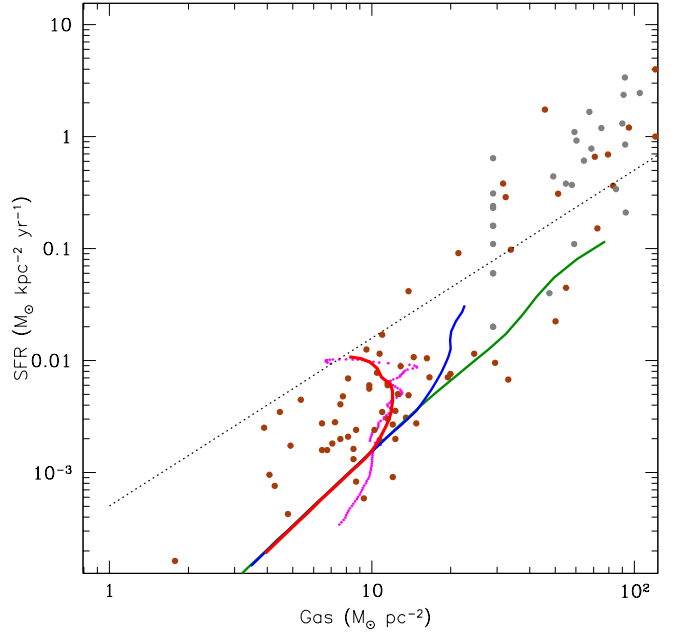}
\caption[]{SFR vs. gas surface density. Model results are displayed for three different epochs, after 4 Gyr (green curve), 8 Gyr (blue) and 12 Gyr (red). The last is compared to the observationally inferred SFR profile of the Milky Way (dotted magenta curve), obtained as discussed in Appendix \ref{App:SFR}. The data points concern extragalactic measurements, compiled by  \cite{Krumholz12} (see also Fig. \ref{Fig:SfrGas}). The dotted line is $\propto\Sigma_{Gas}^{1.5}$.)
}
\label{Fig:SFRvsObs}
\end{center}
\end{figure}

The gaseous profiles are affected by the radial inflow induced by the bar, which is modelled here as shown in Fig.  \ref{Fig:RadFlow}. It is assumed that the bar 
 radius increases from 2 kpc at 6 Gyr to 3 kpc at 12 Gyr and its corotation radius from 2.4 kpc initially to 3.6 kpc in the end. Simulations show indeed that in disks with large gas fractions, the appearance of a bar can be delayed for several Gyr  \citep{Athanassoula2013}, compared to cases with no gas. It then grows with time while slowing down, pushing the corotation resonance outwards, as in our model. The velocity profile has  positive values (towards the Galactic centre) inside corotation  and negative ones (towards  the anti-centre) outside corotation and up to the OLR. Maximum absolute values of the (azimuthally averaged) radial flow velocity are 0.5 km/s. Higher values for the adopted duration of the bar would lead to substantial depletion of the gaseous layer in the inner disk, impossible to replenish by infall. Indeed, the middle and bottom panels of Fig. \ref{Fig:RadFlow} display the net radial flow rate  through the ring of radius $r$ (middle) and the rate of infall onto that same ring (bottom). Around corotation, the former (calculated as $f_r=2 \pi \ r \ \upsilon_r \ \Sigma_g(r)$)  is higher than the latter. Although our 1D prescription for radial inflow is inadequate for describing an intrinsically 2D effect  and none of the adopted parameters can be observationally determined at present, the range of the velocity values we find is  similar to the ones adopted in \cite{Portinari2000} for the case of a simulated bar (their figure 13), i.e. $\upsilon_r<$0.5 km/s. 

The current profiles of gas and SFR reflect short-term features of the Milky Way and  do not constitute strong constraints; indeed, on a timescale of 10$^7$ yr, comparable to the timestep of our model,
they may change considerably. Constraints obtained through 
time-integrated profiles (with negligible
late variations within short timescales) are more severe in that respect. The profiles of stars and metallicity belong to this class.
As already discussed (see Sec. \ref{subsec:rad_gas_flow} and references therein) one of the main effects of the radial flow induced by a bar is to flatten the abundance gradients in the region around the corotation (the other one being to fuel star formation in the central regions). This is clearly seen in the final oxygen profile (bottom left panel of Fig. \ref{Fig:TheorObsProf}), which displays a flattening in the region 3-5 kpc. The iron profile   (bottom middle panel of Fig. \ref{Fig:TheorObsProf}) is less affected, because the sources of Fe (mainly SNIa) are less affected by those of O (CCSN) in that region. The latter are distributed as the gas, which has a flat profile in that region, whereas the former - belonging to an older population - have a steeper distribution. The evolution of the metallicity profiles is similar to the one obtained in many other studies (e.g. \cite{BP99,Hou2000}) and reflects the inside-out formation of the disk, with the profiles flattening with time. Those profiles are widely used in Sec. \ref{sec:LocalEvol} to study the main observables in the solar neighbourhood. A detailed study of the evolution of the abundance profiles of all the  elements of our model, both in the gas and in the stellar populations, is presented elsewhere \citep{Kubryk2014b}.

Figure \ref{Fig:SFRvsObs} displays the star formation profiles of our models as a function of the local  gas surface density  for three different epochs: at $t$=4, 8, and 12 Gyr. Comparison is made to the data compiled by \cite{Krumholz12} for a large number of star-forming galaxies. It can be seen that our values lie in the low range of the observed SFR values for a given gas surface density. In particular, the rate at 12 Gyr presents the steep rise at $\Sigma_{Gas}\sim$10 \mpc \ that is "observed" in the MW disk; however, it is slightly lower than "observed" in the inner disk and slightly higher than "observed" in the outer disk. These discrepancies are also displayed in the SFR profile presented in Figs. \ref{Fig:TheorObsProf} (below the top right panel) and  \ref{Fig:SfrGas}.
The key point here is the absence of  a correlation between the SFR and the local gas surface density in both the data and the model. It implies that some other factor is in play, namely molecular gas, as discussed in Appendix \ref{App:SFR} and in the recent review by \cite{Krumholz2014}.

\subsection{Stellar profiles}
\label{subsec:StarProfile}

The model stellar profiles result from the combined history of star formation and radial migration. The former is discussed in the previous section. Here we analyse the impact of stellar radial migration.

The adopted radial velocity dispersion profile $\sigma_{v,r}$ appears in Fig. \ref{Fig:BlurChurn} (top panel) for two epochs, at $t$=2 Gyr and 12 Gyr\footnote{We calculated the profile of radial velocity dispersion and the corresponding epicyclic amplitudes at each time step; here we show the results for only two ages.}. The corresponding local values (at $r$=8 kpc) are 21 km/s and 40 km/s, respectively.  We then calculated the probabilities of radial displacement due to epicyclic motion. 
The results for the same two epochs and three different radial zones are displayed in the middle panel. The curves are asymmetric about the birth radius $r'$ despite the symmetry of the adopted Gaussian function for blurring (Eq. \ref{eq:prob}), because the corresponding velocity dispersion $\sigma$ is taken at the final radius $r$. The thick solid curves correspond to an age of 12 Gyr, i.e.  to $\sigma_{v,r}$=40 km/s for the stars in the solar neighbourhood, and it may  be compared to  the corresponding curve obtained by \cite{SB2009} (dotted curve in middle panel).
Our formulation of the epicyclic motion produces similar distributions to those of \cite{SB2009} in the solar neighbourhood and beyond and somewhat narrower distributions in the innermost disk, but the overall agreement is quite good.

\begin{figure}
\begin{center}
\includegraphics[width=0.49\textwidth]{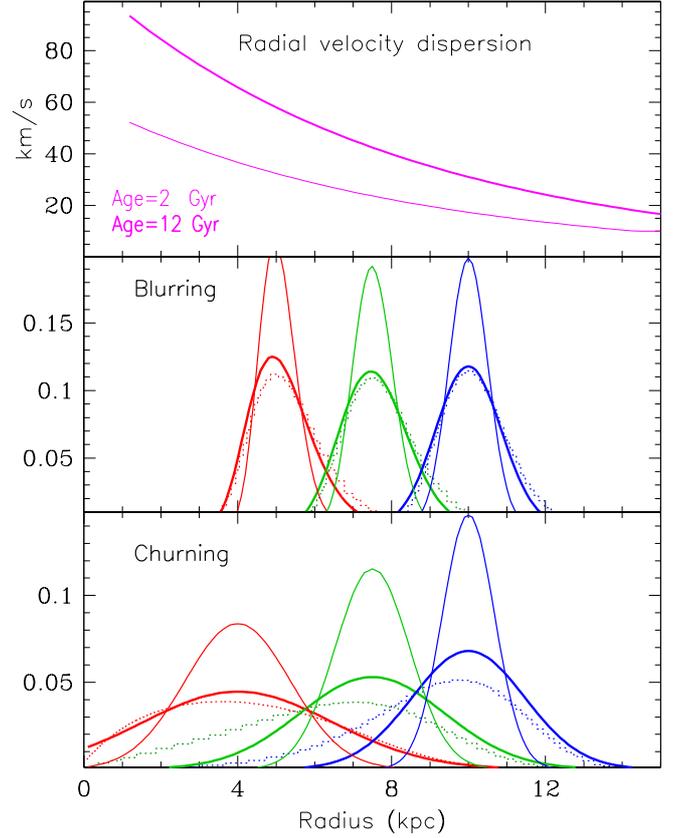}
\caption[]{{\it Top:} Model radial velocity dispersion at 2 Gyr (thin curve) and 12 Gyr (thick curve). {\it Middle:} Probabilities of blurring. {\it Bottom:} Probabilities of churning.  In all panels our results (solid curves) are displayed for  stars of home radius  r=5, 7.6 and 10 kpc, and for 2 Gyr (thin) and 12 Gyr (thick curves). The latter should be compared to the corresponding quantities reported in  \cite{SB2009} ({\it dotted histograms}).
}
\label{Fig:BlurChurn}
\end{center}
\end{figure}

The corresponding probability functions  for churning appear in the lower panel of Fig. \ref{Fig:BlurChurn}. They are evaluated as discussed in Sec. \ref{subsub:churning} and are broader than those of blurring. In the inner disk, they are not very different for the ones adopted in \cite{SB2009}, but they are clearly narrower than the latter in the outer disk. 
There is clearly a difference in the amount of radial migration between the inner and outer disks, both in \cite{SB2009} and in our case, but this difference is more important in our case than in theirs. We think that this can be attributed to the action of the bar, which plays an imortant role in the former simulation, while the latter only considered the effect of transient spirals, which are more uniformly distributed over the disk.

 In the top panel of Fig.  \ref{Fig:Probabilities} we present the total probabilities (blurring + churning) of a stellar population born in a given radius (here: 4, 8 and 12 kpc) to be found in some other disk radius $r$ after times 4, 8 and 12 Gyr. We recall that the radial bin in our simulation has a size $\Delta r$=0.5 kpc. As discussed in Sec. \ref{subsec:starmigr} (see also Fig. \ref{Fig:BlurChurn}) for the blurring and churning probabilities separately, the probability of finding a star from the inner disk to the outer one is greater than the inverse, because of the larger variations in the gravitational potential perturbations in the inner disk than in the outer disk.

\begin{figure}
\begin{center}
\includegraphics[width=0.49\textwidth]{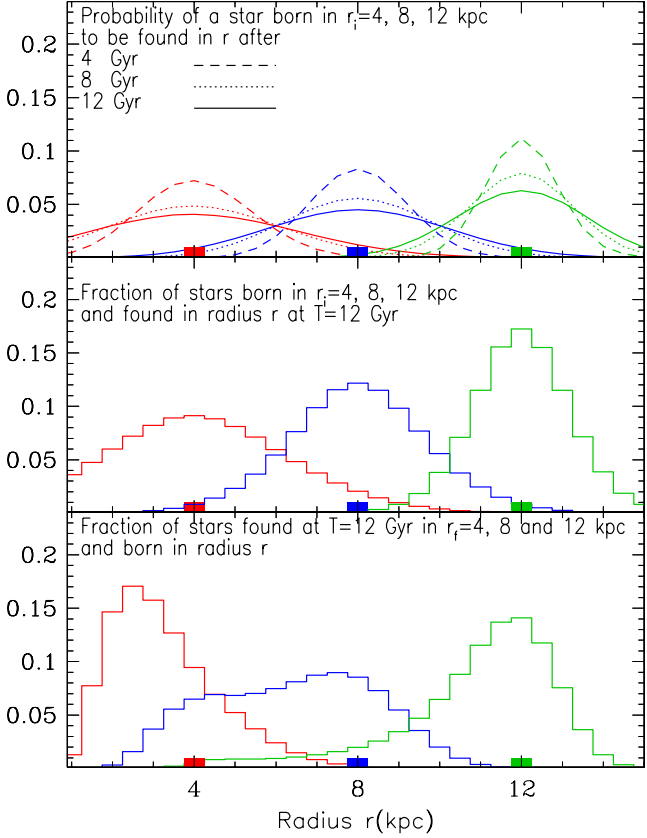}
\caption[]{{\it Top}: Probabilities of stellar migration (blurring + churning) adopted in this work (see text); they are shown for three birth radii (4, 8 and 12 kpc) and for three snapshots (after 4, 8 and 12 Gyr). {\it Middle}: Fractions of stars which are born in radii 4, 8 and 12 kpc and are found in galactocentric radius r in the end of the simulation, at T=12 Gyr. {\it Bottom}: Fractions of stars found at the end of the simulation in radii 4, 8 and 12 kpc according to their birth radius $r$. In all panels, radial bins are $\Delta r$=0.5 kpc wide, as indicated by the bin width of the histograms and the size of the coloured boxes.
}
\label{Fig:Probabilities}
\end{center}
\end{figure}

The middle panel of that figure displays the fractions of stars born in those same radii, which are  found at radius $r$ at the end of the simulation; those fractions are the time-integrated probabilities of the upper panel, weighted by the corresponding SFR($r,t$) history. This explains why the distribution of the outer region is more peaked and less wide than the inner one in the middle panel than in the top one:  stars are formed in the inner disk earlier than in the outer one, on average, because of the inside-out formation scheme adopted here. As a result, inner disk stars have on average  more time to migrate than those formed in the outer disk. Even if the probability distributions in the top panel were identical at all birth  radii (assuming, for instance, the same potential inhomogeneities over the whole disk), those in the middle panel would still be wider in the inner disk, simply because of the inside-out star formation.

The bottom panel in Fig.  \ref{Fig:Probabilities} displays the fractions of stars found at the end of the simulation in
final galactocentric radii $r_f$=4, 8 and 12 kpc, respectively, and born in other positions $r$ in the disk. These fractions integrate not only the probability distributions  (upper panel) and the SFR histories (as in the middle panel), but also 
the fact that there is more mass in the inner disk than in the outer one, because the surface density profile decreases exponentially outwards. For that reason, the resulting fractions are asymmetric in each radius, with more originating in the inner disk than in the outer one. Thus, from all the stars presently found in the bin at $r$=8$\pm$0.25 kpc, only $\sim$10\% are formed in that bin; if the "extended solar neighbourhood" (defined as $r$=8$\pm$1 kpc)  is considered, the corresponding percentage rises to 38\%. Only 7\% of the remaining stars originate outside the extended solar neighbourhood (at radii $r>$9. kpc), while 55\% of the stars of that region are formed in $r<$7 kpc and have migrated here: $\sim$12 \% of its stars have migrated from $r<$4 kpc. In contrast, a negligible amount of stars presently found in $r$=13 kpc originates from $r$=8 kpc.

\begin{figure}
\begin{center}
\includegraphics[width=0.49\textwidth]{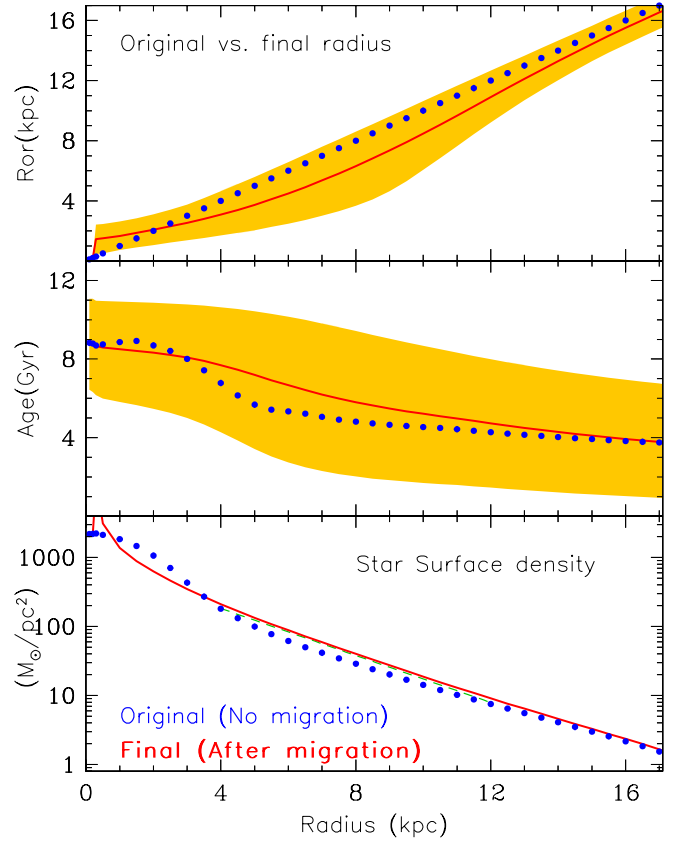}
\caption[]{Average radius of origin of stars ({\it top}), average stellar age ({\it middle})  and star surface density profile ({\it  bottom}), as a function of galactocentric radius. In all panels, the {\it continuous} (red) curve corresponds
to the results of the model with radial migration (churning+blurring), while the {\it dotted} (blue) curve shows the results with no radial migration.
In the {\it top} and {\it middle} panels, the {\it shaded area} contains the $\pm$1$\sigma$ values - containing $\pm$34\% of the stars - around the average. In the bottom panel, a best fit exponential with  scalelength $r_d$=2of the exponentially decreasing outwards.25 kpc in the 3 to 13 kpc region and with $\Sigma$($r$=8 kpc)= 38 \mpc \  is also displayed (green dashed line).} 
\label{Fig:MigrProf}
\end{center}
\end{figure}

The impact of radial migration (blurring+churning) on some radial properties of the disk appears in Fig. \ref{Fig:MigrProf}. 
It can be seen that stars presently found in the region 5$<r_f<$10 kpc, originate in regions located on average $\sim$1.-2. kpc inwards. On the other hand, stars presently found in the bulge (here taken to be the region $r<$2 kpc) have a significant fraction of them originating from the inner disk, from up to 3 kpc.
Finally, stars beyond $r$=13 kpc are affected very little by radial migration on average (at least with the adopted scheme for churning). We note that \cite{Minchev2013} find a bimodal distribution function for the average birth radius of stars presently in the solar neighbourhood (their Fig. 3, right):  the maximum is located in the region $\sim$5.6 kpc and a secondary maximum is found at $\sim$7 kpc, compared to a single maximum at $\sim$6.2 kpc in our model.
It is difficult to compare the two models, since they are based on different N-body simulations and different descriptions of the radial migration. We suspect that it is the stronger bar and larger OLR radius in their case that produces a more extended mixing of stars from the inner regions to the solar neighbourhood, while we have "calibrated" our bar to the one of the MW size.

The aforementioned results also explain the average age of stars in each galactocentric radius (middle panel in Fig. \ref{Fig:MigrProf}). The average age of stars formed in the bulge ($r<$2 kpc)  is $\sim$9 Gyr, in the solar neighbourhood $\sim$5 Gyr and beyond 12 kpc it is 4 Gyr. Stars presently found in the region 5$<r_f<$11 kpc are, on average, 1-2 Gyr older than those formed {\it in situ}, because their population has been altered by the radial migration of stars from the inner disk. Again, beyond $r$=13 kpc, the average stellar age is not affected. It should be noted that  a uniform age dispersion of $\sim\pm$3 Gyr is found at all radii. 


\section{Local evolution}
\label{sec:LocalEvol}

The results discussed in the previous section  for the whole disk help for understanding the results
obtained for the solar annulus, defined here as the radial bin at $r$=8$\pm$0.25 kpc. 
We stress - as already done in Sec. 2.5 - that this region is not neccessarily
representative of the solar  neighbourhood, and any strict comparison to observational
results should take the corresponding observational biases into: account.

Figure  \ref{Fig:XvsAge} displays the average birth radius of stars found in this zone at the end of the simulation, as a function of their age. In the absence of any radial migration, the birth radius would be the 
horizontal line at $r$=8 kpc. 
As expected - in view of the discussion in sec. \ref{subsec:StarProfile} - the average birth radius when radial migration is considered, is close to 8 kpc for the youngest stars, but it decreases steadily for older stars. The reason for that decrease is twofold: first, the older stars  have more time
to migrate from other regions and, in particular, from the inner disk; second, 
the SFR at $r$=8 kpc is low in the first few Gyr, while it is quite high in the inner disk during that same period (see top right panel in Fig. \ref{Fig:Evol_new}). 
Thus, the old stars migrating here from the inner disk
overwhelm by number the few old stars formed locally. 
 A local "average star" of age=4.5 Gyr originates at $r\sim$7 kpc, whereas stars older than 8 Gyr were born inwards of 5.5 kpc, on average. Equally interesting is the spread in birth radii that increases as a function of age (the associated 1-$\sigma$ dispersion in the
birth radius is indicated by the shaded area). Thus, about two thirds of the local stars of solar age have birth radii
ranging between 5.5 and 8.5 kpc, while two thirds of those with an age of 8 Gyr were born between 2.5 and 8 kpc. 
Their birth place affects  their chemical composition and  the properties
of the solar annulus.
The same behaviour of birth radius vs. age, not only qualitatively but also quantitatively,  is obtained in the N-body simulation of e.g. \cite{Loebman2011,Brook2012}, which are not tuned to reproducing a Milky Way disk: the oldest stars of their simulation presently in the "solar cylinder" (between 7 and 9 kpc from the centre), have home radii as small as 2 kpc.

\subsection{The local age-metallicity relation and its dispersion}
\label{subsec:AgeMet}

\begin{figure}
\begin{center}
\includegraphics[width=0.49\textwidth]{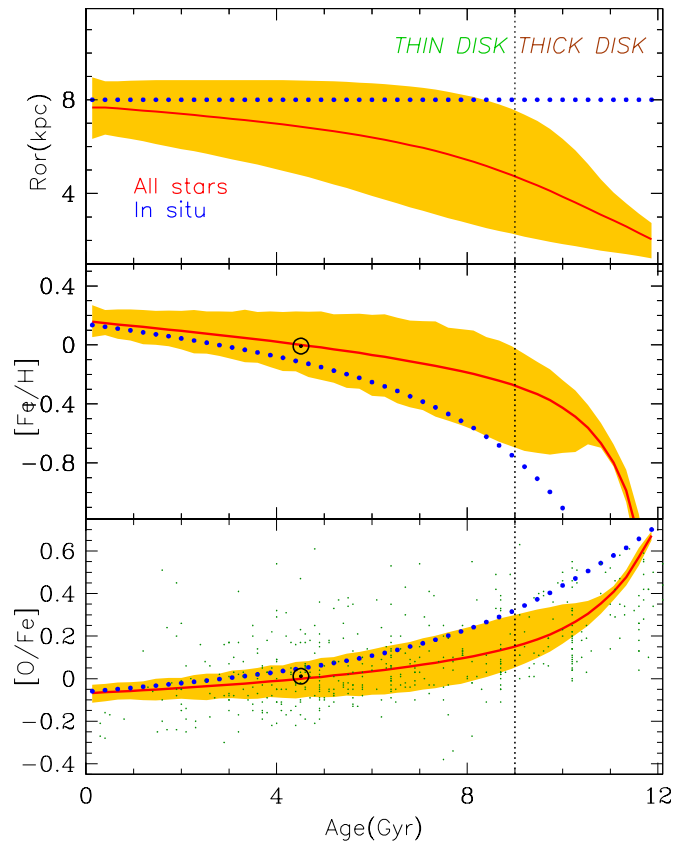}
\caption[]{Solar neighbourhood: Average birth radius ({\it top}), average [Fe/H]  ({\it middle}) and average [O/Fe] 
of stars ({\it bottom}) as function of their age. In all panels, the {\it dotted} (blue) curves display the results for stars formed {\it in situ} and the {\it solid} (red) curves  the results for all stars found at T=12 Gyr in radius $r$=8$\pm$0.25 kpc. The {\it shaded areas} in all  panels enclose $\pm$1$\sigma$ values around the corresponding averages. Data points are from the survey of \cite{Bensby2014}. The dotted vertical line at 9 Gyr separates the thin from thick disk stars, according to the assumption made here (see Sec. \ref{subsec:AgeMet}. The symbol $\odot$ denotes the position of the Sun in the corresponding panel, as well as in Figs. \ref{Fig:XvsAge_Disp}, \ref{Fig:LocalMDFa} and  \ref{Fig:LocalMDFb}.
}
\label{Fig:XvsAge}
\end{center}
\end{figure}

The impact of radial migration on the chemical evolution of the local disk appears in the middle and bottom panels of Fig. \ref{Fig:XvsAge}. The local average 
age-metallicity relation (middle panel, red solid curve ) is flatter than the one for the locally born stars; the latter represents the metallicity evolution of the local ISM. 
The flattening   depends on both the adopted radial migration scheme (the churning coefficients) and the whole history of the disk (the inside-out formation and the resulting abundance profiles). 
Since the gas is well mixed locally today - and presumably at earlier times as well -  the fact that the metallicity of the Sun 4.5 Gyr ago is larger than the corresponding gas value in our model, implies that the Sun was not born locally, but migrated from inner galactic regions as suggested  in \cite{Wielen1996}. 
Although it is rather early to draw definitive conclusions, it appears that the hypothesis that the Sun was formed inwards (by 1-2 kpc) of its present galactic position, is the most convenient one for  explaining several observational facts
(see \cite{Nieva2012} for an updated discussion and references).

The bottom panel of Fig. \ref{Fig:XvsAge} displays the O/Fe vs age relation. It starts at values around [O/Fe]$\sim$0.5 for the oldest stars - where only CCSN enrich the ISM - and then decreases smoothly to [O/Fe]$\sim$0 for the youngest stars, because of the steady Fe input from SNIa. 
This distinctive behaviour in the decrease in [O/Fe] for old and young stars is also apparent in the work of \cite{Haywood2013} who reanalysed a sample of local stars with high quality abundance determinations (their Figs. 6 and 7). However, they interpret their data 
as two clearly differing regimes with different slopes of [$\alpha$/Fe] vs age, while we find just a gradual decline of the [$\alpha$/Fe] ratio, albeit with a strong reduction of the corresponding slope with time.
The dispersion in the [O/Fe] relation is quite small ($\sim$0.1 dex) at every age; the reason is the quasi-similar evolution of the SNIa/CCSN ratio in all radial zones of our model (right above-bottom panel of Fig. \ref{Fig:Evol_new}), except  those of the bulge. This fact justifies the use of [O/Fe] as a proxy for age, as suggested e.g. in \cite{Bovy2012}.


The top panel of Fig.\ref{Fig:XvsAge_Disp}, displays  the fraction of stars born in the solar annulus  as a function of stellar age and the  corresponding fraction of 
all stars currently present in the solar annulus (i.e. born anywhere but found in 8$\pm$0.25 kpc at the end of the simulation).
The fraction of stars born in-situ is a decreasing fraction of stellar age, because of the SFR rate history (see Fig. \ref{Fig:Evol_new}, top right panel). On the other hand, the fraction of all stars currently in the solar neighbourhood increases between 0 and 8 Gyr because large numbers of old stars have migrated  to the solar neighbourhood, shifting the average age to higher values (see Fig. \ref{Fig:MigrProf}). In particular, almost all stars older than 10 Gyr have been formed inwards of $r$=5 kpc, on average. 
(see also \cite{Roskar2008}).

This result has some important implications: It implies that radial migration makes it impossible to try to infer the past local star formation rate through star counts as function of stellar  age (e.g. \citet{Rocha-Pinto2000}): most of the old stars presently found here were formed elsewhere, whereas the opposite holds for the younger stars. For that same reason,  it becomes impossible to use the method of the luminosity function of white dwarfs  to infer the star formation history of the solar neighbourhood, e.g. \cite{Isern2013} and references therein.

\begin{figure}
\begin{center}
\includegraphics[width=0.49\textwidth]{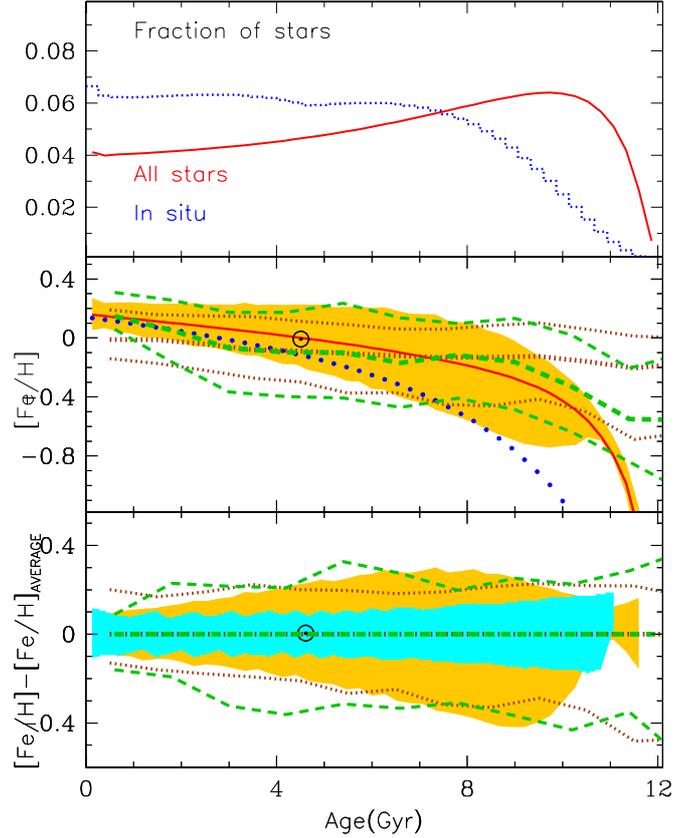}
\caption[]{Solar annulus:  {\it Top}: Fraction of all stars ever born {\it in situ} in the solar annnulus (blue {\it dotted}) and fraction of all  stars found in the solar annulus at T=12 Gyr (red {\it solid}) as a function of stellar age. Average [Fe/H] 
of stars ({\it middle}) and 1-$\sigma$ dispersion around the average ({\it bottom}) as function of stellar age, compared to observations.  The {\it solid} (red) curve in the middle panel displays  the results for all stars found at T=12 Gyr in radius $r$=8$\pm$0.25 kpc.  In both panels, the {\it shaded (yellow) areas} represent the  $\pm$1$\sigma$ limits of the model; the  brown ({\it dotted}) and green ({\it dashed}) curves represent the corresponding averages and $\pm$1$\sigma$ limits of the observations of \cite{casagrande2011} and \cite{Bensby2014}, respectively. The narrower {\it blue shaded area }
in the {\it bottom} panel shows the results of a calculation with blurring alone.  
}
\label{Fig:XvsAge_Disp}
\end{center}
\end{figure}

In the middle panel of Fig. \ref{Fig:XvsAge_Disp} we compare the model [Fe/H] vs age relation to data of two recent major surveys. We display observational data from a re-analysis of the Geneva-Copenhagen survey  \citep{casagrande2011} and from  the survey of \cite{Bensby2014}.  In both cases, we determine average values  and $\pm$1-$\sigma$ widths around those averages in the [Fe/H] vs age relation, by using age bins of 1 Gyr and apply the same statistics to our model stars.  
To better compare the metallicity dispersion vs age between observations and the model, we show in the bottom  panel of Fig. \ref{Fig:XvsAge_Disp} the 1-$\sigma$ dispersion around the average value. 
 We display our results for blurring only  and for blurring plus churning , and we compare with the corresponding dispersions of the data in \cite{Bensby2014}  and \cite{casagrande2011}. We note some differences between the two data sets, the
 one of \cite{Bensby2014} having a slightly larger dispersion than the one of \cite{casagrande2011}. 
This may be due to kinematic biases and to magnitude selection effects.
     
The epicyclic motion, as calculated  in our model, produces a uniform dispersion of $\pm$0.1 dex for most stellar ages. Only  the older ones - age $>$9 Gyr - have a slightly larger dispersion, up to $\pm$0.2 dex.  The epicyclic motion is not sufficient to explain the observed dispersion in the
age-metallicity relation. 
When churning is also considered, dispersion increases  steadily, up to $\pm$0.3-0.4 dex. For thick disk stars (assumed here to be older than 9 Gyr), the model dispersion starts decreasing. The reason is that these old local stars are formed almost exclusively in the innermost disk regions, where the very rapid evolution produced a quasi-identical age-metallicity relation; as a result, these stars
show smaller dispersion in their metallicities. Our analysis of the data of \cite{casagrande2011} and \cite{Bensby2014} shows no such decrease for old stars. We notice however, that a re-analysis of the data of the GCS survey and of \cite{Adibekyan2011} by \cite{Haywood2013}, leads to a decreasing dispersion in [Fe/H] for the older stars, in line with our model (at least qualitatively).

In summary, the  dispersion in the age-metallicity relation stems from the extent of churning
(which is related to the inhomogeneities of the gravitational potential) and the
SFR history of the disk. Comparison of theoretical dispersions to observed ones constrains   
the combination of those processes, not the just churning (or churning+blurring). Comparison to 
other data, e.g. past and present metallicity gradients, will be necessary to disentangle the various effects.

\subsection{The local metallicity distribution }
\label{subsec:MetDist}


\begin{figure}
\begin{center}
\includegraphics[width=0.49\textwidth]{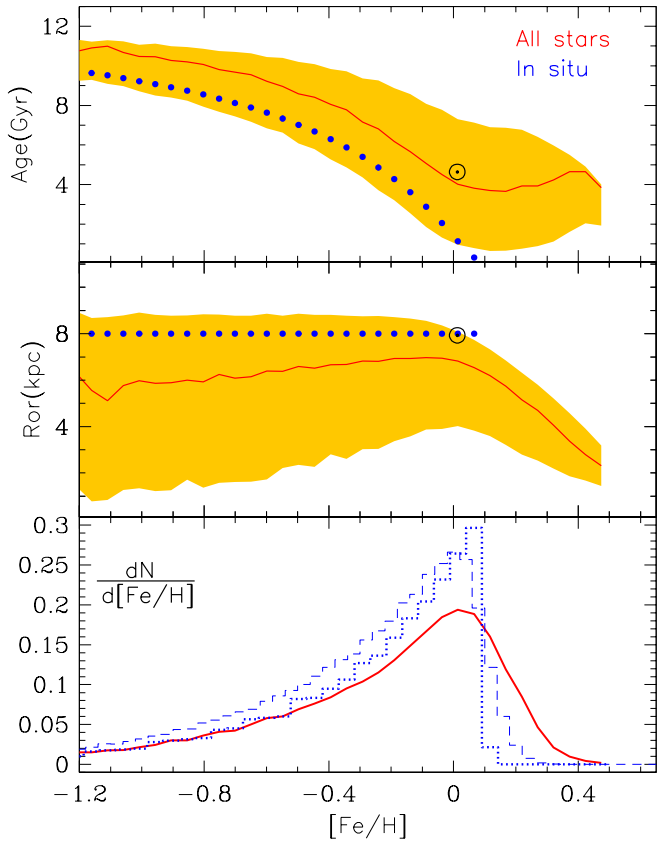}
\caption[]{Solar neighbourhood: Average age of stars ({\it top}), average birth radius ({\it middle}) and number distribution
of stars (metallicity distribution, {\it bottom}) as function of their metallicity. In all panels, the {\it dotted} (blue) curves display the results for stars formed {\it in situ} and the {\it solid} (red) curves  the results for the average values of all stars found at T=12 Gyr in radius $r$=8$\pm$0.250 kpc. The {\it shaded areas} in the {\it top } and {\it middle} panels enclose $\pm$1$\sigma$ values around the corresponding averages. The {\it dashed} histogram in the lower panel is calculated with the effect of epicycles only.}
\label{Fig:LocalMDFa}
\end{center}
\end{figure}

In Fig. \ref{Fig:LocalMDFa} we present our results for the local  metallicity distribution (MD). The lower panel displays the MD of stars formed {\it in situ} (dotted histogram), which peaks at 0.08 dex and terminates abruptly at 0.1 dex. It also displays the MD obtained with only the epicyclic motion considered, which peaks at the same metallicity and is slightly broader, extending up to 0.15 dex.
Finally, the total MD - including blurring and churning -  is considerably broader than the previous two distributions and extends up to [Fe/H]=0.4. It was already pointed out by \cite{Chiappini2009} that the local evolution cannot produce stars as metal-rich as [Fe/H]=0.4. Here we show that the conclusion holds even if epicyclic motion is considered. Only radial migration can 
account for such stars in the solar neighbourhood.

The upper panel of Fig. \ref{Fig:LocalMDFa} shows the average age of the local stars (i)  for those formed {\it in situ} and (ii) for all stars. The latter are always older than the former, by 1 Gyr for the less metallic ones and by 2.5 Gyr for those of [Fe/H]=0. For higher metallicities, there are no stars formed {\it in situ} and the average age of those present in the solar neighbourhood - coming from inner regions - is around 3-4 Gyr. 
 The middle  panel of Fig. \ref{Fig:LocalMDFa} shows the average birth radius of the local stars.
 The average  birth  radius of all local stars is at $r\sim$6.5 kpc,
 as stated in Sec. \ref{subsec:starmigr}. Star more metallic than the Sun have  birth  radii increasingly closer to the galactic centre, with those of [Fe/H]=0.4 coming from the region  around 2-3 kpc. 
 
 The results of the previous paragraph are presented  in greater detail in  Fig. \ref{Fig:LocalMDFb}.
 The upper panel displays the MDs of four age ranges, namely 0-3 Gyr, 3-6 Gyr, 6-9 Gyr and $>$9 Gyr.
The younger the stars,  the narrower  their MD (because metallicity increases less rapidly at late times) and the more it peaks to higher [Fe/H] values.  The lower panel displays the MDs of five radial ranges, namely $<$3 kpc, 3-5 kpc, 5-7 kpc, 7-9 kpc(="extended solar annulus") and 9-11 kpc. There are practically no stars entering the solar neighbourhood from beyond 11 kpc in our model.
Most stars originate in the 7-9 kpc range and their MD peaks at [Fe/H]=0. The smaller the average birth radius of stars (down to 3 kpc), the more they contribute to the most metallic stars at [Fe/H]=0.4. Stars from the 9-11 kpc range do not contribute to metallicities higher than solar.
  

\begin{figure}
\begin{center}
\includegraphics[width=0.49\textwidth]{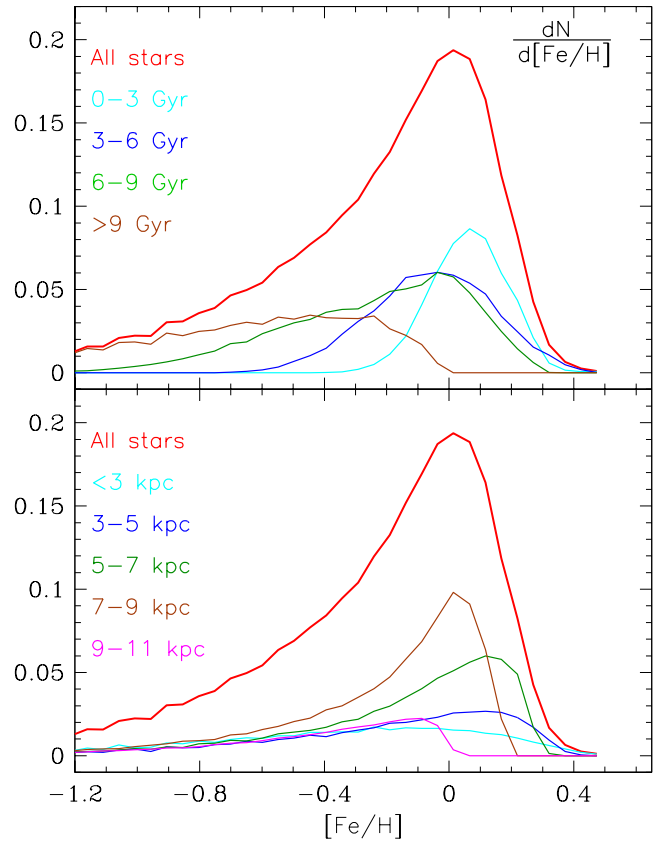}
\caption[]{Solar annulus: {\it Top:} Metallicity distribution in the region of radius $r$=8$\pm$0.25 kpc, with the age ranges of the corresponding stellar populations.
{\it Bottom:} Metallicity distribution in the region of radius $r$=8$\pm$0.25 kpc, with the radial
 ranges of the birth radii of the corresponding stellar populations.
}
\label{Fig:LocalMDFb}
\end{center}
\end{figure}

\subsection{Thin vs. thick disk: age, composition, structure}
\label{subsec:ThinThick}

A large number of formation scenarios of the thick disk have been proposed up to now, involving either external influences or secular evolution through radial mixing, e.g. \cite{Sales2009,Minchev2013,Rix2013} and references therein.
In particular, radial migration has been suggested as a potential thick disk formation mechanism by \cite{SB2009}, who developed a semi-analytical model (similar in some respects to the  present one) and investigated  the resulting chemical and kinematic properties extensively. They showed that for some values in the parameter space, their model can reproduce several key properties of the thick disk just by secular processes:
morphology (i.e. the two-slope  vertical star density profile in the solar neighbourhood), the vertical abundance gradient and the puzzling "two-branch" behaviour of O/Fe vs Fe/H, with the thick disk displaying higher O/Fe values than the thin disk for the same  range of metallicities. Some of those results were also found in N-body simulations, e.g. \cite{Loebman2011}, but others, especially the claimed dynamical behaviour of thick disk stars, were not confimed  and are still  debated
(see e.g.  \cite{Minchev2013}) and references therein).

In this work we do not consider any properties of the disk vertical to the plane (either chemical or kinematic) and we focus on its radial properties, in particular concerning the solar neighbourhood.
In view of the uncertainties as to the best defining feature of the thick disk, we adopt  the simplest possible criterion, namely age. Following  \cite{Binney_sanders2013_arxiv} and \cite{Haywood2013}, we assume
that the thick disk is simply the early period of the MW disk formation, lasting from -12 to -9 Gyr, whereas the thin disk corresponds to the subsequent 9 Gyr of evolution. There is no physical ingredient in our model that marks the transition between the two eras: star formation, infall and radial migration all over the disk are continuous functions of time and space. Still, the consequences of that simple assumption are quite important, as can be seen in Fig. \ref{Fig:LocalMDFc}, where various data are plotted as a function of  [Fe/H] and compared to relevant observations.

\begin{figure}
\begin{center}
\includegraphics[width=0.49\textwidth]{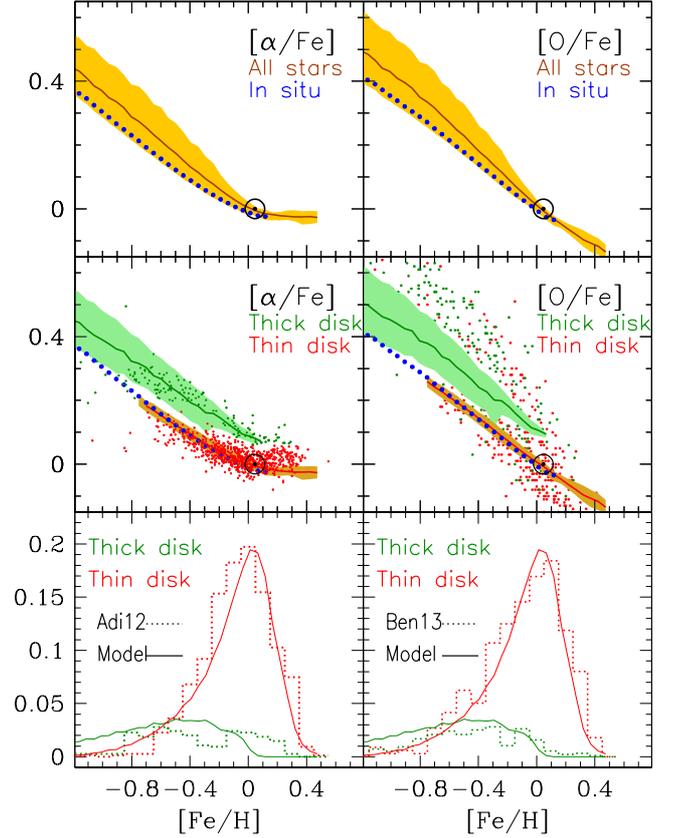}
\caption[]{Solar annulus: {\it Top:} \afe \  ({\it left}) and  [O/Fe] ({\it right}) vs. metallicity, for all stars present today ({\it solid} brown curve) and for all stars born in situ ({\it dotted} blue curve). {\it Middle:} [$\alpha$]/Fe ({\it left}) and  [O/Fe] ({\it right}) vs. metallicity, for stars of the thick disk ($>$9 Gyr, {\it green} curve) and for stars of the thin disk ($<$9 Gyr, {\it red} curve). They are compared to corresponding data from  Adibekyan et al. (2011) (left) and from \cite{Bensby2014} (right).
In {\it top} and {\it middle} panels the {\it shaded areas} enclose the $\pm$1$\sigma$ values around the corresponding model averages.  
The MD data for the thick disk of \cite{Bensby2014} are arbitrarily reduced by a factor of 3 (see text)
}
\label{Fig:LocalMDFc}
\end{center}
\end{figure}

In the upper panels of Fig. \ref{Fig:LocalMDFc} we display  the evolution of \afe \ (left)\footnote{We assume that \afe=([Mg/Fe]+[Si/Fe])/2, although \cite{Adibekyan2011} assume that  \afe=([Mg/Fe]+[Si/Fe]+[Ti/Fe])/3; however, the adopted massive star yields of \cite{Nomoto2013} - but also other yields, e.g. \cite{WW95} - fail to reproduce the observed behaviour of Ti as an $\alpha$ element in the halo of the MW. For that reason, we adopt two $\alpha$ elements with well behaved yields.}   and [O/Fe] (right) vs [Fe/H]
for all the stars currently present in the radial bin $r$=8$\pm$0.25 kpc  and for those born in situ.
As expected, the curve of the \afe \ evolution of the in situ stars lies lower than the one of all the stars present in the solar cylinder. The latter is affected by old stars migrated from inner regions, which have higher \afe for the same value of [Fe/H]. The same holds  for [O/Fe]. 

In the middle panels of Fig. \ref{Fig:LocalMDFc} we display separately the evolutions of the "old" stars ($>$9 Gyr, aka thick disk) and of the "young" stars ($<$9 Gyr, aka thin disk). As expected from the results already presented  in Fig. \ref{Fig:XvsAge}, thick disk stars cover a wide range of metallicities extending to approximately  solar, while
thin disk stars appear around [Fe/H]$\sim$-0.8 and extend up to [Fe/H]$\sim$0.4. 
Our average trends of \afe \ vs [Fe/H] are in good agreement with the data of \cite{adibekian2013}\footnote{We classify stars in the sample of \cite{Adibekyan2011} into thin and thick disks by applying the criterion (dividing line in the [O/Fe] vs [Fe/H] plane) suggested in \cite{adibekian2013}.}, both for the thick and the thin disk and we consider this to be a successful test of the idea that the thick disk can be identified with the component of the disk older than 9 Gyr. We note that the model 1-$\sigma$ dispersion of \afe \ of the thick disk is larger than the corresponding one of the thin disk. 

In the case of the thin disk, the evolution of \afe \ corresponds practically to the one of the {\it in situ} formed stars (the dotted curve, same as in the upper panel), an important feature already identified in the work of \cite{SB2009}. 
However, this does not imply that thin disk stars are mostly formed in situ: a large fraction of thin disk stars are formed in other galactic regions and migrated in the solar neighbourhood, during the last 9 Gyr. But in the beginning of that period, the value of \afe \ {\it all over the inner disk} was  already  reduced to $\sim$0.2 - because of the previous activity of SNIa -  and it evolved very slowly afterwards (see Fig.  \ref{Fig:XvsAge}, bottom panel).

Our results for [O/Fe] vs [Fe/H] are displayed in the right top and middle panels of Fig. \ref{Fig:LocalMDFc} and they are compared with the data of \cite{Bensby2014} \footnote{We use the criterion suggested by \cite{Bensby2014} for the classification into thin and thick disk, namely probabilities higher than 2 for the former and smaller than 0.5 for the latter.}. The behaviour is qualitatively, but not quantitatively, similar to the one of \afe. Although oxygen is an $\alpha$ element, its evolution is slightly different from the one of \afe \ here, because we included Si in the definition of \afe. A non-negligible amount of Si  is produced by SNIa, while this is not the case for oxygen. As a result, [O/Fe] varies more than \afe \ extending both to higher and lower values. However, despite the resulting large variation 
in [O/Fe], our results do not satisfactorily match the data of \cite{Bensby2014} which extend from [O/Fe]$>$0.6 to values lower than -0.2. The decline of [O/Fe] in the data is much steeper than in our model. It is difficult to say whether this discrepancy is due to the model (inadequate yields or SNia rates) or to the data (uncertainties in the oxygen abundances). 

In the bottom panel of Fig. \ref{Fig:LocalMDFc} we present the resulting metallicity distributions for the thin and thick disks, and we compare them to the data of \cite{adibekian2013} \ (left) and \cite{Bensby2014} (right).
The metallicity distribution of the thick disk is already presented in Fig. \ref{Fig:LocalMDFb} (top panel): it is the curve corresponding to stars old than 9 Gyr. It is quite broad, it extends up to [Fe/H]$\sim$0 and it peaks at  [Fe/H]$\sim$-0.6 to -0.5. the thin disk MD of the model is much narrower, it extends up to [Fe/H]$\sim$0.45 and
it peaks at [Fe/H]$\sim$0. 

The thin disk MD is in excellent agreement with the data of both \cite{adibekian2013} and \cite{Bensby2014}. 
As for the thick disk MD, it extends only up to [Fe/H]=0 and thus  does not reach the metallicity range of {\it high-alpha metal-rich stars} ($hamr$) in the sample of \cite{adibekian2013}, which appears to be bi-modal in that respect; this bi-modality, which still needs to be confirmed, cannot be explained in the framework of our model. 
The sample of \cite{Bensby2014} is biased in favour of thick disk stars (it contains 234 stars in a total of 629), so we reduced the proportion of its thick disk a factor of 3, to bring it in agreement with our results in Fig. \ref{Fig:LocalMDFc}.

The results of Fig. \ref{Fig:LocalMDFc}  confirm the suggestion  of \cite{SB2009} that radial migration can explain the  two branches observed in the [O/Fe] vs [Fe/H] relation. They also provide further support to the idea that the local thick disk was formed by that process, since they siccessfully reproduce the local distributions of both the thick and the thin disk. However, the absence of any features in our model (either morphological or kinematic) concerning the direction vertical to the Galactic plane, does not allow us to explore this issue further.

We now turn to the radial distribution of the thick and thin disks in our scenario. In Fig. \ref{Fig:ThickThin_Prof} we show (upper panel) the radial surface density profiles of stars older and younger than 9 Gyr, respectively,  after radial migration. The main results concerning local surface densities and scalelengths ar displayed in Table \ref{Tab:ThinThickDisk}.

\begin{figure}
\begin{center}
\includegraphics[width=0.49\textwidth]{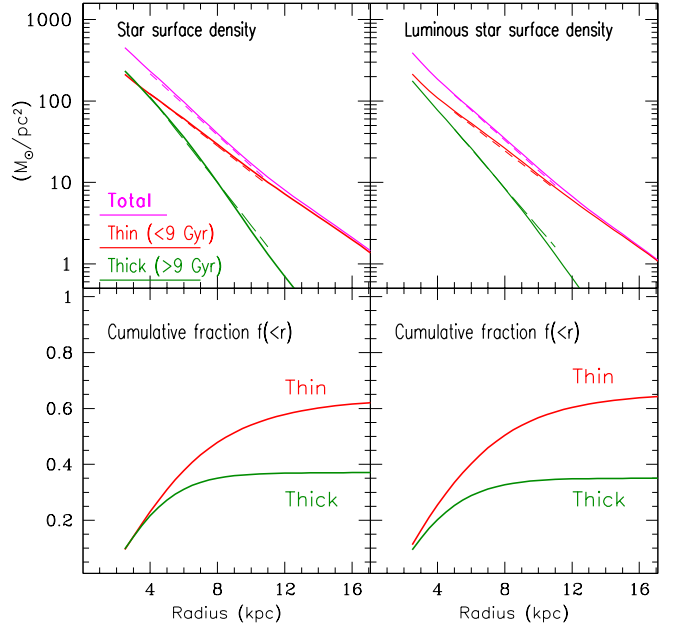}
\caption[]{Thins vs. thick disk: {\it Left:} Stars plus stellar remnants; {\it Right:} Luminous stars only. {\it Top:} Present-day stellar profiles: {\it magenta}: Total;  {\it green}: (thick disk);   {\it red}: thin disk. In all curves, there is an attached best-fit exponential (dashed line of same colour) in the 5-11 kpc range, with scalelengths and column densities at $R_0$=8 kpc as
reported in Table \ref{Tab:ThinThickDisk}. {\it Bottom:} Cumulative fraction of the thin and thick disk masses with respect to the corresponding total disk mass $M_{disk}$: $M_{thin}(<r)/M_{disk}$ (red curve) and $M_{thick}(<r)/M_{disk}$ (green curve) as a function of radius $r$. }
\label{Fig:ThickThin_Prof}
\end{center}
\end{figure}

The recent work of \cite{Bovy2013} who  analysed the dynamics of 16 269 G-type dwarfs from SEGUE (sampling the radial range
5 kpc $< r <$ 12 kpc) leads to a 	dynamically determined  surface density of stars+remnants $\Sigma_*$=38$\pm$4 \mpc at $r$=8 kpc. In our case, stellar remnants contribute  5 \mpc, and stars still shining  33 \mpc (our IMF extending from 0.1 \ms \ to 100 \ms, we included no brown dwarfs).  We note that in their analysis of
the local surface density contributed by mono-abundance populations,
\cite{Bovi2012} find a total surface density contributed by stars of  
$\Sigma_*$=30$\pm$1 \mpc, which depends slightly on the adopted IMF: for a \cite{Kroupa2001} IMF, they obtain  $\Sigma_*$=32 \mpc, in excellent agreement with our results.
Our total baryonic surface density in the solar neighbourhood, 
 is 51.2 \mpc, including 13.2 \mpc for the total gas. This is in agreement with the local baryonic contribution of $\Sigma_{gas}+\Sigma_*$=55$\pm$5 \mpc \ estimated in \cite{Zhang2013} or  $\Sigma_{gas}+\Sigma_*$=51$\pm$4 \mpc \ estimated in \cite{Bovy2013}.

\begin{table}
\caption{\label{Tab:ThinThickDisk}{Properties of the Galactic disks ($r$>2 kpc).}
}
\begin{tabular}{lcccl}
 \hline \hline
   &   Thin &    Thick & Total & \\
   \hline
$\Sigma$(8kpc): Stars  & 24.5  & 8.5 & 33  & \mpc   \\
$\Sigma$(8kpc): Stars+Remnants  & 28  & 10 & 38  & \mpc \\
$\Sigma$(8kpc): Remnants  &  3.5 & 1.5 & 5 & \mpc \\
Scalelength: Stars  & 2.7  & 1.8  & 2.35 & kpc \\
Scalelength: Stars+Remnants  & 2.7   &  1.65 & 2.25 &  kpc \\
Mass: Stars+Remnants  & 2.2 & 1.2 & 3.4 &  10$^{10}$ \ms \\
\hline
 
\end{tabular}  
 \end{table} 


The scalelength of the total (stars plus stellar remnants) disk of our model (2.25 kpc) is in excellent agreement with the recent dynamical estimate of \cite{Bovy2013} for the Milky Way:
2.15$\pm$0.14 kpc. We note that their analysis of the combination of constraints from
 rotation-curve shape  and surface-density measurements leads to a local value
for the rotational velocity of $V_c$ = 218 $\pm$ 10 km/s, which is not very different from
 the
 value of 212 km/s of our model (see Fig. \ref{Fig:TheorObsProf}). The total stellar mass of our disk is lower than the value of 4.6 10$^{10}$ \ms \ obtained by \cite{Bovy2013}, because
they obviously include the region inside 2 kpc in their calculation of the disk mass, while
 we do not include it. For those 
 reasons, our estimate of the  total stellar disk mass of 3.4  10$^{10}$ \ms (end point of our calculations in Fig. \ref{Fig:Evol0}) corresponds to the total stellar mass outside $r$=2 kpc. If we add  the final stellar mass 
1.3  10$^{10}$ \ms \ that we find inside 2 kpc, we obtain a total stellar mass of 4.7 10$^{10}$ \ms, again in agreement with the estimates of \cite{Bovy2013}.

We systematically find that the  thick disk is more centrally condensed than the thin disk, i.e. with a scalelength  shorter by $\sim$1 kpc. The reason is  the inside-out star formation of the Galaxy: the old stellar population is more centrally condensed than the young one and radial migration does not change  that feature (although it certainly attenuates it).

The scalelength of the stellar thick disk in our model (1.8 kpc) is 
in fair agreement with recent estimates, e.g. \cite{Bensby2011},  
\cite{Cheng2012}, 
\cite{Bovy2012} 
who find 2., 1.8$^{+2.1}_{-0.5} $ and 2.01$\pm$0.05 kpc, respectively.
However, the scalelength of our thin disk (2.35 kpc) is substantially shorter than those obtained in these studies (respectively, 3.8, 3.4,  and 3.6$\pm$0.2 kpc).

As already discussed extensively in the literature (e.g. \cite{Rix2013} and references therein)
the radial extension of the thin and thick disks can be used
to constrain various scenarios for thick disk formation. In particular,
the lack of high-$\alpha$ stars at $r>$10 kpc and large distances from the plane, constrains
the strength of migration due to transient spiral
structure: it cannot be very efficient beyond that distance. In our case, the thick disk  has a short scalelength and the efficiency of radial migration is indeed small beyond $r$=11-12 kpc, as seen in  the profiles in the top and middle panels of 
Fig. \ref{Fig:MigrProf}. However, the thin disk appears  to be  shorter than in the observations.

The lower panels of Fig. \ref{Fig:ThickThin_Prof} display the corresponding cumulative fractions of the thin and thick disks as a function of distance $r$. As already stated in Sec. \ref{sec:GlobEvol} and Fig. \ref{Fig:Evol_new}, we define here the disk as the region of radius $r>$2 kpc, the region inside belonging to the bulge. The resulting fractions differ very little in the two cases (stars+ remnants and stars only): the thick disk contributes almost as much as the thin disk in the inner Galaxy (where a large fraction of the stars was formed in the first 3 Gyr) and its contribution gradually lags behind the one of the thin disk with radius. In total, the thick disk weighs about half the mass of the thin disk, i.e. their stellar masses are 1.2 10$^{10}$ \ms \ and 2.2 10$^{10}$ \ms, respectively.

The derived mass ratio  of the thick and thin disks of the MW M$_{thick}$/M$_{thin}\sim$1/2 is substantially higher than expected from observations of external disk galaxies by  \cite{Yoachim2006}. They find that M$_{thick,*}$/M$_{thin,*}$ is a decreasing function of the rotational velocity of disks, ranging from 1-2 at $V_C\sim$70 km/s to 0.2 at $V_C\sim$200 km/s; the ratio we find is almost 3 times larger than the latter value. On the other hand, based on {\it Spitzer} observations of a sample of $\sim$30 
galaxies, \cite{Comeron2011} find that thick and thin disks have, typically, similar masses.
Our value of 0.5 lies between those two results.

\begin{figure}
\begin{center}
\includegraphics[width=0.49\textwidth]{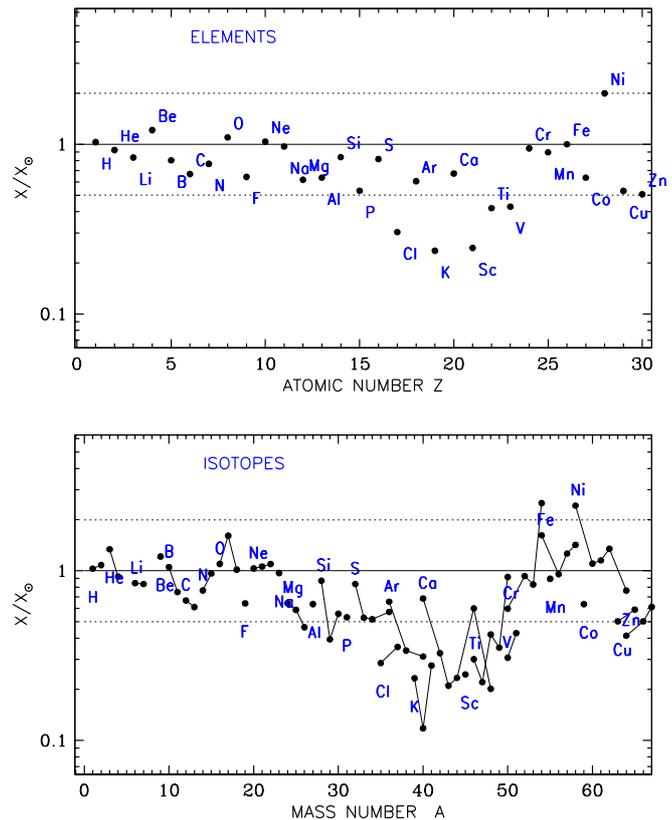}
\caption[]{Average composition of 4.5 Gyr old stars in the solar neighbourhood, elemental ({\it top}) and isotopic ({\it bottom})  compared to the observed solar composition. Most elements and isotopes are co-produced within a factor of two of their solar value. The corresponding values for the local ISM 4.5 Gyr
ago are $\sim$0.1-0.15 dex lower for all elements but H and He. No normalisation of the results is made here.
}
\label{Fig:SolarComp}
\end{center}
\end{figure}

\subsection{Abundances in local thin and thick disks }
 \label{subsec:Abundances}
 
Up to now, we used only two elements, namely O and Fe, to study the chemical evolution of the MW disks.
 These are key elements in galactic chemical evolution studies, because they are abundant and easy to measure. However, the evaluation of oxygen abundances in stars is not straightforward (see e.g. the monography by Stasinska 2012 and references therein), while the evolution of Fe is affected by uncertainties on its massive star yields and on the rate of SNIa.

\begin{figure}
\begin{center}
\includegraphics[width=0.49\textwidth]{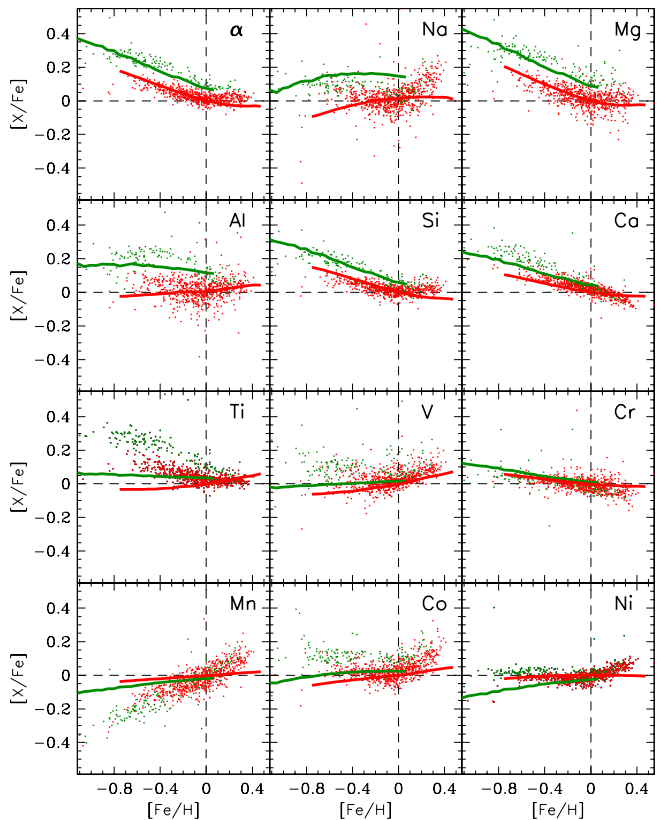}
\caption[]{[X/Fe] vs [Fe/H] for various intermediate mass elements in the solar neighbourhood: comparison of the average stellar abundances of the model to data of \cite{adibekian2013}. The data are split into thick disk (green) and thin disk (red) according to the prescription of \cite{adibekian2013},  in the middle panel of their Fig. 2). The model results are normalised such as the average abundance over all stars of age=4.5 Gyr that are present today in the solar neighbourhood is solar (see text).}

\label{Fig:ThickThin_Adi}
\end{center}
\end{figure}

\begin{figure}
\begin{center}
\includegraphics[width=0.49\textwidth]{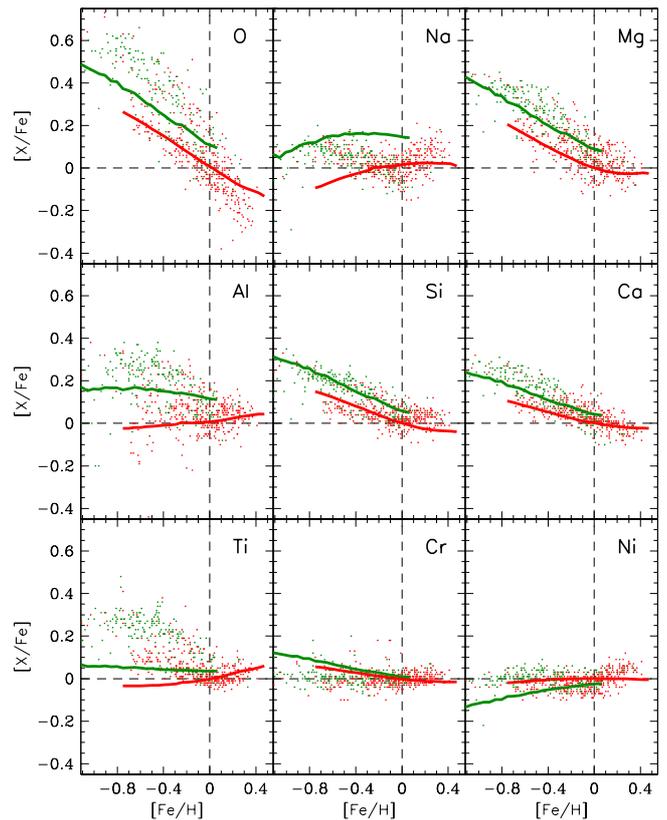}
\caption[]{ [X/Fe] vs [Fe/H] for various intermediate mass elements in the solar neighbourhood: comparison of model to data of \cite{Bensby2014} The data are split into thick disk (green, $P>$2) and thin disk (red, $P<$0.5) according to the probabilities $P$ given in \cite{Bensby2014}. The model results are normalized so that the average abundance over all  stars of age=4.5 Gyr that are  present today in the solar neighbourhood is solar.
 }
\label{Fig:ThickThin_Ben}
\end{center}
\end{figure}

 A wealth of data from recent and forthcoming surveys will allow one to explore the diagnostic potential  of many more elements - produced mainly in massive stars - thereby complementing  the information obtained from O and Fe.
 Such observations will also help to constrain the stellar yields, which suffer still from large uncertainties in some cases (see \cite{Romano2010} for a recent study of the impact of stellar yields on the chemical evolution of the MW). 
 
Figure \ref{Fig:SolarComp} presents our results for the average composition of local stars of 4.5 Gyr. As discussed in previous sections, the O and Fe abundances are very close to solar, making  the analysis of Sec. \ref{sec:LocalEvol} possible. For other elements, however, the situation is less satisfactory: this is the case for all elements between (and including) P and Ti and most of their isotopes, which are severely underproduced. Obviously, if the yields of \cite{Nomoto2013} were taken at face value, the observed evolution of those elemental ratios  vs. Fe/H would not be reproduced.


We assume here that the isotopic yields of \cite{Nomoto2013} differ from the "real life" yields by various factors (of the order of unity), but their metallicity dependence is correct. We correct for those factors by normalising the average composition of local stars 4.5 Gyr old to be exactly solar and we apply the derived correction factor for each isotope to its evolution at all places. 
In that way,  we force all isotopic abundances and ratios to be exactly solar for an average local star of age 4.5 Gyr, because we assume that the Sun is such an average star. We proceed then with the analysis of the thin vs thick disks as in the previous section for O vs Fe.

We display our results in Fig. \ref{Fig:ThickThin_Adi}, and compare them with data from the survey of \cite{adibekian2013} for the local thin and thick disks. It can be seen that
the observations of the $\alpha$ elements are nicely reproduced in the framework of our model: the model accounts quantitatively for the evolution of Mg, Si, Ca and Cr in both the thin and thick disks. A qualitative agreement is also obtained for the cases of Na and Al, while Co and Ni reproduce  the data for the thin disk alone. In contrast,  the model fails to reproduce Ti (which behaves observationally as an $\alpha$ element), V, Mn and Co. The failure of the adopted yields to reproduce these observations does not necessarily imply that
there is something  wrong with the yields, at least not in all cases.
After all, the solar abundances of a large number of elements still suffer from systematic uncertainties and are subject to revision  (see e;g. \citealt{Scott2014a} for the Fe group elements). Cross-checks with respect to other observational data (concerning e.g. the halo stars) should be made and the role of the IMF investigated before concluding. We note here that there is certainly a problem with the
\cite{Nomoto2013} yields for the cases of Ti and V, as can be seen from Fig. 10 in that work, where comparison is made of a simple GCE evolution model with halo and local disk data: it is obvious that, even after normalisation to solar values, the observed evolution of Ti and V cannot be reproduced.

In a similar vein, we compare our results to the data of the \cite{Bensby2014} survey in Fig. \ref{Fig:ThickThin_Ben}.
The agreement of the model to observations of $\alpha$-elements Mg, Si and Ca is excellent, as in Fig. \ref{Fig:ThickThin_Adi}; in contrast, the evolution of oxygen is poorly reproduced, as discussed in Sec. \ref{subsec:ThinThick}. The agreement with Al and Na is qualitatively, but not quantitatively, good and the failure in the case of Ti as bad as before.

We note that the separation into thin and thick disks is made on the basis of different criteria in \cite{adibekian2013} and \cite{Bensby2014}: chemical vs kinematic, respectively.
Despite that, there is rather good agreement between the two data sets, making a comparison of model to the data meaningful.
Here we attempted such a comparison for the first time, showing the strong potential of such detailed observations to constrain both stellar yields and evolutionary models for the MW disk.

\section{Summary}
\label{sec:summary}

In this work we presented a model for the evolution of the MW disk, involving radial motions of gas and stars. 
We considered the epicyclic motion of stars (blurring) separately from the true variation in their guiding radius (churning). 
Our parametrisation of radial migration corresponds to a barred disk galaxy, like the MW, in contrast  to the work of \cite{SellwoodBinney2002} or \cite{SB2009}.
We compared our results to an extended set of recent observational data, concerning the global evolution of the MW, the present-day radial profiles of various quantities and the solar cylinder. 

Our model reproduces the present-day values of all the main "global" observables of the MW disk and bulge (here taken to correspond to regions outside and inside $r$=2 kpc, respectively): stellar, atomic and molecular gas masses, rates of infall, SFR, CCSN and SNIa (Fig. \ref{Fig:Evol0}). 
 We obtain a very good agreement between the model evolution of the bulge and disk stellar mass and the corresponding observations of the "stacked evolution" of MW-type disks of \cite{Dokkum2013} (Fig. \ref{Fig:Evol01}) and in the light of these data, the MW appears as a fairly average disk galaxy. 
The present-day profiles of stars, gas (\hatm \ and \hmol), SFR, and rotational velocity are also reproduced well by the model (Fig. \ref{Fig:TheorObsProf}). 

The main focus of this study was the impact of stellar radial migration on the properties
of the Galactic disk. We find that, with the adopted scheme for radial migration (blurring+churning), the regions mostly affected  are those in the range 4-12 kpc and in particular the zone between 5 and 9 kpc. Stars in those regions are formed on average $\sim$1-2 kpc inwards of their current position and are, on average, $\sim$1-2 Gyr older than locally formed stars (Fig. \ref{Fig:MigrProf}).  This implies, in particular, that the Sun
was probably formed $\sim$1.2 kpc inwards of its present position of $R_0$=8 kpc from the Galactic centre. 

As already shown in previous studies, we find that radial migration brings mostly old and metal-poor stars into the solar neighbourhood, thus flattening the age-metallicity relation. It also considerably increases the dispersion in metallicity at every age, making it larger with age, as found in \cite{SellwoodBinney2002}. 

 We emphasize that the   local observables of  our model concern the so-called "solar cylinder", that is all stars found in the end of the simulation in a cylinder of radius 0.25 kpc (half the size of our radial bin), perpendicular to the Galactic plane and centered on the solar position, at Galactocentric distance $R_0$=8 kpc. We do not apply any selection biases on those
results in order to compare with specific  observations. For that reason, a successful comparison to observations does not  imply that the model is necessarily correct, only that it possesses potentially interesting features. 

 We show quantitatively that - at least in the framework of this model and with the caveat of the previous paragraph - epicyclic motions  cannot produce the observed metallicity dispersion; in contrast, our adopted radial migration scheme reproduces available observations of dispersion in the age-metallicity relation (Fig. \ref{Fig:XvsAge_Disp}). We argue that this observable provides one of the most powerful probes of the extent of radial migration in the MW and that it should be scrutinized in future observational and theoretical studies. On the other hand, the local [O/Fe] abundance ratio is found to have very little dispersion with age, making it a much better  proxy for age than [Fe/H] and this is  true for other [$\alpha$/Fe] ratios.

 We analyse the origin of the stellar populations presently found in the solar annulus as a function of their metallicity (Fig. \ref{Fig:LocalMDFa}). We find that, at all metallicities, stars are $\sim$1 Gyr older, on average,  than locally formed stars of the same metallicity; they also display an age dispersion of $\sim$1-3 Gyr around the average age. Cleary, radial migration affects  the relation between age and metallicity dramatically, allowing for young stars of low metallicity to co-exist with old stars of high metallicity. In  our model, the less metallic local stars are $\sim$11 Gyr old,  and they originate, on average, in
the region at $r\sim$5-6 kpc, while stars of solar metallicity are $\sim$4.5 Gyr old and originate at $r\sim$7 kpc. The most metallic local  stars ([Fe/H]$\sim$0.3-0.4 \zs) are 3-4 Gyr old and originate in the inner Galaxy, at $r\sim$2-3 kpc. We find that stars of different ages and different  birth  radii contribute to some extent to all metallicity bins (Fig. \ref{Fig:LocalMDFb}).}

 To handle the issue of the thin vs thick disks we adopt a simple criterion that has already been suggested in the literature, namely age (another reason for that choice being the absence of the z-direction in our model, making it impossible to use kinematic or spatial quantities in that direction).  We assume that stars older than 9 Gyr belong to the thick disk and younger ones to the thin disk. That simple criterion allows us to reproduce fairly well the observed [$\alpha$/Fe]  vs. Fe/H behaviour of stars  classified as belonging to the thin and thick disks on the basis of different criteria (chemical or kinematic). At the same time, we reproduce the corresponding metallicity distributions, very well for the thin disk and satisfactorily for the thick disk (Fig. \ref{Fig:LocalMDFc}).
Both results provide strong support to the idea that the thick disk is simply the early part of the MW disk (corresponding to the first few Gyr of its formation) and that the local thick disk results largely from the radial migration of stars from the inner disk.

 We evaluate quantitatively the radial structure of the thin and thick disks in our model. We find that, because of the inside-out formation
 adopted, the thick disk has a considerably shorter scalelength than the thin disk (Fig. \ref{Fig:ThickThin_Prof}), even though it has undergone a much more important radial migration than the latter. The local surface densities of both disks are in excellent  agreement with recent evaluations,  and this is also true for the scalelength of the thick disk; however, we obtain a thin disk with a scalelength shorter by $\sim$1 kpc than the one observationally determined. Overall, the thick disk accounts for one-third of the total stellar disk  and for one-fourth of the local stellar disk surface density. 
 
 Finally, we investigate  the evolution of several elemental abundances in the local thin and thick disk and we compare our results to the large  data sets obtained from recent surveys. We argue that such a study requires  fine and homogeneous grids of stellar yields (such as the one provided by \cite{Nomoto2013}, adopted here). We find that several observed features, i.e. the behaviour of [$\alpha$/Fe] vs Fe/H, can be reproduced fairly well for both the thick and thin disks, but the situation is much less satisfactory for other elements. Current and forthcoming data in this field will provide powerful constraints on stellar nucleosynthesis and the overall evolution of the MW disk, especially when combined with kinematic and spatial information.

\medskip
\noindent
{\it Acknowledgments}: We are grateful to  T. Bensby for providing data before their publication and advice on their use. EA acknowledges financial support to the DAGAL network from the People Programme  (Marie Curie Actions) of the European Union's Seventh
Framework Programme FP7/2007-2013/ under REA grant agreement number
PITN-GA-2011-289313 and from the CNES (Centre National d'Etudes
Spatiales - France). We also acknowledge partial support from the PNCG
(Programme National Cosmologie et Galaxies - France) and from the European Science Foundation (ESF)  project EuroGENESIS/MASCHE.  The simulation used for the modelling in this paper was run using the HPC resources from GENCI - TGCC/CINES (Grants x2013047098 and x2014047098)

\begin{appendix}
\section{Gas in the MW disk}
\label{App:Gas}

\begin{figure}
\begin{center}
\includegraphics[width=0.49\textwidth]{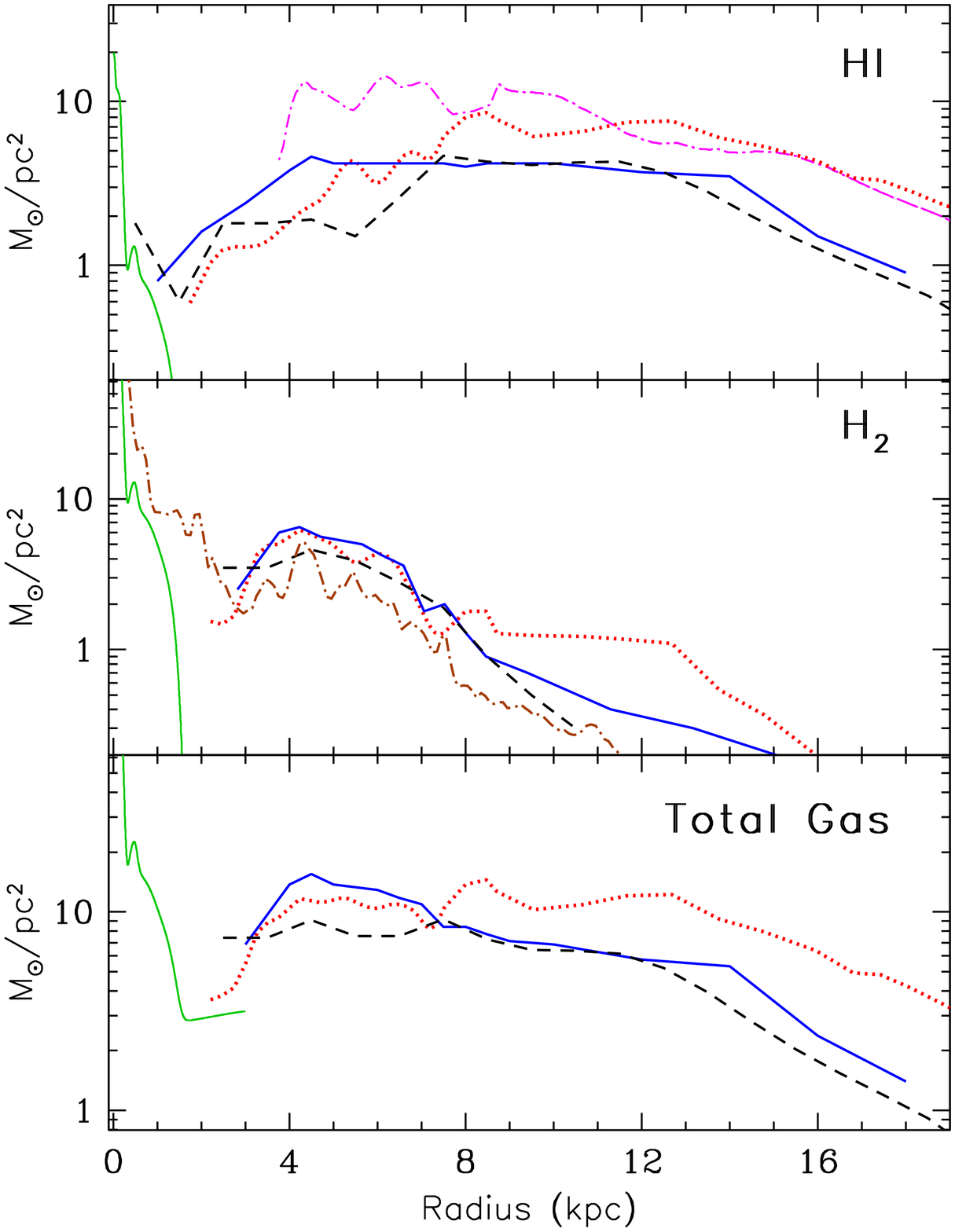}
\caption[]{Surface density profiles of atomic hydrogen \hatm (top), molecular hydrogen \hmol (middle) and
total gas $\Sigma_{G}$=1.4 (\hatm+\hmol) (bottom). Disk data 
(beyond 2 kpc) are from: \cite{Dame93}, solid; \cite{OllingMer01}, dotted ; and \cite{NakaSofue03,NakaSofue06}, dashed. The dot-dashed curve in the \hatm \ panel corresponds to data from \cite{Kalberla08}
and in the \hmol \ panel to data from \cite{Pohl08}.
 Bulge data (inner 2 kpc) in all panels are from Ferri\`ere (2001).
}
\label{Fig:GasProf}
\end{center}
\end{figure}

\begin{figure}
\begin{center}
\includegraphics[width=0.49\textwidth]{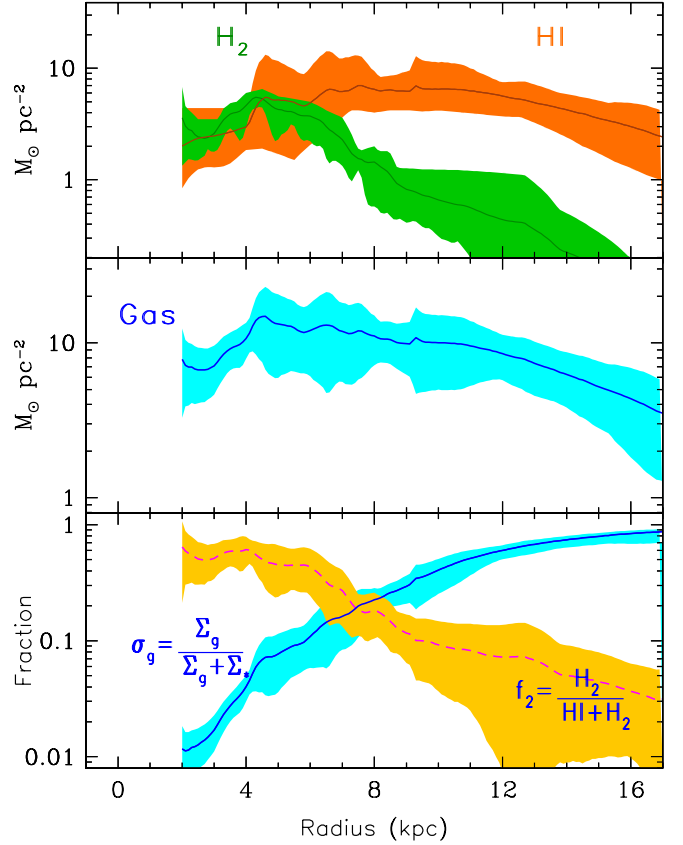}
\caption[]{Average profiles of the previous figure. {\it Top}: atomic and molecular  hydrogen; {\it middle}: total gas ; {\it bottom}: fractions of gas (solid) and molecular gas (dashed). For the latter panel,
an exponential stellar density profile with a scale-length of 2.3 kpc normalised to $\Sigma_*$(R=8 kpc)=38 \mpc is adopted, while the molecular fraction is evaluated with the formalism of Sec. \ref{App:SFR}.
}
\label{Fig:GasFrac}
\end{center}
\end{figure}

In Fig. \ref{Fig:GasProf} we present the data adopted for the gas profiles of the MW. Atomic hydrogen (\hatm) \ is displayed in the upper panel and molecular hydrogen (\hmol)  in the middle panel. In both cases, data are from \cite{Dame93},\cite{OllingMer01} and \cite{NakaSofue03}. The data have been rescaled to a distance of R$_0$=8 kpc of the Sun from the Galactic centre. 
The bottom panel shows the total gas surface density $\Sigma_G$=1.4(\hatm+\hmol), where the factor 1.4 accounts for the presence of $\sim$28\% of He.
In all panels, the data for the inner Galaxy (R$<$2 kpc) are from \cite{Ferriere2007} and they are provided for completeness, since the evolution of the bulge is not studied here.

Despite systematic differences in the data, the gas profiles in the three panels of Fig. \ref{Fig:GasProf} have some common features:

- The \hatm \ profile is essentially flat in the 4-12 kpc region.

- The \hmol \  profile displays the well known "molecular ring" in the 4-5 kpc region and declines rapidly outwards.

- The total gas profile is approximately constant (or slowly declining outwards) in the 4-12 kpc region; it  declines more rapidly inside the molecular ring, as well as outside 13 kpc, where it has a scalelength of 3.75 kpc \citep{Kalberla08}.

To minimize systematic uncertainties in the following, we adopt
 the averages of the aforementioned observed profiles as a function of galactocentric radius as the "reference gaseous profiles" for the MW disk. They appear in Fig. \ref{Fig:GasFrac}, top panel for HI and H$_2$ and middle panel for the total gas. 
 As a typical uncertainty in each radius, we adopt either
 50\% of the average value (typical statistical uncertainty in e.g.
 \cite{NakaSofue06} or half the difference between the minimum and maximum values in each radial bin (whichever is larger). The resulting values for the total galactic content of \hatm, \hmol \ and gas appear in Table \ref{Tab:TableGas}. The derived masses appear lower than
 the ones obtained with the mass model of the ISM in the Milky Way of \cite{Missi06},
 who find total masses of  M$_{H_2}$=1.3 x 10$^9$ \ms \ and M$_{HI}$=8.2 x 10$^9$,  but they match the FIR emission of the whole
 Galaxy, whereas we quote here results for the 2-19 kpc range. There
 are considerable amounts of \hatm \ in the outer disk, as discussed
 in e.g. \cite{Kalberla08}. 
 
\begin{table}
{\small 
\caption{\label{Tab:TableGas}{Gas in the MW disk (2-19 kpc) in 10$^9$ \ms.}
}
\begin{tabular}{lccc}
 \hline \hline
   &
   \hatm &
   \hmol &
   Total$^1$ \\
   \hline 
\cite{Dame93} & 3.25  & 1.12 & 6.10\\
\cite{OllingMer01} & 5.85 & 1.21 & 9.90  \\
\cite{NakaSofue03,NakaSofue06} & 2.75 & 0.71 & 4.86\\
\cite{Kalberla08} & 7.7  &  & \\
\cite{Pohl08} &   &  0.55& \\
Adopted average &4.90$\pm$2.2  & 0.9$\pm$0.35 &  8.2$\pm$3.5 \\
\hline
 
\end{tabular}  

1: Total includes 0.28 of He by mass fraction.}
\end{table}

We find then that the Galactic disk has a total gaseous content of $\sim$8.2$\pm$3.5 10$^{9}$ \ms  in the 2-19 kpc range. 
Assuming an exponential stellar profile with a scalelength \rd=2.3 kpc for the MW disk, normalised to a local (\rsol=8 kpc) surface density $\Sigma_{*,0}$=38 \mpc \ \citep{Flynn2006}, we find a total
stellar mass of 3.2 10$^{10}$ \ms \ and
obtain the radial profile of the gas fraction $\sigma_G$(R)=$\frac{\Sigma_G}{\Sigma_G+\Sigma_*}$.
It is displayed in the bottom panel of Fig. \ref{Fig:GasFrac} and it is a monotonically increasing function of radius. 
Integrating as before the gaseous and stellar profiles over the disk region between 2 and 19 kpc we find an average gas fraction
of $\sigma_G$=0.20$\pm$0.05 for the disk. If the  bulge (of stellar mass $\sim$1.5 10$^{10}$ \ms \ and negligible gas) is also included, the gas fraction of the MW is found to be $\sim$14\%.

In that same panel we display the molecular fraction 
$\rm f_2$(R)=$\frac{\Sigma_{H2}}{\Sigma_{HI}+\Sigma_{H2}}$, which shows a plateau of $\rm f_2\sim$0.65 in the region of the molecular ring (3-6 kpc) and decreases strongly outwards, down to a few per cent (although with large uncertainties).

\section{Star formation in the MW disk}
\label{App:SFR}

\begin{figure}
\begin{center}
\includegraphics[width=0.49\textwidth]{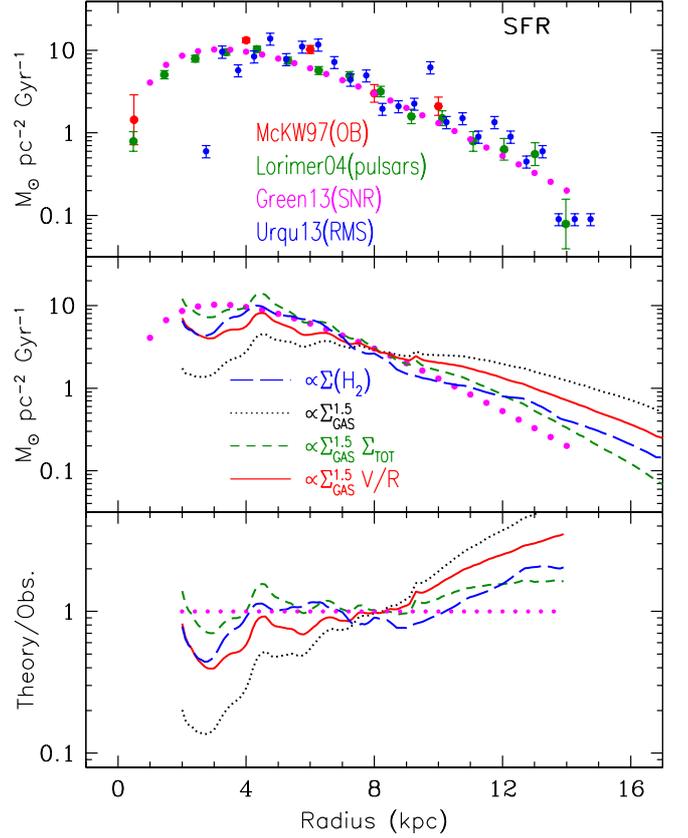}
\caption[]{{\it Top}: Observed surface density profiles of various SFR tracers (see text); the dotted curve - with no error bars- is the analytical form suggested by Green (2013) and it is here adopted as representative of the MW SFR profile. {\it Middle}: Theoretical or empirical SF rates compared to the adopted  profile of SFR tracers (the dotted curve from the upper panel); all profiles are normalised to the same value in R$_0$=8 kpc. {\it Bottom}: ratio of the theoretical or empirical profiles to the adopted observed one.  
}
\label{Fig:SfrProf}
\end{center}
\end{figure}

\begin{figure}
\begin{center}
\includegraphics[width=0.49\textwidth]{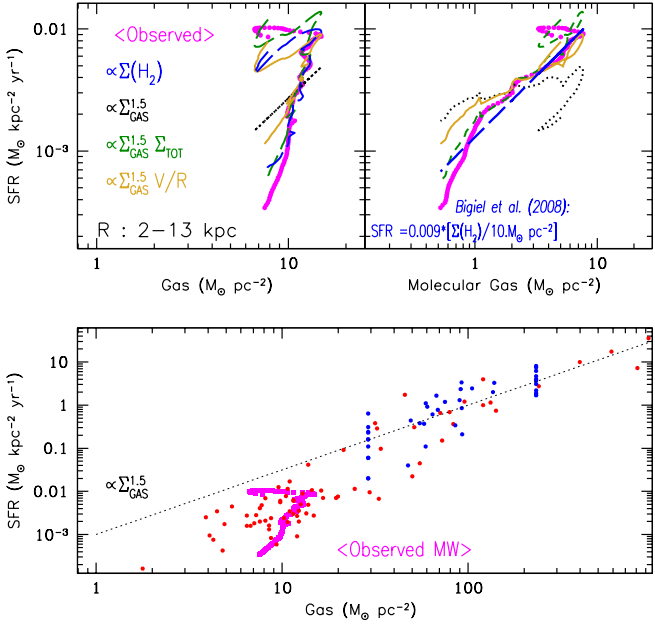}
\caption[]{{\it Top:} SFR surface density vs total gas surface density ({\it left}) and
vs. \hmol \ ({\it right}) for the MW disk, in the region between 2 and 13 kpc. 
In both panels, curves 
represent the same quantities as in Fig. \ref{Fig:SfrProf}; the long dashed curve
(SFR$\propto\Sigma({\rm H_2})$) corresponds quantitatively to the fit of \cite{Bigiel08} to extragalactic data and fits  the adopted "observed" SFR profile in the MW disk quite well.
{\it Bottom}: The "observed" SFR vs. gas relation in the MW is compared to a compilation of extragalactic
data from \cite{Krumholz12}.
}
\label{Fig:SfrGas}
\end{center}
\end{figure}

 Star formation  in the MW  is discussed in the recent review of 
 \cite{KenniEvans12}. They discuss only "traditional" tracers of star formation, also used in extragalactic studies, like FIR emission.
In their Fig. 7, they display a radial distribution of the SFR, claimed to be based
on data from \cite{Missi06}, who made a full 3D model of the Galactic
FIR and NIR emission observed with  COBE. 
\cite{Missi06} found that their modelling of the IR emission corresponds to a SFR$\propto \Sigma_G^{2}$ and they compared their findings to a compilation of old star formation tracers for the MW provided in \cite{BP99}, who considered
HII regions, but also pulsars and supernova remnants. Here we 
consider such tracers related to  massive, short-lived, stars and their residues: pulsars, supernova remnants (SNR),  and OB associations.
Those objects have ages up to a few My for SNR and up to a few tens of My for OB associations and isolated radio pulsars, so they can probe recent star formation.

Galactic radial distributions of luminous massive stars, pulsars, SNR and OB associations
appear in Fig. \ref{Fig:SfrProf}.The distribution of luminous massive stars is from \cite{Urqu13}; the one displayed on Fig. \ref{Fig:SfrProf} is the average between the southern and northern hemispheres and is considered to be complete for luminisities above 2 10$^4$ \ls.
The pulsar distribution {\citep{Lorimer04,Lorimer06} contains more than 1000 pulsars. 
All distributions have a broad peak in the region of the molecular ring and their radial variation
agrees well with the much sparser data of \cite{McKee97} on OB associations. \cite{Lorimer06} proposed 
an analytical fit to the pulsar distribution, which is in
perfect agreement with the one proposed for the distribution of 
Galactic SNR in the recent compilation of \cite{Green13}, who used 56 bright SNR (to avoid selection effects).  The latter distribution (thick dotted curve, corresponding to model C of that work) is also 
displayed in the upper panel of Fig. \ref{Fig:SfrProf} and is given by
\begin{equation}
\rm \Psi(R) \ = \ A \ \left(\frac{R}{R_0}\right)^B \ exp\left[-C(\frac{R-R_0}{R_0})\right]
\label{Eq:SFR}
\end{equation}
where $R_0$=8 kpc and parameters B=2. and C=5.1 (\citet{Lorimer06} give
B=1.9 and C=5.).

There is reasonably good agreement between all the SFR tracers of Fig. \ref{Fig:SfrProf}, with the exception of the one for luminous massive stars of the RMS survey \citep{Urqu13}, which differs considerably from the others inside 4 kpc (where selection biases are expected to be more important, even for such luminous stars) and displays an unexpected enhancement in the region around 9 kpc. Baring those differences, 
we  adopt Eq. \ref{Eq:SFR}  as the
expression for the radial dependence of the SFR in the MW disk, after normalising it (by putting A=3.5 \ms/kpc$^2$/y) to the total SFR rate of the Milky Way $\Psi_{MW}$=2 \ms/y \citep{Chomiuk11}:

\begin{equation}
2 \pi \int \Psi(R) R dR \ = \ \Psi_{MW}
\end{equation}

In the middle panel of Fig. \ref{Fig:SfrProf}, we compare the "observed" SFR profile with various theoretical or empirical profiles in the literature, which make use of the corresponding profiles of total or molecular gas discussed in this section. All of them are normalised to the adopted observed value of the SFR in the solar neighbourhood. 

The most widely used SFR prescription is the so-called
"Schmidt-Kennicut" law, based on observations of quiescent and active disk galaxies: $\Psi \propto \Sigma_G^{k}$, where the gas surface density $\Sigma_G$ runs over three orders of magnitude and the data suggest $k$=1.5.
It turns out that this form of the SFR, with $k$=1.5 is too flat to fit the MW data: \cite{Missi06} find $k$=2 from their modelling of the IR emission of the MW. Also, it is well known that a steeper function is required in galactic chemical evolution models  to explain the observed abundance gradients in the MW disk. 

\cite{BP99} adopted a
law of the form $\Psi(R) \propto \Sigma_G^{1.5} V(R)/R$; the factor
$V(R)/R$ is $\propto 1/R$ for a flat curve of the rotational velocity $V(R)$ and is
attributed to spiral waves inducing star formation with that
frequency \citep{Wyse86,WyseSilk89}. On the other hand, models by Chiappini (2001) adopt $\Psi(R) \propto \Sigma_{TOT} \Sigma_G^{1.5}$,
where $\Sigma_{TOT}$ is the total disk surface density (dominated in the inner Galaxy by the rapidly increasing stellar profile), and they introduced a cut-off in the SFR efficiency, below 2 \mpc. Both SFR laws  also appear in the middle and lower panels of Fig. \ref{Fig:SfrProf}: they fit relatively well the "observed" SFR profile and it turns out that the corresponding models reproduce  
several   key properties of the MW disk relatively well.

The aforementioned laws make use of the total gaseous profile of the disk. Based on a detailed, sub-kpc scale, observations of a large sample of disk galaxies, \cite{Bigiel08}  have found that the SFR appears to follow the \hmol \  surface density, rather than the \hatm \ or the total gas surface density. In a companion paper, \cite{Leroy08} argue that the observed radial decline in star formation efficiency is too steep to be reproduced only by increases in the free-fall time or orbital time and they find no clear indications of a cut-off in the SFR.

Following these studies, we checked whether such a correspondence 
between the adopted SFR and molecular gas profiles also holds in the MW disk. The comparison, presented in the middle and lower panels
of Fig. \ref{Fig:SfrProf}, is favourable to that idea: the SFR follows the \hmol \ profile to better than 30\% in the 3-13 kpc range.

Figure \ref{Fig:SfrGas} displays the data in  a different way, with SFR surface density vs gas (total or molecular) surface densities. 
In the top left panel, it appears that the "observed" SFR varies too steeply with the total gas density and it cannot be fit with a simple
Schmidt-Kennicutt" law $\Psi\propto \Sigma_G^{1.5}$; a strong radial dependence of the SF efficiency, such as the aforementioned ones, is required to improve the situation. In the top right panel it is seen that
the "observed" SFR increases almost linearly with \hmol, and that the SFR proposed by \cite{Bigiel08}, $\Psi$=0.009 $\Sigma_{\rm H2}$/(10 \mpc),
reproduces  the data quite well. The measurement of \cite{Bigiel08} concern \hmol \ surface densities above 3 \mpc, i.e. they correspond to the upper half of the figure. It appears that, at least in the case of the MW, that dependence is prolonged to even smaller surface densities. In the bottom panel of Fig. \ref{Fig:SfrGas} the "observed" SFR vs. gas relation in the MW disk is
compared to data of a sample of disk galaxies from \cite{Krumholz12}.
It is clearly seen that the MW disk SFR corresponds to a narrow range of gas surface densities, so the extragalactic data of SFR vs. gas cannot be used as guide to the MW SFR; in contrast, data on MW SFR vs. \hmol \ cover a wider dynamical range of \hmol \ and offer convincing evidence of a linear relationship between the two quantities.

In view of this observational support, both for the MW disk (this work) and for external galaxies \citep{Bigiel08, Leroy08}, we adopt here a star formation law depending on the 
\hmol \ surface density. In order to calculate it in the model of chemical evolution we adopt the semi-emi-empirical prescription of \cite{BlitzRos06} for the ratio $R_{mol}$=\hmol/\hatm \
and we find  rather good agreement between the observed and theoretically calculated  molecular fractions 
in the MW disk
\begin{equation}
f_2 \ = \ \frac{R_{mol}}{R_{mol}+1}
\label{eq:fMol}
\end{equation}
as seen in the top panel of Fig. \ref{Fig:MolFrac}. The resulting 
radial profiles \hmol$(R)=f_2(R) \ \Sigma_G(R)$  and \hatm$(R)=[1-f_2(R)]  \Sigma_G(R)$ also compare favourably to the observed ones (middle panel). Finally, the corresponding  SFR (Eq. 3 in \cite{Bigiel08}, but with a coefficient 0.0016 instead of 0.008)
\begin{equation}
\Psi(R) \ = 0.0016  \ f_2(R) \ \left(\frac{\Sigma_G(R)}{\rm M_{\odot}/pc^2} \right) \ {\rm M_{\odot}/kpc^2/yr}
\end{equation}
reproduces  well the "observed" SFR profile of the MW disk in the 4-12 kpc region, and somewhat less successfully outside that region (bottom panel). This agreement has already found in \cite{BlitzRos06},
but with older data for the gas and SFR profiles of the MW.

\begin{figure}
\begin{center}
\includegraphics[width=0.49\textwidth]{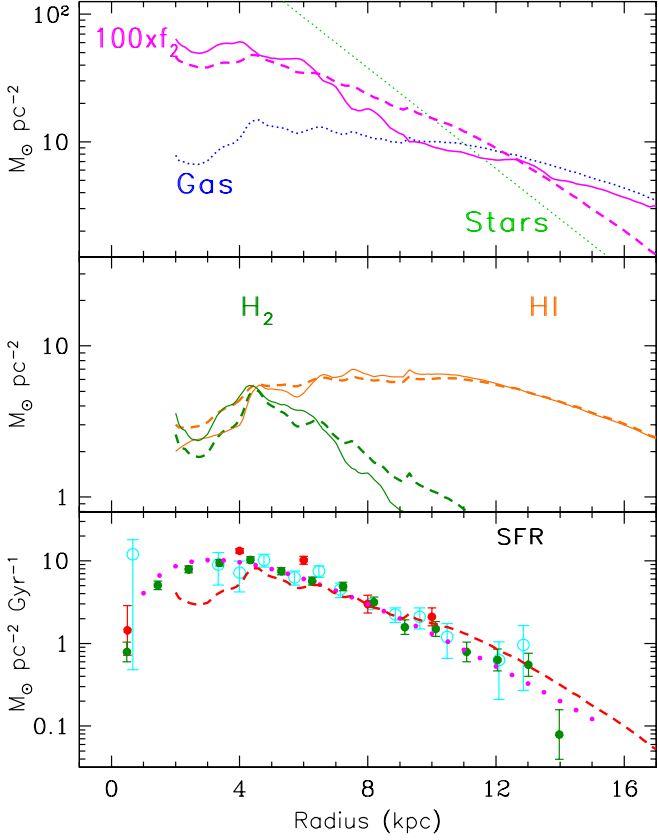}
\caption[]{{\it Top:} Observed  molecular fraction $f_2$=$\frac{\rm H_2}{\rm HI+H_2}$ 
({\it solid} curve) and theoretical one ({\it dashed}), obtained with the prescription of Eq. B.6
from the displayed  gaseous and stellar profiles ({\it dotted}). {\it Middle}:
Observed ({\it solid} curves) vs. theoretical ({\it dashed} curves) profiles of
\hatm \ and \hmol; the latter are obtained from the observed gas profile 
and the theoretically evaluated molecular fraction $f_2$ (both in the top panel), as
$\Sigma$(\hmol)=$f_2$x $\Sigma_G$/1.4 and  $\Sigma$(\hatm)=(1-$f_2$)x $\Sigma_G$/1.4.
{\it Bottom}: Observed SFR profiles in the MW disk, vs a theoretical SFR ({\it dashed}) curve), obtained from the theoretical profile of the molecular gas in the middle panel.
}
\label{Fig:MolFrac}
\end{center}
\end{figure}

\section{Chemical evolution}
\label{App:Chem}

\begin{figure}
\begin{center}
\includegraphics[width=0.49\textwidth]{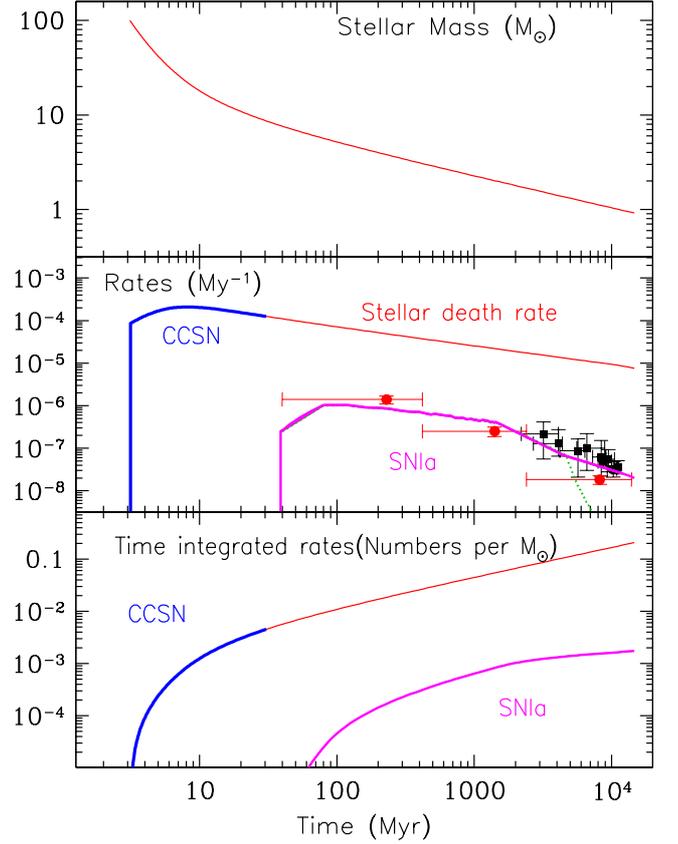}
\caption[]{{\it Top}: Stellar mass vs. lifetime. {\it Middle:}
Stellar death rate after an initial "burst" forming 1 M$_{\odot}$ 
 of  stars: $dN/dt=dN/dM \ x dM/dt$, where $dN/dM$ is the stellar IMF and $dM/dt$ the derivative of the curve in the top panel. The thick portion of the curve (up to $\sim$35 My, corresponding to a star of 8 \ms) is the rate of CCSN. The bottom right part of the middle panel displays the corresponding SNIa rate (the time delay distribution or TDD) adopted in this work (thick curve);
 it is a mixture of the \cite{Greggio05} formulation for the SD scenario up to 4.5 Gyr and an extrapolation $\propto t^{-1}$ after that time, in order to fit  the data points from \citet{Maoz2012} (filled circles) and from \cite{Maoz10} (squares). {\it Bottom}: Time-integrated numbers of CCSN (thick portion of upper curve), single stars of mass M$<$8 \ms (thin portion of upper curve) and SNIa (lower curve), as a function of time, for an initial "burst" of 1 \ms.
}
\label{Fig:StellarRates}
\end{center}
\end{figure}
}

In studies of galactic chemical evolution,
the changes in the chemical composition of the system are described by 
a system of integro-differential equations. The  mass of element/isotope i in the gas is $m_i=m_GX_i$, (where $m_G$ is the mass of the gas and $X_i$ is the mass fraction of $i$) and its evolution  is given by:

\begin{equation}
{{d(m_G \ X_i)}\over{dt}} \ = \ - \Psi X_i \ + \ E_i \ (+ {\rm infall} \
{\rm or} {\rm \ outflow }) \ {\rm terms} 
\label{eq:chem1}
\end{equation}
i.e. star formation at a rate $\Psi$ removes element $i$ from the ISM at a rate $\Psi X_i$, while at the same time stars inject in the ISM that element
at a rate $E_i(t)$. The {\it rate of ejection of element i by stars} is given by:

\begin{equation}
E_i(t) \ = \ \int_{M_t}^{M_U} \ Y_i(M) \ \Psi(t-\tau_M) \ \Phi(M) \ dM
\label{eq:chem2}
\end{equation}
where the star of mass $M$, created at the time $t-\tau_M$, dies at time $t$ (if its lifetime $\tau_M$  is lower than $t$) and releases a mass
$Y_i(M)$ in the form  of element/isotope i ({\it stellar yield} of i from mass  $M$). Here,
$\Phi(M)=dN/dM $ is the initial mass function (IMF), assumed to be independent of time $t$, $M_t$ is the mass of the smallest star that has lifetime  $\tau_M=t$ and $M_U$ is the most massive star of the IMF;

Because of the presence of the term $\Psi(t-\tau_M)$, Eqs.\ref{eq:chem1} and \ref{eq:chem2} have to be solved numerically (except if specific assumptions, like the instantaneous recycling approximation  -IRA - - are made). The integral \ref{eq:chem2}
is evaluated over the stellar masses, properly weighted by the term $\Psi(t-\tau_M)$
corresponding in each mass  $M$. It is explicitly assumed in that case that
 all the stellar masses created in a given place,  release their ejecta in that same place.

This assumption does not hold anymore if stars are allowed to travel away from their
birth places before dying. In that case, the mass $E_i(t)$ released in a given place of spatial coordinate $R$ and at time $t$ is the sum of the ejecta of stars born in various places $R'$ and times $t-t'$, with different star formation rates $\Psi(t',R')$ for all stellar masses $M$ with lifetimes $\tau_M<t-t'$
Instead of Eq. \ref{eq:chem2}, the isochrone formalism, concerning instantaneous "bursts" of star formation or single stellar populations (SSP),  has to be used then.
Eq.\ref{eq:chem2} is rewritten as
 
\begin{equation}
E_i(t) =  \int_{\tau_{M_U}}^{t} \Psi(t')dt' \ \left(\frac{Y_i(M) dN}{dt'}\right)_{t-t'} 
\label{eq:psitr}
\end{equation}
where $dN=\Phi(M)dM$ is the number of stars between $M$ and $M+dM$ and
$\Psi(t')dt'$ is the mass of stars (in \ms) created in time interval $dt'$
at time $t'$. 
The term $(dN/dt')_{t-t'}$ represents the stellar death rate (by number) at time $t$
of a unit mass of stars born in an instantaneous burst at  time $t-t'$. The term $Y_i(M)dN/dt$
represents the corresponding rate of release of element $i$ in \ms/yr.  

Expression \ref{eq:psitr} is equivalent to expression \ref {eq:chem2} and it is used in N-body+SPH simulations (see \citet{Lia2002} or \citet{Wiersma09}), since it allows one to account for the ejecta released in a given place by "star particles" produced with different star formation rates in other places (see main text). It naturally incorporates  the metallicity dependence of the stellar yields and of the stellar lifetimes, both found in the term $Y_i(M,Z)(dN/dt)(Z)$.

\begin{figure}
\begin{center}
\includegraphics[width=0.49\textwidth]{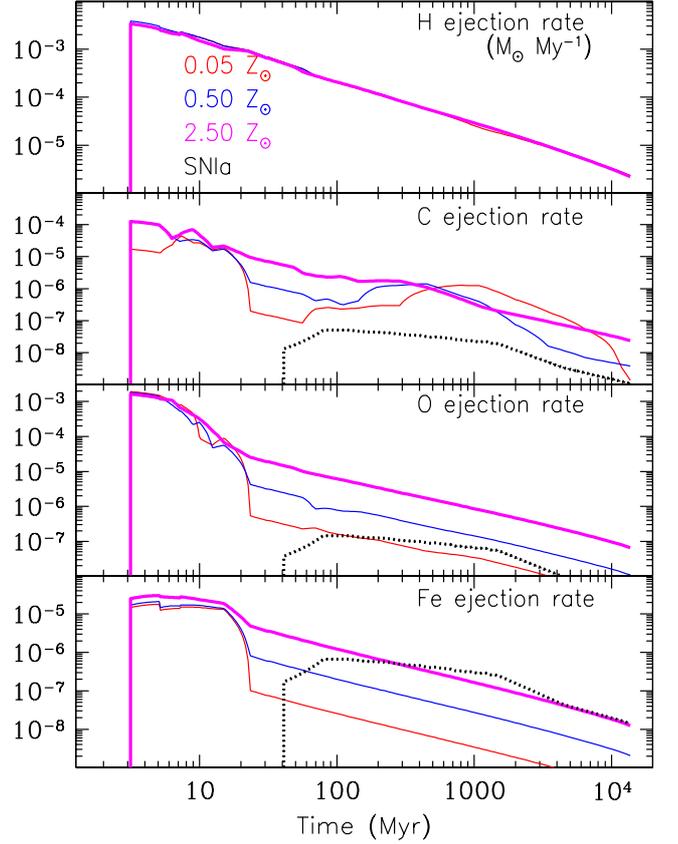}
\caption[]{Ejection rates of hydrogen, carbon, oxygen and Fe from a stellar population of 1 \ms as a function of time. Yields are from \cite{Nomoto2013}.The curves represent different metallicities, as indicated in the top panel . The dotted curves  show the contribution of SNIa (resulting from a SSP of 1 \ms)
to the production of those elements.
}
\label{Fig:EjectaRates}
\end{center}
\end{figure}

As in \cite{BP99}, we use here the stellar lifetimes $\tau(M,Z)$
of \cite{Schaller92} for stars in the mass range 0.1-120 \ms and for two metallicities \zs and 0.05 \zs (covering the metallicity of the MW disk during its whole evolution except, perhaps, its earliest phases).

We adopt the stellar IMF of \cite{Kroupa02} in the mass range 0.1-100 \ms, but with a  slope $x$=-2.7 in the range 1-100 \ms (the "Scalo slope"), since the "Salpeter slope" of -2.35 overproduces metals in the evolution of the Solar neighbourhood. Moreover, the "Scalo" slope
is similar to the one of the integrated galactic IMF (IGIMF) 
suggested in e.g. \cite{Kroupa2008}, which is appropriate for galactic evolution studies .
 
We adopt the yields provided by \cite{Nomoto2013}\footnote{Particular care is required from users of those yields. The ones for low mass stars are "net yields" for all metallicities except Z=0. The ones of massive stars are "total yields" coming out through the final explosion. One has then to include  the composition of the envelope (calculated as the mass of the star minus the mass of the remnant minus the sum of the explosion ejecta),
assumed to have the initial composition.}  They concern both low and intermediate mass stars in the mass range 0.9 to 3.5 \ms (calculations from \cite{Karakas2010}) and for massive stars in the range 11-40 \ms. They are particularly adapted to the study of the galactic disk, because they cover a uniform grid of 6 initial metallicities  ($Z$=0, 0.001, 0.002, 0.004, 0.008, 0.02 and 0.05, i.e. from 0 to about 3 times solar), allowing for a detailed study of both the inner and the outer MW regions. This constitutes a clear advantage over other sets of widely used yields (e.g. \cite{WW95}).  The massive star models of \cite{Nomoto2013} have mass loss but no rotation; yields of hypernovae (very energetic explosions) are also provided but we do not  use them  here.
We include all 82 stable isotopic species from H to Ge. We calculate their evolution and we sum up at each time step to obtain the corresponding evolution of their elemental abundances. We  note
that the use of the yields in chemical evolution calculations requires some interpolation in the mass range of the super-AGB stars (6 or 8 to 11 \ms). 

We force the sum of the ejected masses of all isotopes of a star to be equal to the original stellar mass  minus the one of the compact residue (white dwarf, neutron star or black hole). This is important in order to ensure mass conservation in the system during the evolution.
We interpolate logarithmically  the yields in metallicity and in the mass range between 3.5 and 11 \ms, and we include a detailed treatment for the production of the light nuclides Li,Be and B by cosmic rays \citep{Prantzos12}. 

For the rate of SNIa we adopt a semi-empirical approach: the observational data of recent surveys are described well by a power-law in time, of the form $\propto t^{-1}$,
e.g. \cite{MaozMannucci12} and references therein.
At the earliest times, the DTD is unknown/uncertain, but a cut-off must certainly exist
before the formation of the first white dwarfs ($\sim$35-40 Myr after the birth of the SSP). We adopt then the formulation of \cite{Greggio05} for the single-degenerate
(SD) scenario of SNIa. That formulation reproduces, in fact, l the observations up to $\sim$4-5 Gyr (see Fig.\ref{Fig:StellarRates}, middle panel) quite ell. For longer timescales, where the SD scenario fails, we simply adopt the $t^{-1}$ power law.
The corresponding SNIa rate at time $t$ from all previous SSP is
obtained as
\begin{equation}
R_{SNIa}(t) = \int_0^t  \Psi(t') DTD(t-t') dt'
\end{equation}
  
As in GP2000 we adopt the SNIa yields of \cite{Iwamoto99} 
for Z=0 and Z=\zs,
interpolating logarithmically in metallicity between those values.

In Fig. \ref{Fig:StellarRates} we show the results for a SSP concerning the stellar death rates, including the CCSN rate, the SNIa rate (middle panel) and the corresponding time-integrated rates. With the adopted IMF, there are about 4.5 10$^{-3}$ CCSN and 1.2
10$^{-3}$ SNIa for every \ms \ of stars formed.

In Fig. \ref{Fig:EjectaRates}
we present the ejection rates for H, C, O and Fe as a function of time, for single stars (3 initial metallicities) and for SNIa. 
SNIa produce more than half of solar Fe and they make minor contributions  to
the production of "light" metals like C or O (but substantial ones to the
production of Si or Ca).

 We made extensive tests of the implemented SSP formalism (Eq. \ref{eq:psitr}) against the "classical" one (Eq. \ref{eq:chem1}) and  found excellent agreement in all cases.

\end{appendix}

\bibliographystyle{aa} 
\bibliography{Paper_Kubryk} 

\begin{thebibliography}{157}
\expandafter\ifx\csname natexlab\endcsname\relax\def\natexlab#1{#1}\fi

\bibitem[{{Adibekyan} {et~al.}(2013){Adibekyan}, {Figueira}, {Santos},
  {Hakobyan}, {Sousa}, {Pace}, {Delgado Mena}, {Robin}, {Israelian}, \&
  {Gonz{\'a}lez Hern{\'a}ndez}}]{adibekian2013}
{Adibekyan}, V.~Z., {Figueira}, P., {Santos}, N.~C., {et~al.} 2013, \aap, 554,
  A44

\bibitem[{{Adibekyan} {et~al.}(2011){Adibekyan}, {Santos}, {Sousa}, \&
  {Israelian}}]{Adibekyan2011}
{Adibekyan}, V.~Z., {Santos}, N.~C., {Sousa}, S.~G., \& {Israelian}, G. 2011,
  \aap, 535, L11

\bibitem[{{Afflerbach} {et~al.}(1997){Afflerbach}, {Churchwell}, \&
  {Werner}}]{Afflerbach1997}
{Afflerbach}, A., {Churchwell}, E., \& {Werner}, M.~W. 1997, \apj, 478, 190

\bibitem[{{Athanassoula}(1992)}]{Athanassoula1992}
{Athanassoula}, E. 1992, \mnras, 259, 345

\bibitem[{{Athanassoula} {et~al.}(1982){Athanassoula}, {Bosma}, {Creze}, \&
  {Schwarz}}]{Athanassoula1982}
{Athanassoula}, E., {Bosma}, A., {Creze}, M., \& {Schwarz}, M.~P. 1982, \aap,
  107, 101

\bibitem[{{Athanassoula} {et~al.}(2013){Athanassoula}, {Machado}, \&
  {Rodionov}}]{Athanassoula2013}
{Athanassoula}, E., {Machado}, R.~E.~G., \& {Rodionov}, S.~A. 2013, \mnras,
  429, 1949

\bibitem[{{Baba} {et~al.}(2013){Baba}, {Saitoh}, \& {Wada}}]{Baba2013}
{Baba}, J., {Saitoh}, T.~R., \& {Wada}, K. 2013, \apj, 763, 46

\bibitem[{{Babusiaux} \& {Gilmore}(2005)}]{Babusiaux2005}
{Babusiaux}, C. \& {Gilmore}, G. 2005, \mnras, 358, 1309

\bibitem[{{Bekki} \& {Tsujimoto}(2011)}]{Bekki2011}
{Bekki}, K. \& {Tsujimoto}, T. 2011, \apj, 738, 4

\bibitem[{{Bensby}(2013)}]{Bensby2013a}
{Bensby}, T. 2013, ArXiv e-prints

\bibitem[{{Bensby} {et~al.}(2011){Bensby}, {Alves-Brito}, {Oey}, {Yong}, \&
  {Mel{\'e}ndez}}]{Bensby2011}
{Bensby}, T., {Alves-Brito}, A., {Oey}, M.~S., {Yong}, D., \& {Mel{\'e}ndez},
  J. 2011, \apjl, 735, L46

\bibitem[{{Bensby} {et~al.}(2014){Bensby}, {Feltzing}, \& {Oey}}]{Bensby2014}
{Bensby}, T., {Feltzing}, S., \& {Oey}, M.~S. 2014, \aap, 562, A71

\bibitem[{{Bhattacharjee} {et~al.}(2014){Bhattacharjee}, {Chaudhury}, \&
  {Kundu}}]{Bhattar2013}
{Bhattacharjee}, P., {Chaudhury}, S., \& {Kundu}, S. 2014, \apj, 785, 63

\bibitem[{{Bigiel} {et~al.}(2008){Bigiel}, {Leroy}, {Walter}, {Brinks}, {de
  Blok}, {Madore}, \& {Thornley}}]{Bigiel08}
{Bigiel}, F., {Leroy}, A., {Walter}, F., {et~al.} 2008, \aj, 136, 2846

\bibitem[{{Bilitewski} \& {Sch{\"o}nrich}(2012)}]{Bilitewski2012}
{Bilitewski}, T. \& {Sch{\"o}nrich}, R. 2012, \mnras, 426, 2266

\bibitem[{{Binney} \& {Sanders}(2014)}]{Binney_sanders2013_arxiv}
{Binney}, J. \& {Sanders}, J.~L. 2014, in IAU Symposium, Vol. 298, IAU
  Symposium, ed. S.~{Feltzing}, G.~{Zhao}, N.~A. {Walton}, \& P.~{Whitelock},
  117--129

\bibitem[{{Binney} \& {Tremaine}(2008)}]{binney_tremaine2008}
{Binney}, J. \& {Tremaine}, S. 2008, {Galactic Dynamics: Second Edition}
  (Princeton University Press)

\bibitem[{{Bird} {et~al.}(2013){Bird}, {Kazantzidis}, {Weinberg}, {Guedes},
  {Callegari}, {Mayer}, \& {Madau}}]{Bird2013}
{Bird}, J.~C., {Kazantzidis}, S., {Weinberg}, D.~H., {et~al.} 2013, \apj, 773,
  43

\bibitem[{{Blitz} \& {Rosolowsky}(2006)}]{BlitzRos06}
{Blitz}, L. \& {Rosolowsky}, E. 2006, \apj, 650, 933

\bibitem[{{Blitz} \& {Spergel}(1991)}]{Blitz1991}
{Blitz}, L. \& {Spergel}, D.~N. 1991, \apj, 379, 631

\bibitem[{{Bobylev} {et~al.}(2014){Bobylev}, {Mosenkov}, {Bajkova}, \&
  {Gontcharov}}]{Bobylev2014}
{Bobylev}, V.~V., {Mosenkov}, A.~V., {Bajkova}, A.~T., \& {Gontcharov}, G.~A.
  2014, ArXiv e-prints

\bibitem[{{Boissier} \& {Prantzos}(1999)}]{BP99}
{Boissier}, S. \& {Prantzos}, N. 1999, \mnras, 307, 857

\bibitem[{{Bovy} \& {Rix}(2013)}]{Bovy2013}
{Bovy}, J. \& {Rix}, H.-W. 2013, \apj, 779, 115

\bibitem[{{Bovy} {et~al.}(2012{\natexlab{a}}){Bovy}, {Rix}, \&
  {Hogg}}]{Bovi2012}
{Bovy}, J., {Rix}, H.-W., \& {Hogg}, D.~W. 2012{\natexlab{a}}, \apj, 751, 131

\bibitem[{{Bovy} {et~al.}(2012{\natexlab{b}}){Bovy}, {Rix}, {Liu}, {Hogg},
  {Beers}, \& {Lee}}]{Bovy2012}
{Bovy}, J., {Rix}, H.-W., {Liu}, C., {et~al.} 2012{\natexlab{b}}, \apj, 753,
  148

\bibitem[{{Brook} {et~al.}(2012){Brook}, {Stinson}, {Gibson}, {Kawata},
  {House}, {Miranda}, {Macci{\`o}}, {Pilkington}, {Ro{\v s}kar}, {Wadsley}, \&
  {Quinn}}]{Brook2012}
{Brook}, C.~B., {Stinson}, G.~S., {Gibson}, B.~K., {et~al.} 2012, \mnras, 426,
  690

\bibitem[{{Brunetti} {et~al.}(2011){Brunetti}, {Chiappini}, \&
  {Pfenniger}}]{Brunetti2011}
{Brunetti}, M., {Chiappini}, C., \& {Pfenniger}, D. 2011, \aap, 534, A75

\bibitem[{{Cabrera-Lavers} {et~al.}(2007){Cabrera-Lavers}, {Hammersley},
  {Gonz{\'a}lez-Fern{\'a}ndez}, {L{\'o}pez-Corredoira}, {Garz{\'o}n}, \&
  {Mahoney}}]{Cabrera2007}
{Cabrera-Lavers}, A., {Hammersley}, P.~L., {Gonz{\'a}lez-Fern{\'a}ndez}, C.,
  {et~al.} 2007, \aap, 465, 825

\bibitem[{{Casagrande} {et~al.}(2011){Casagrande}, {Sch{\"o}nrich}, {Asplund},
  {Cassisi}, {Ram{\'{\i}}rez}, {Mel{\'e}ndez}, {Bensby}, \&
  {Feltzing}}]{casagrande2011}
{Casagrande}, L., {Sch{\"o}nrich}, R., {Asplund}, M., {et~al.} 2011, \aap, 530,
  A138

\bibitem[{{Cheng} {et~al.}(2012){Cheng}, {Rockosi}, {Morrison}, {Lee}, {Beers},
  {Bizyaev}, {Harding}, {Malanushenko}, {Malanushenko}, {Oravetz}, {Pan},
  {Schlesinger}, {Schneider}, {Simmons}, \& {Weaver}}]{Cheng2012}
{Cheng}, J.~Y., {Rockosi}, C.~M., {Morrison}, H.~L., {et~al.} 2012, \apj, 752,
  51

\bibitem[{{Chiappini}(2009)}]{Chiappini2009}
{Chiappini}, C. 2009, in IAU Symposium, Vol. 254, IAU Symposium, ed.
  J.~{Andersen}, {Nordstr{\"o}ara}, B.~{m}, \& J.~{Bland-Hawthorn}, 191--196

\bibitem[{{Chiappini} {et~al.}(1997){Chiappini}, {Matteucci}, \&
  {Gratton}}]{Chiappini1997}
{Chiappini}, C., {Matteucci}, F., \& {Gratton}, R. 1997, \apj, 477, 765

\bibitem[{{Chomiuk} \& {Povich}(2011)}]{Chomiuk11}
{Chomiuk}, L. \& {Povich}, M.~S. 2011, \aj, 142, 197

\bibitem[{{Comer{\'o}n} {et~al.}(2011){Comer{\'o}n}, {Elmegreen}, {Knapen},
  {Salo}, {Laurikainen}, {Laine}, {Athanassoula}, {Bosma}, {Sheth}, {Regan},
  {Hinz}, {Gil de Paz}, {Men{\'e}ndez-Delmestre}, {Mizusawa},
  {Mu{\~n}oz-Mateos}, {Seibert}, {Kim}, {Elmegreen}, {Gadotti}, {Ho},
  {Holwerda}, {Lappalainen}, {Schinnerer}, \& {Skibba}}]{Comeron2011}
{Comer{\'o}n}, S., {Elmegreen}, B.~G., {Knapen}, J.~H., {et~al.} 2011, \apj,
  741, 28

\bibitem[{{Courteau} {et~al.}(2014){Courteau}, {Cappellari}, {de Jong},
  {Dutton}, {Emsellem}, {Hoekstra}, {Koopmans}, {Mamon}, {Maraston}, {Treu}, \&
  {Widrow}}]{Courteau2013}
{Courteau}, S., {Cappellari}, M., {de Jong}, R.~S., {et~al.} 2014, Reviews of
  Modern Physics, 86, 47

\bibitem[{{Daflon} \& {Cunha}(2004)}]{Daflon2004}
{Daflon}, S. \& {Cunha}, K. 2004, \apj, 617, 1115

\bibitem[{{Dame}(1993)}]{Dame93}
{Dame}, T.~M. 1993, in American Institute of Physics Conference Series, Vol.
  278, Back to the Galaxy, ed. S.~S. {Holt} \& F.~{Verter}, 267--278

\bibitem[{{Di Matteo} {et~al.}(2013){Di Matteo}, {Haywood}, {Combes},
  {Semelin}, \& {Snaith}}]{DiMatteo2013}
{Di Matteo}, P., {Haywood}, M., {Combes}, F., {Semelin}, B., \& {Snaith}, O.~N.
  2013, \aap, 553, A102

\bibitem[{{Drimmel} \& {Spergel}(2001)}]{Drimmel2001}
{Drimmel}, R. \& {Spergel}, D.~N. 2001, \apj, 556, 181

\bibitem[{{Dutil} \& {Roy}(1999)}]{Dutil1999}
{Dutil}, Y. \& {Roy}, J.-R. 1999, \apj, 516, 62

\bibitem[{{Ferri{\`e}re} {et~al.}(2007){Ferri{\`e}re}, {Gillard}, \&
  {Jean}}]{Ferriere2007}
{Ferri{\`e}re}, K., {Gillard}, W., \& {Jean}, P. 2007, \aap, 467, 611

\bibitem[{{Ferrini} {et~al.}(1994){Ferrini}, {Molla}, {Pardi}, \&
  {Diaz}}]{Ferrini1994}
{Ferrini}, F., {Molla}, M., {Pardi}, M.~C., \& {Diaz}, A.~I. 1994, \apj, 427,
  745

\bibitem[{{Flynn} {et~al.}(2006){Flynn}, {Holmberg}, {Portinari}, {Fuchs}, \&
  {Jahrei{\ss}}}]{Flynn2006}
{Flynn}, C., {Holmberg}, J., {Portinari}, L., {Fuchs}, B., \& {Jahrei{\ss}}, H.
  2006, \mnras, 372, 1149

\bibitem[{{Forbes} {et~al.}(2012){Forbes}, {Krumholz}, \&
  {Burkert}}]{Forbes2012}
{Forbes}, J., {Krumholz}, M., \& {Burkert}, A. 2012, \apj, 754, 48

\bibitem[{{Friedli} \& {Benz}(1993)}]{Friedli1993}
{Friedli}, D. \& {Benz}, W. 1993, \aap, 268, 65

\bibitem[{{Friedli} {et~al.}(1994){Friedli}, {Benz}, \&
  {Kennicutt}}]{Friedli1994}
{Friedli}, D., {Benz}, W., \& {Kennicutt}, R. 1994, \apjl, 430, L105

\bibitem[{{Fu} {et~al.}(2013){Fu}, {Kauffmann}, {Huang}, {Yates}, {Moran},
  {Heckman}, {Dav{\'e}}, {Guo}, \& {Henriques}}]{Fu2013}
{Fu}, J., {Kauffmann}, G., {Huang}, M.-l., {et~al.} 2013, \mnras, 434, 1531

\bibitem[{{Genovali} {et~al.}(2013){Genovali}, {Lemasle}, {Bono}, {Romaniello},
  {Primas}, {Fabrizio}, {Buonanno}, {Fran{\c c}ois}, {Inno}, {Laney},
  {Matsunaga}, {Pedicelli}, \& {Th{\'e}venin}}]{Genovali2013}
{Genovali}, K., {Lemasle}, B., {Bono}, G., {et~al.} 2013, \aap, 554, A132

\bibitem[{{Grand} {et~al.}(2012){Grand}, {Kawata}, \& {Cropper}}]{Grand2012}
{Grand}, R.~J.~J., {Kawata}, D., \& {Cropper}, M. 2012, \mnras, 421, 1529

\bibitem[{{Grand} {et~al.}(2014){Grand}, {Kawata}, \& {Cropper}}]{Grand2014}
{Grand}, R.~J.~J., {Kawata}, D., \& {Cropper}, M. 2014, \mnras, 439, 623

\bibitem[{{Green}(2014)}]{Green13}
{Green}, D.~A. 2014, in IAU Symposium, Vol. 296, IAU Symposium, ed. A.~{Ray} \&
  R.~A. {McCray}, 188--196

\bibitem[{{Greggio}(2005)}]{Greggio05}
{Greggio}, L. 2005, \aap, 441, 1055

\bibitem[{{Hammer} {et~al.}(2007){Hammer}, {Puech}, {Chemin}, {Flores}, \&
  {Lehnert}}]{Hammer2007}
{Hammer}, F., {Puech}, M., {Chemin}, L., {Flores}, H., \& {Lehnert}, M.~D.
  2007, \apj, 662, 322

\bibitem[{{Haywood} {et~al.}(2013){Haywood}, {Di Matteo}, {Lehnert}, {Katz}, \&
  {G{\'o}mez}}]{Haywood2013}
{Haywood}, M., {Di Matteo}, P., {Lehnert}, M.~D., {Katz}, D., \& {G{\'o}mez},
  A. 2013, \aap, 560, A109

\bibitem[{{Holmberg} {et~al.}(2007){Holmberg}, {Nordstr{\"o}m}, \&
  {Andersen}}]{Holmberg2007}
{Holmberg}, J., {Nordstr{\"o}m}, B., \& {Andersen}, J. 2007, \aap, 475, 519

\bibitem[{{Hou} {et~al.}(2000){Hou}, {Prantzos}, \& {Boissier}}]{Hou2000}
{Hou}, J.~L., {Prantzos}, N., \& {Boissier}, S. 2000, \aap, 362, 921

\bibitem[{{Isern} {et~al.}(2013){Isern}, {Artigas}, \&
  {Garc{\'{\i}}a-Berro}}]{Isern2013}
{Isern}, J., {Artigas}, A., \& {Garc{\'{\i}}a-Berro}, E. 2013, in European
  Physical Journal Web of Conferences, Vol.~43, European Physical Journal Web
  of Conferences, 5002

\bibitem[{{Iwamoto} {et~al.}(1999){Iwamoto}, {Brachwitz}, {Nomoto},
  {Kishimoto}, {Umeda}, {Hix}, \& {Thielemann}}]{Iwamoto99}
{Iwamoto}, K., {Brachwitz}, F., {Nomoto}, K., {et~al.} 1999, \apjs, 125, 439

\bibitem[{{Kalberla} \& {Dedes}(2008)}]{Kalberla08}
{Kalberla}, P.~M.~W. \& {Dedes}, L. 2008, \aap, 487, 951

\bibitem[{{Kang} {et~al.}(2012){Kang}, {Chang}, {Yin}, {Hou}, {Zhang}, {Zhang},
  \& {Han}}]{Kang2012}
{Kang}, X., {Chang}, R., {Yin}, J., {et~al.} 2012, \mnras, 426, 1455

\bibitem[{{Karakas}(2010)}]{Karakas2010}
{Karakas}, A.~I. 2010, \mnras, 403, 1413

\bibitem[{{Kennicutt} \& {Evans}(2012)}]{KenniEvans12}
{Kennicutt}, R.~C. \& {Evans}, N.~J. 2012, \araa, 50, 531

\bibitem[{{Kordopatis} {et~al.}(2013){Kordopatis}, {Gilmore}, {Wyse},
  {Steinmetz}, {Siebert}, {Bienaym{\'e}}, {McMillan}, {Minchev}, {Zwitter},
  {Gibson}, {Seabroke}, {Grebel}, {Bland-Hawthorn}, {Boeche}, {Freeman},
  {Munari}, {Navarro}, {Parker}, {Reid}, \& {Siviero}}]{Kordopatis2013}
{Kordopatis}, G., {Gilmore}, G., {Wyse}, R.~F.~G., {et~al.} 2013, \mnras, 436,
  3231

\bibitem[{{Kroupa}(2001)}]{Kroupa2001}
{Kroupa}, P. 2001, \mnras, 322, 231

\bibitem[{{Kroupa}(2002)}]{Kroupa02}
{Kroupa}, P. 2002, Science, 295, 82

\bibitem[{{Kroupa}(2008)}]{Kroupa2008}
{Kroupa}, P. 2008, in Astronomical Society of the Pacific Conference Series,
  Vol. 390, Pathways Through an Eclectic Universe, ed. J.~H. {Knapen}, T.~J.
  {Mahoney}, \& A.~{Vazdekis} (Astronomical Society of the Pacific), 3

\bibitem[{{Krumholz}(2014)}]{Krumholz2014}
{Krumholz}, M.~R. 2014, ArXiv e-prints

\bibitem[{{Krumholz} {et~al.}(2012){Krumholz}, {Dekel}, \&
  {McKee}}]{Krumholz12}
{Krumholz}, M.~R., {Dekel}, A., \& {McKee}, C.~F. 2012, \apj, 745, 69

\bibitem[{{Kubryk} {et~al.}(2013){Kubryk}, {Prantzos}, \&
  {Athanassoula}}]{KPA2013}
{Kubryk}, M., {Prantzos}, N., \& {Athanassoula}, E. 2013, \mnras, 436, 1479

\bibitem[{{Kubryk} {et~al.}(2014){Kubryk}, {Prantzos}, \&
  {Athanassoula}}]{Kubryk2014b}
{Kubryk}, M., {Prantzos}, N., \& {Athanassoula}, E. 2014, ArXiv e-prints

\bibitem[{{Lacey} \& {Fall}(1985)}]{Lacey1985}
{Lacey}, C.~G. \& {Fall}, S.~M. 1985, \apj, 290, 154

\bibitem[{{Lagos} {et~al.}(2011){Lagos}, {Baugh}, {Lacey}, {Benson}, {Kim}, \&
  {Power}}]{Lagos2011}
{Lagos}, C.~D.~P., {Baugh}, C.~M., {Lacey}, C.~G., {et~al.} 2011, \mnras, 418,
  1649

\bibitem[{{Lehner} \& {Howk}(2011)}]{Lehner2011}
{Lehner}, N. \& {Howk}, J.~C. 2011, Science, 334, 955

\bibitem[{{Lemasle} {et~al.}(2013){Lemasle}, {Fran{\c c}ois}, {Genovali},
  {Kovtyukh}, {Bono}, {Inno}, {Laney}, {Kaper}, {Bergemann}, {Fabrizio},
  {Matsunaga}, {Pedicelli}, {Primas}, \& {Romaniello}}]{Lemasle2013}
{Lemasle}, B., {Fran{\c c}ois}, P., {Genovali}, K., {et~al.} 2013, \aap, 558,
  A31

\bibitem[{{L{\'e}pine} {et~al.}(2003){L{\'e}pine}, {Acharova}, \&
  {Mishurov}}]{Lepine2003}
{L{\'e}pine}, J.~R.~D., {Acharova}, I.~A., \& {Mishurov}, Y.~N. 2003, \apj,
  589, 210

\bibitem[{{Leroy} {et~al.}(2008){Leroy}, {Walter}, {Brinks}, {Bigiel}, {de
  Blok}, {Madore}, \& {Thornley}}]{Leroy08}
{Leroy}, A.~K., {Walter}, F., {Brinks}, E., {et~al.} 2008, \aj, 136, 2782

\bibitem[{{Lewis} \& {Freeman}(1989)}]{Lewis1989}
{Lewis}, J.~R. \& {Freeman}, K.~C. 1989, \aj, 97, 139

\bibitem[{{Li} {et~al.}(2007){Li}, {Mo}, {van den Bosch}, \& {Lin}}]{Li2007}
{Li}, Y., {Mo}, H.~J., {van den Bosch}, F.~C., \& {Lin}, W.~P. 2007, \mnras,
  379, 689

\bibitem[{{Lia} {et~al.}(2002){Lia}, {Portinari}, \& {Carraro}}]{Lia2002}
{Lia}, C., {Portinari}, L., \& {Carraro}, G. 2002, \mnras, 330, 821

\bibitem[{{Liu} \& {van de Ven}(2012)}]{Liu2012}
{Liu}, C. \& {van de Ven}, G. 2012, \mnras, 425, 2144

\bibitem[{{Loebman} {et~al.}(2011){Loebman}, {Ro{\v s}kar}, {Debattista},
  {Ivezi{\'c}}, {Quinn}, \& {Wadsley}}]{Loebman2011}
{Loebman}, S.~R., {Ro{\v s}kar}, R., {Debattista}, V.~P., {et~al.} 2011, \apj,
  737, 8

\bibitem[{{Lorimer}(2004)}]{Lorimer04}
{Lorimer}, D.~R. 2004, in IAU Symposium, Vol. 218, Young Neutron Stars and
  Their Environments, ed. F.~{Camilo} \& B.~M. {Gaensler}, 105

\bibitem[{{Lorimer} {et~al.}(2006){Lorimer}, {Faulkner}, {Lyne}, {Manchester},
  {Kramer}, {McLaughlin}, {Hobbs}, {Possenti}, {Stairs}, {Camilo}, {Burgay},
  {D'Amico}, {Corongiu}, \& {Crawford}}]{Lorimer06}
{Lorimer}, D.~R., {Faulkner}, A.~J., {Lyne}, A.~G., {et~al.} 2006, \mnras, 372,
  777

\bibitem[{{Luck} \& {Lambert}(2011)}]{Luck2011}
{Luck}, R.~E. \& {Lambert}, D.~L. 2011, \aj, 142, 136

\bibitem[{{Maoz} \& {Mannucci}(2012)}]{MaozMannucci12}
{Maoz}, D. \& {Mannucci}, F. 2012, \pasa, 29, 447

\bibitem[{{Maoz} {et~al.}(2012){Maoz}, {Mannucci}, \& {Brandt}}]{Maoz2012}
{Maoz}, D., {Mannucci}, F., \& {Brandt}, T.~D. 2012, \mnras, 426, 3282

\bibitem[{{Maoz} {et~al.}(2010){Maoz}, {Sharon}, \& {Gal-Yam}}]{Maoz10}
{Maoz}, D., {Sharon}, K., \& {Gal-Yam}, A. 2010, \apj, 722, 1879

\bibitem[{{Marasco} {et~al.}(2012){Marasco}, {Fraternali}, \&
  {Binney}}]{Marasco2012}
{Marasco}, A., {Fraternali}, F., \& {Binney}, J.~J. 2012, \mnras, 419, 1107

\bibitem[{{Martin} \& {Roy}(1994)}]{Martin1994}
{Martin}, P. \& {Roy}, J.-R. 1994, \apj, 424, 599

\bibitem[{{Mart{\'{\i}}nez-Serrano} {et~al.}(2009){Mart{\'{\i}}nez-Serrano},
  {Serna}, {Dom{\'e}nech-Moral}, \& {Dom{\'{\i}}nguez-Tenreiro}}]{Martinez2009}
{Mart{\'{\i}}nez-Serrano}, F.~J., {Serna}, A., {Dom{\'e}nech-Moral}, M., \&
  {Dom{\'{\i}}nguez-Tenreiro}, R. 2009, \apjl, 705, L133

\bibitem[{{Matteucci}(2012)}]{Matteucci2012}
{Matteucci}, F. 2012, {Chemical Evolution of Galaxies} (Springer Verlag)

\bibitem[{{Mayor} \& {Vigroux}(1981)}]{Mayor1981}
{Mayor}, M. \& {Vigroux}, L. 1981, \aap, 98, 1

\bibitem[{{McMillan}(2011)}]{McMillan2011}
{McMillan}, P.~J. 2011, \mnras, 414, 2446

\bibitem[{{Minchev} {et~al.}(2013){Minchev}, {Chiappini}, \&
  {Martig}}]{Minchev2013}
{Minchev}, I., {Chiappini}, C., \& {Martig}, M. 2013, \aap, 558, A9

\bibitem[{{Minchev} {et~al.}(2014){Minchev}, {Chiappini}, \&
  {Martig}}]{Minchev2014}
{Minchev}, I., {Chiappini}, C., \& {Martig}, M. 2014, \aap, 572, A92

\bibitem[{Minchev \& Famaey(2010)}]{minchevfamaey2010}
Minchev, I. \& Famaey, B. 2010, The Astrophysical Journal, 722, 112

\bibitem[{{Minchev} {et~al.}(2011){Minchev}, {Famaey}, {Combes}, {Di Matteo},
  {Mouhcine}, \& {Wozniak}}]{Minchev2011}
{Minchev}, I., {Famaey}, B., {Combes}, F., {et~al.} 2011, \aap, 527, A147

\bibitem[{{Minchev} {et~al.}(2012{\natexlab{a}}){Minchev}, {Famaey}, {Quillen},
  {Dehnen}, {Martig}, \& {Siebert}}]{Minchev2012b}
{Minchev}, I., {Famaey}, B., {Quillen}, A.~C., {et~al.} 2012{\natexlab{a}},
  \aap, 548, A127

\bibitem[{{Minchev} {et~al.}(2012{\natexlab{b}}){Minchev}, {Famaey}, {Quillen},
  {Di Matteo}, {Combes}, {Vlaji{\'c}}, {Erwin}, \&
  {Bland-Hawthorn}}]{Minchev2012a}
{Minchev}, I., {Famaey}, B., {Quillen}, A.~C., {et~al.} 2012{\natexlab{b}},
  \aap, 548, A126

\bibitem[{{Misiriotis} {et~al.}(2006){Misiriotis}, {Xilouris},
  {Papamastorakis}, {Boumis}, \& {Goudis}}]{Missi06}
{Misiriotis}, A., {Xilouris}, E.~M., {Papamastorakis}, J., {Boumis}, P., \&
  {Goudis}, C.~D. 2006, \aap, 459, 113

\bibitem[{{Moll{\'a}} \& {D{\'{\i}}az}(2005)}]{Molla2005}
{Moll{\'a}}, M. \& {D{\'{\i}}az}, A.~I. 2005, \mnras, 358, 521

\bibitem[{{Monari} {et~al.}(2013){Monari}, {Antoja}, \& {Helmi}}]{Monari2013}
{Monari}, G., {Antoja}, T., \& {Helmi}, A. 2013, ArXiv e-prints

\bibitem[{{Monari} {et~al.}(2014){Monari}, {Helmi}, {Antoja}, \&
  {Steinmetz}}]{Monari2014}
{Monari}, G., {Helmi}, A., {Antoja}, T., \& {Steinmetz}, M. 2014, \aap, 569,
  A69

\bibitem[{{Nakanishi} \& {Sofue}(2003)}]{NakaSofue03}
{Nakanishi}, H. \& {Sofue}, Y. 2003, \pasj, 55, 191

\bibitem[{{Nakanishi} \& {Sofue}(2006)}]{NakaSofue06}
{Nakanishi}, H. \& {Sofue}, Y. 2006, \pasj, 58, 847

\bibitem[{{Navarro} {et~al.}(2011){Navarro}, {Abadi}, {Venn}, {Freeman}, \&
  {Anguiano}}]{Navarro2011}
{Navarro}, J.~F., {Abadi}, M.~G., {Venn}, K.~A., {Freeman}, K.~C., \&
  {Anguiano}, B. 2011, \mnras, 412, 1203

\bibitem[{{Nieva} \& {Przybilla}(2012)}]{Nieva2012}
{Nieva}, M.-F. \& {Przybilla}, N. 2012, \aap, 539, A143

\bibitem[{{Nomoto} {et~al.}(2013){Nomoto}, {Kobayashi}, \&
  {Tominaga}}]{Nomoto2013}
{Nomoto}, K., {Kobayashi}, C., \& {Tominaga}, N. 2013, \araa, 51, 457

\bibitem[{{Olling} \& {Merrifield}(2001)}]{OllingMer01}
{Olling}, R.~P. \& {Merrifield}, M.~R. 2001, \mnras, 326, 164

\bibitem[{{Pagel}(2009)}]{Pagel2009}
{Pagel}, B.~E.~J. 2009, {Nucleosynthesis and Chemical Evolution of Galaxies}
  (Cambridge University Press)

\bibitem[{{Picaud} \& {Robin}(2004)}]{Picaud2004}
{Picaud}, S. \& {Robin}, A.~C. 2004, \aap, 428, 891

\bibitem[{{Pohl} {et~al.}(2008){Pohl}, {Englmaier}, \& {Bissantz}}]{Pohl08}
{Pohl}, M., {Englmaier}, P., \& {Bissantz}, N. 2008, \apj, 677, 283

\bibitem[{{Porcel} {et~al.}(1998){Porcel}, {Garzon}, {Jimenez-Vicente}, \&
  {Battaner}}]{Porcel1998}
{Porcel}, C., {Garzon}, F., {Jimenez-Vicente}, J., \& {Battaner}, E. 1998,
  \aap, 330, 136

\bibitem[{{Portinari} \& {Chiosi}(2000)}]{Portinari2000}
{Portinari}, L. \& {Chiosi}, C. 2000, \aap, 355, 929

\bibitem[{{Prantzos}(2009)}]{Prantzos09}
{Prantzos}, N. 2009, in IAU Symposium, Vol. 254, IAU Symposium, ed.
  J.~{Andersen}, B.~{Nordstr{\"o}m}, \& J.~{Bland-Hawthorn}, 381--392

\bibitem[{{Prantzos}(2012)}]{Prantzos12}
{Prantzos}, N. 2012, \aap, 542, A67

\bibitem[{{Prantzos} \& {Aubert}(1995)}]{PranAub1995}
{Prantzos}, N. \& {Aubert}, O. 1995, \aap, 302, 69

\bibitem[{{Prantzos} {et~al.}(1996){Prantzos}, {Aubert}, \&
  {Audouze}}]{Prantzos_AubAud1996}
{Prantzos}, N., {Aubert}, O., \& {Audouze}, J. 1996, \aap, 309, 760

\bibitem[{{Prantzos} {et~al.}(2011){Prantzos}, {Boehm}, {Bykov}, {Diehl},
  {Ferri{\`e}re}, {Guessoum}, {Jean}, {Knoedlseder}, {Marcowith}, {Moskalenko},
  {Strong}, \& {Weidenspointner}}]{Prantzos2011}
{Prantzos}, N., {Boehm}, C., {Bykov}, A.~M., {et~al.} 2011, Reviews of Modern
  Physics, 83, 1001

\bibitem[{{Rashkov} {et~al.}(2013){Rashkov}, {Pillepich}, {Deason}, {Madau},
  {Rockosi}, {Guedes}, \& {Mayer}}]{Rashkov13}
{Rashkov}, V., {Pillepich}, A., {Deason}, A.~J., {et~al.} 2013, \apjl, 773, L32

\bibitem[{{Rix} \& {Bovy}(2013)}]{Rix2013}
{Rix}, H.-W. \& {Bovy}, J. 2013, \aapr, 21, 61

\bibitem[{{Robitaille} \& {Whitney}(2010)}]{Robitaille2010}
{Robitaille}, T.~P. \& {Whitney}, B.~A. 2010, \apjl, 710, L11

\bibitem[{{Rocha-Pinto} {et~al.}(2000){Rocha-Pinto}, {Scalo}, {Maciel}, \&
  {Flynn}}]{Rocha-Pinto2000}
{Rocha-Pinto}, H.~J., {Scalo}, J., {Maciel}, W.~J., \& {Flynn}, C. 2000, \aap,
  358, 869

\bibitem[{{Romano} {et~al.}(2010){Romano}, {Karakas}, {Tosi}, \&
  {Matteucci}}]{Romano2010}
{Romano}, D., {Karakas}, A.~I., {Tosi}, M., \& {Matteucci}, F. 2010, \aap, 522,
  A32

\bibitem[{{Roskar}(2010)}]{Roskar2010}
{Roskar}, R. 2010, PhD thesis, University of Washington

\bibitem[{{Ro{\v s}kar} {et~al.}(2013){Ro{\v s}kar}, {Debattista}, \&
  {Loebman}}]{Roskar2013}
{Ro{\v s}kar}, R., {Debattista}, V.~P., \& {Loebman}, S.~R. 2013, \mnras, 433,
  976

\bibitem[{{Ro{\v s}kar} {et~al.}(2008){Ro{\v s}kar}, {Debattista}, {Quinn},
  {Stinson}, \& {Wadsley}}]{Roskar2008}
{Ro{\v s}kar}, R., {Debattista}, V.~P., {Quinn}, T.~R., {Stinson}, G.~S., \&
  {Wadsley}, J. 2008, \apjl, 684, L79

\bibitem[{{Sales} {et~al.}(2009){Sales}, {Helmi}, {Abadi}, {Brook},
  {G{\'o}mez}, {Ro{\v s}kar}, {Debattista}, {House}, {Steinmetz}, \&
  {Villalobos}}]{Sales2009}
{Sales}, L.~V., {Helmi}, A., {Abadi}, M.~G., {et~al.} 2009, \mnras, 400, L61

\bibitem[{{S{\'a}nchez} {et~al.}(2012){S{\'a}nchez}, {Rosales-Ortega},
  {Marino}, {Iglesias-P{\'a}ramo}, {V{\'{\i}}lchez}, {Kennicutt},
  {D{\'{\i}}az}, {Mast}, {Monreal-Ibero}, {Garc{\'{\i}}a-Benito},
  {Bland-Hawthorn}, {P{\'e}rez}, {Gonz{\'a}lez Delgado}, {Husemann},
  {L{\'o}pez-S{\'a}nchez}, {Cid Fernandes}, {Kehrig}, {Walcher}, {Gil de Paz},
  \& {Ellis}}]{Sanchez2012}
{S{\'a}nchez}, S.~F., {Rosales-Ortega}, F.~F., {Marino}, R.~A., {et~al.} 2012,
  \aap, 546, A2

\bibitem[{{S{\'a}nchez-Bl{\'a}zquez} {et~al.}(2009){S{\'a}nchez-Bl{\'a}zquez},
  {Courty}, {Gibson}, \& {Brook}}]{Sanchez2009}
{S{\'a}nchez-Bl{\'a}zquez}, P., {Courty}, S., {Gibson}, B.~K., \& {Brook},
  C.~B. 2009, \mnras, 398, 591

\bibitem[{{Schaller} {et~al.}(1992){Schaller}, {Schaerer}, {Meynet}, \&
  {Maeder}}]{Schaller92}
{Schaller}, G., {Schaerer}, D., {Meynet}, G., \& {Maeder}, A. 1992, \aaps, 96,
  269

\bibitem[{{Sch{\"o}nrich} \& {Binney}(2009{\natexlab{a}})}]{SB2009}
{Sch{\"o}nrich}, R. \& {Binney}, J. 2009{\natexlab{a}}, \mnras, 396, 203

\bibitem[{{Sch{\"o}nrich} \& {Binney}(2009{\natexlab{b}})}]{SB09b}
{Sch{\"o}nrich}, R. \& {Binney}, J. 2009{\natexlab{b}}, \mnras, 399, 1145

\bibitem[{{Scott} {et~al.}(2014){Scott}, {Asplund}, {Grevesse}, {Bergemann}, \&
  {Sauval}}]{Scott2014a}
{Scott}, P., {Asplund}, M., {Grevesse}, N., {Bergemann}, M., \& {Sauval}, A.~J.
  2014, ArXiv e-prints

\bibitem[{Sellwood \& Binney(2002)}]{SellwoodBinney2002}
Sellwood, J. \& Binney, J. 2002, MNRAS, 336, 785

\bibitem[{{Shevchenko}(2011)}]{Shevchenko2011}
{Shevchenko}, I.~I. 2011, \apj, 733, 39

\bibitem[{{Simpson} {et~al.}(1995){Simpson}, {Colgan}, {Rubin}, {Erickson}, \&
  {Haas}}]{Simpson1995}
{Simpson}, J.~P., {Colgan}, S.~W.~J., {Rubin}, R.~H., {Erickson}, E.~F., \&
  {Haas}, M.~R. 1995, \apj, 444, 721

\bibitem[{{Smartt} \& {Rolleston}(1997)}]{Smartt1997}
{Smartt}, S.~J. \& {Rolleston}, W.~R.~J. 1997, \apjl, 481, L47

\bibitem[{{Sofue}(2012)}]{Sofue2012}
{Sofue}, Y. 2012, \pasj, 64, 75

\bibitem[{{Sofue}(2013)}]{Sofue2013}
{Sofue}, Y. 2013, \pasj, 65, 118

\bibitem[{{Steinmetz}(2012)}]{Steinmetz2012}
{Steinmetz}, M. 2012, Astronomische Nachrichten, 333, 523

\bibitem[{{Sygnet} {et~al.}(1988){Sygnet}, {Tagger}, {Athanassoula}, \&
  {Pellat}}]{Sygnet1988}
{Sygnet}, J.~F., {Tagger}, M., {Athanassoula}, E., \& {Pellat}, R. 1988,
  \mnras, 232, 733

\bibitem[{{Tinsley} \& {Larson}(1978)}]{Tinsley1978}
{Tinsley}, B.~M. \& {Larson}, R.~B. 1978, \apj, 221, 554

\bibitem[{{Urquhart} {et~al.}(2013){Urquhart}, {Figura}, {T.~J.}, {Moore},
  {Hoare}, {Lumsden}, {Mottram}, {Thompson}, \& {Oudmaijer}}]{Urqu13}
{Urquhart}, J.~S., {Figura}, C.~C., {T.~J.}, T., {et~al.} 2013, ArXiv e-prints

\bibitem[{{van Dokkum} {et~al.}(2013){van Dokkum}, {Leja}, {Nelson}, {Patel},
  {Skelton}, {Momcheva}, {Brammer}, {Whitaker}, {Lundgren}, {Fumagalli},
  {Conroy}, {F{\"o}rster Schreiber}, {Franx}, {Kriek}, {Labb{\'e}},
  {Marchesini}, {Rix}, {van der Wel}, \& {Wuyts}}]{Dokkum2013}
{van Dokkum}, P.~G., {Leja}, J., {Nelson}, E.~J., {et~al.} 2013, \apjl, 771,
  L35

\bibitem[{{Wakker} {et~al.}(1999){Wakker}, {Howk}, {Savage}, {van Woerden},
  {Tufte}, {Schwarz}, {Benjamin}, {Reynolds}, {Peletier}, \&
  {Kalberla}}]{Wakker1999}
{Wakker}, B.~P., {Howk}, J.~C., {Savage}, B.~D., {et~al.} 1999, \nat, 402, 388

\bibitem[{{Wang} \& {Zhao}(2013)}]{Wang2013}
{Wang}, Y. \& {Zhao}, G. 2013, \apj, 769, 4

\bibitem[{{Wielen} {et~al.}(1996){Wielen}, {Fuchs}, \& {Dettbarn}}]{Wielen1996}
{Wielen}, R., {Fuchs}, B., \& {Dettbarn}, C. 1996, \aap, 314, 438

\bibitem[{{Wiersma} {et~al.}(2009){Wiersma}, {Schaye}, {Theuns}, {Dalla
  Vecchia}, \& {Tornatore}}]{Wiersma09}
{Wiersma}, R.~P.~C., {Schaye}, J., {Theuns}, T., {Dalla Vecchia}, C., \&
  {Tornatore}, L. 2009, \mnras, 399, 574

\bibitem[{{Williams} \& {McKee}(1997)}]{McKee97}
{Williams}, J.~P. \& {McKee}, C.~F. 1997, \apj, 476, 166

\bibitem[{{Wilson} {et~al.}(2011){Wilson}, {Helmi}, {Morrison}, {Breddels},
  {Bienaym{\'e}}, {Binney}, {Bland-Hawthorn}, {Campbell}, {Freeman},
  {Fulbright}, {Gibson}, {Gilmore}, {Grebel}, {Munari}, {Navarro}, {Parker},
  {Reid}, {Seabroke}, {Siebert}, {Siviero}, {Steinmetz}, {Williams}, {Wyse}, \&
  {Zwitter}}]{Wilson2011}
{Wilson}, M.~L., {Helmi}, A., {Morrison}, H.~L., {et~al.} 2011, \mnras, 413,
  2235

\bibitem[{{Woosley} \& {Weaver}(1995)}]{WW95}
{Woosley}, S.~E. \& {Weaver}, T.~A. 1995, \apjs, 101, 181

\bibitem[{{Wyse}(1986)}]{Wyse86}
{Wyse}, R.~F.~G. 1986, \apjl, 311, L41

\bibitem[{{Wyse} \& {Silk}(1989)}]{WyseSilk89}
{Wyse}, R.~F.~G. \& {Silk}, J. 1989, \apj, 339, 700

\bibitem[{{Yoachim} \& {Dalcanton}(2006)}]{Yoachim2006}
{Yoachim}, P. \& {Dalcanton}, J.~J. 2006, \aj, 131, 226

\bibitem[{{Zaritsky} {et~al.}(1994){Zaritsky}, {Kennicutt}, \&
  {Huchra}}]{Zaritsky1994}
{Zaritsky}, D., {Kennicutt}, Jr., R.~C., \& {Huchra}, J.~P. 1994, \apj, 420, 87

\bibitem[{{Zhang} {et~al.}(2013){Zhang}, {Rix}, {van de Ven}, {Bovy}, {Liu}, \&
  {Zhao}}]{Zhang2013}
{Zhang}, L., {Rix}, H.-W., {van de Ven}, G., {et~al.} 2013, \apj, 772, 108

\end{thebibliography}


\end{document}